\documentclass[aps,amsmath,floats,floatfix, twocolumn,
superscriptaddress,nofootinbib,showpacs]{revtex4-1}

\usepackage[colorlinks, pdfborder={0 0 0}, plainpages=false]{hyperref}
\usepackage{graphicx}
\usepackage{xspace}
\usepackage[usenames,dvipsnames]{color}
\usepackage{amssymb}
\usepackage{mathrsfs}
\usepackage[normalem]{ulem} 
\usepackage{comment}

\DeclareMathOperator*{\argmax}{argmax}
\DeclareMathOperator*{\argmin}{argmin}

\newcommand{\Caltech}{\affiliation{Theoretical Astrophysics 350-17,
    California Institute of Technology, Pasadena, CA 91125, USA}}
\newcommand{\Cornell}{\affiliation{Cornell Center for Astrophysics 
    and Planetary Science, Cornell University, Ithaca, NY 14853, USA}}
\newcommand{\UmassD}{\affiliation{Mathematics Department, University of 
  Massachusetts Dartmouth, Dartmouth, MA 02747, USA}}
\newcommand{\LIGOLab}{\affiliation{LIGO Laboratory, California Institute of Technology, MS 100-36, Pasadena, California 91125, USA}}

\begin{document}

\title{
A Surrogate Model of Gravitational Waveforms from Numerical Relativity
Simulations of Precessing Binary Black Hole Mergers
}

\author{Jonathan Blackman} \Caltech
\author{Scott E. Field} \Cornell \UmassD
\author{Mark A. Scheel} \Caltech
\author{Chad R. Galley} \Caltech
\author{Daniel A. Hemberger} \Caltech
\author{Patricia Schmidt} \Caltech \LIGOLab
\author{Rory Smith} \Caltech \LIGOLab

\date{\today}

\begin{abstract}
We present the first
surrogate model for gravitational waveforms from the coalescence
of precessing binary black holes.
We call this surrogate model NRSur4d2s.
Our methodology significantly extends recently introduced reduced-order and 
surrogate modeling techniques, and is capable of directly modeling 
numerical relativity waveforms
without introducing phenomenological assumptions
or approximations to general relativity.
Motivated by GW150914, LIGO's first detection of gravitational waves from
merging black holes, the model is built from
a set of $276$ numerical
relativity (NR) simulations with mass ratios $q \leq 2$, dimensionless
spin magnitudes up to $0.8$, and the restriction that the initial spin
of the smaller black hole lies along the axis of orbital angular momentum.
It produces waveforms which begin $\sim 30$ gravitational wave cycles before merger and
continue through ringdown, and which contain the effects of precession
as well as all $\ell \in \{2, 3\}$ spin-weighted spherical-harmonic modes.
We perform cross-validation studies to compare the model to NR waveforms
\emph{not} used to build the model and find a better agreement
within the parameter range of the model than other,
state-of-the-art precessing waveform models,
with typical mismatches of $10^{-3}$.
We also construct a frequency domain surrogate model
(called NRSur4d2s\_FDROM)
which can be evaluated in $50\, \mathrm{ms}$ and is suitable for performing
parameter estimation studies on gravitational wave detections
similar to
GW150914.
\end{abstract}


\maketitle

\section{Introduction}
With two confident detections of gravitational waves (GWs) from
binary black hole (BBH) systems \cite{LIGOVirgo2016a,Abbott:2016nmj},
an exciting era of gravitational wave astronomy has begun.
Once a 
signal has been detected, the masses and spins of the black
holes (BHs), and their uncertainties, can be determined by comparing
the signal to waveforms predicted by general relativity (GR)
\cite{TheLIGOScientific:2016wfe}. Similarly, 
by comparing the signal to predictions,
tests of GR
can now be performed
in the regime of strong-field dynamics
with relativistic velocities~\cite{TheLIGOScientific:2016src}.

Parameter estimation and tests of GR
typically require the computation of
predicted gravitational waveforms
for a large set of different source parameters 
(e.g. black hole masses and spins).
A typical Bayesian parameter estimation analysis,
for example, evaluates millions of waveforms~\cite{Veitch:2015}.
Therefore, in
order to obtain reliable results on realistic timescales,
the GW model
must be fast to evaluate. 
Additionally, the waveform model must be accurate not only during
the weak-field perturbative binary inspiral, but also in the strong-field,
large-velocity regime. Otherwise the model may introduce
biases in parameter estimation
and inaccuracies in tests of GR.
Waveform accuracy will become increasingly important 
in future
GW measurements, because
higher signal-to-noise-ratio detections are anticipated as
detector technology improves.

Numerical relativity (NR) is now in a sufficiently mature state that
there are a number of codes~\cite{Pretorius2005a, Zlochower:2005bj, SpECwebsite,
einsteintoolkit, Husa2007, Bruegmann2006, Herrmann2007b}
capable of accurately
simulating the late inspiral, merger and ringdown of a BBH system, 
and the resulting GWs, even for
somewhat extreme spins~\cite{Scheel2014, Ruchlin:2014zva} and high mass
ratios~\cite{LoustoZlochower2010, husa2015frequency}.
While the resulting waveforms are quite accurate,
the simulations can take weeks or months, thereby
precluding them from being directly used in most data analysis studies.
Therefore, data analysis studies
currently use approximate NR-tuned waveform models that are fast to
evaluate~\cite{Ajith2009, Santamaria:2010yb, Hannam:2013oca, Khan:2015jqa,
    Husa:2015iqa, Taracchini:2013rva, Pan:2013rra, Bohe:2016gbl}.

For the analysis of GW150914~\cite{TheLIGOScientific:2016wfe, Abbott:2016izl}, 
the first GW detection by Advanced LIGO~\cite{aLIGO2}, waveform
models built within the effective-one-body
(EOB)~\cite{Damour:024009, Damour2009a, Taracchini:2013rva, Purrer:2015tud, Pan:2013rra, Bohe:2016gbl} and
the phenomenological
(Phenom)~\cite{Hannam:2013oca, Khan:2015jqa, Husa:2015}
frameworks were used~\cite{TheLIGOScientific:2016wfe, Abbott:2016izl}. 
All models necessarily introduce
some systematic error, however small, 
which 
are often 
quantified either by comparing to NR simulations
directly~\cite{Jani:2016wkt, Kumar:2015tha,
    Babak:2016tgq, Khan:2015jqa, Abbott:2016wiq} or
by performing
parameter estimation with many different waveform models 
and monitoring the discrepancies.
In the case of GW150914, 
the systematic error for the black hole masses
was estimated to be
smaller than the statistical uncertainty.
However, estimating a model's
systematic error in this way is complicated by the 
fact that the waveform models
make similar simplifications.
For example, the models ignore spin-weighted spherical-harmonic
(SWSH) modes with $\ell > 2$, which may be
significant since the signal's
power is dominated by the late inspiral and merger.
Recent studies continue to investigate this systematic parameter estimation bias
through the use of newer waveform models including additional physics
\cite{Abbott:2016izl} and by comparing to NR waveforms~\cite{Abbott:2016wiq}.

In this paper, we use a surrogate model, which we call NRSur4d2s,
to compute waveforms
approaching the accuracy of NR simulations.
A surrogate 
model~\cite{Blackman:2015pia,Field:2013cfa,Purrer:2014,Purrer:2015tud} 
is a way to substantially accelerate the evaluation of
a slower but accurate waveform model (in our case, NR), 
while largely retaining the accuracy
of the original model.  This is done
by through an expensive \emph{offline} stage, 
where we perform many accurate NR simulations for
different input parameter values 
and subsequently build and validate
the surrogate model on this set of simulations.
The waveforms from these simulations are then ``interpolated"
in parameter space in an inexpensive \emph{online} stage.
The resulting model can be used in place of performing additional NR
simulations.
Surrogates can be used to accelerate other analytical models, and
have been used to successfully speed up 
non-spinning EOB models with multiple SWSH
modes~\cite{Field:2013cfa},
and spin-aligned EOB models 
that include only
the $\ell=2$ modes~\cite{Purrer:2014,Purrer:2015tud}.
Most recently, surrogates have been used to speed up
non-spinning BBH waveforms from NR simulations
including $40$ SWSH modes~\cite{Blackman:2015pia}.

The surrogate model we develop here is based on NR simulations
using the Spectral Einstein Code (SpEC)~\cite{SpECwebsite,
Pfeiffer2003, Lovelace2008, Lindblom2006, Szilagyi:2009qz, Scheel2009,
Szilagyi:2014fna}.  It
extends previous NR surrogate models~\cite{Blackman:2015pia} to include
precessing binaries.
The number of NR simulations required to build a surrogate model
increases with parameter space size,
and NR simulations become more expensive as the mass
ratio and spin magnitudes grow.
To reduce the computational cost, we restrict
to a subspace of the full precessing parameter space.
The initial spin direction of the smaller black hole is restricted
to be parallel to the orbital angular momentum.
We also restrict the mass ratio of the black holes to
    $1 \leq q \leq 2$ and the dimensionless spin magnitudes
    to be at most $0.8$.
The duration of each NRSur4d2s
waveform is equal to that of the NR
simulations, which begin $4500M$ before merger, corresponding to $\sim 30$
gravitational wave cycles.

It has been shown that waveforms from precessing systems closely resemble
waveforms from non-precessing systems 
when viewed in a suitable non-inertial, coprecessing frame
\cite{Schmidt:2012rh, Pekowsky:2013ska}.
We use this relationship to simplify the construction of the
surrogate model by decomposing each precessing waveform
into a simpler waveform measured in a coprecessing frame
\cite{Schmidt2010, OShaughnessy2011, Boyle:2011gg}.
plus a time-dependent rotation
that characterizes the precession.  
Additional simplification is achieved by further decomposing
each waveform into a set of functions that are slowly varying in
parameter space and thus easier to model (cf. Fig.~\ref{fig:DecomposeData}).
The model is evaluated by ``interpolating'' these slowly-varying
functions to a desired point in parameter space, and then using the
interpolated functions to reconstruct the 
waveform in the inertial source frame of the binary.

The NRSur4d2s surrogate model just described produces a waveform in the
time domain, and takes approximately one second
to evaluate.
While this is much faster than computing a waveform using NR,
it is still too slow for many applications; furthermore many
LIGO analyses are more easily performed
in the frequency
domain rather than the time domain.  Therefore, we build a
second surrogate model in the frequency domain, called NRSur4d2s\_FDROM, using
NRSur4d2s as input.
NRSur4d2s\_FDROM does not employ complicated decompositions of its input
waveforms, so it requires significantly more waveforms to build (an
offline cost), but because of its simplicity it is significantly
faster, and can be evaluated in about $50\, \mathrm{ms}$.

We compute errors in both our time-domain and frequency-domain
surrogate models by comparing the resulting waveforms with selected
NR waveforms that were \emph{not} used to build the models; 
see Section~\ref{sec:Assessment}
for details.  While these errors are larger than the numerical
truncation error of the underlying NR simulations, we find that
the agreement between NR and our surrogate models is better than
that between NR and other precessing waveform models. The accuracy
of the surrogate models could be further improved by incorporating
additional NR waveforms.

Section~\ref{sec:surrogatemethod} describes the surrogate modeling
methods that have been used previously, and our modifications to
them for this work.
The NR simulations, as well as their parameters and
waveforms, are described in Section~\ref{sec:popul-train-set}.
Section~\ref{sec:decomposition} describes how the NR waveforms are
decomposed into simple pieces, and surrogate models for each
piece are built in Section~\ref{sec:DataSurrogate}.
The errors of NRSur4d2s
are analyzed and compared to
other waveform models in Section~\ref{sec:Assessment}.
Section~\ref{sec:freqSurrogates} describes the construction of
NRSur4d2s\_FDROM from NRSur4d2s,
which reduces the computational
cost by over an order
of magnitude without sacrificing accuracy.
Finally, Section~\ref{sec:discussion} summarizes this work
and discusses potential modifications and improvements.

\section{Surrogate modeling methods}
\label{sec:surrogatemethod}

Compared to previous 
work~\cite{Field:2013cfa,Purrer:2014,Cannon:2012gq,Cannon:2011rj,brown_sc_2013_13,Purrer:2015tud,Blackman:2015pia}, 
which focused on surrogates of analytical waveform models or on surrogates 
of simpler NR waveforms,
surrogate models of precessing numerical relativity (NR) waveforms pose 
a number of new, unique challenges. First, the
complicated waveform morphologies characteristic
of precessing systems~\cite{Apostolatos1994, Kidder:1995zr} suggest that
a substantially larger training set may
be necessary for these systems than for simpler cases considered previously.
On the other hand, NR waveforms 
require the solution of computationally intensive 
time-dependent partial differential equations; 
current hardware and binary black hole evolution codes
are capable of producing only roughly
${\cal O}(1\,,000)$ simulations in about a year.

In this section we outline our method for the construction of precessing NR
waveform surrogates, briefly summarizing existing techniques 
while focusing on solutions to the new challenges. 
A dimensionless, complex gravitational-wave
strain\footnote{More precisely, we work with the distance-independent
dimensionless strain $R h/M$, where
$R$ is the distance from the binary's center-of-mass
and $M$ is the total Christodoulou mass~\cite{Christodoulou71}
measured after
the initial burst of junk radiation~\cite{Aylott:2009ya} has passed.
In this paper we choose units so that $c=G=1$.}
\begin{align} \label{eq:strain}
h(t,\theta,\phi;\pmb{\lambda}) =  h_+(t,\theta,\phi;\pmb{\lambda}) - 
                                i h_\times (t,\theta,\phi;\pmb{\lambda}) \, ,
\end{align}
can be expressed in terms of its two fundamental polarizations $h_+$ and $h_\times$.
Here, $t$ denotes time, $\theta$ and $\phi$ are the polar and azimuthal angles
for the direction of gravitational wave propagation away from the source,
and $\pmb{\lambda}$ is
a set of parameters that characterize the waveform.
For concreteness, the
parameters $\pmb{\lambda}$ we will use in Sec.~\ref{sec:DataSurrogate}
will be the initial mass ratio and
spin vectors of the black holes, but the discussion in this section applies
to a general set of parameters.
Gravitational waveforms considered in this paper are parameterized
through their dependence on the initial data, and we shall focus on the
the five-dimensional subspace described in Sec.~\ref{sec:alignment}.

When numerically generating a waveform
by solving partial differential equations,
one solves an initial-boundary value problem for a fixed $\pmb{\lambda}$,
thereby generating
a waveform on a dense temporal grid. In this paper we seek to build an accurate
and fast-to-evaluate surrogate gravitational-wave strain model
$h_{\rm S}(t,\theta,\phi;\pmb{\lambda})$ by numerically solving the Einstein
equations
for judicious choices of $\pmb{\lambda}$. Surrogate evaluations require only
simple function evaluations, matrix-vector products and coordinate
transformations.
In Sec.~\ref{sec:freqSurrogates}
we also build a frequency-domain surrogate model, using our time-domain
surrogate model as input data, with the purpose of accelerating the
evaluation of model waveforms. Evaluation of
the frequency-domain model is
about $20$ times faster than the corresponding time-domain surrogate.
Except for Sec.~\ref{sec:freqSurrogates} our
discussion will focus exclusively on time-domain surrogates.

The complex gravitational-wave strain can be written in terms of
SWSHs
${}_{-2}Y_{\ell m} \left(\theta, \phi \right)$ via
\begin{align} \label{eq:strain_mode}
h(t,\theta,\phi;\pmb{\lambda}) = 
\sum_{\ell=2}^{\infty} \sum_{m=-\ell}^{\ell} h^{\ell m}(t;\pmb{\lambda}) {}_{-2}Y_{\ell m} \left(\theta, \phi \right) \, ,
\end{align}
where the sum includes
all SWSH modes $h^{\ell m}(t;\pmb{\lambda})$.
In many data analysis applications, however, one often requires only
the most dominant SWSH modes.
The NRSur4d2s surrogate model will include all $\ell \leq 3$ modes,
while our
assessment of the model's error will compare to
NR waveforms with all $\ell \leq 5$ modes.
Including modes in the NR waveforms which are not included in our model
ensures our error studies are sensitive to the effect of neglecting higher
order modes.
We find that including $\ell=4$ and $\ell=5$ modes in our model
does not significantly reduce the surrogate errors, but it increases
the evaluation cost of the model.
As seen in Table~\ref{tab:component_errs}, however,
neglecting all $\ell=3$ modes would significantly increase
the surrogate errors,
which is why we include $\ell \leq 3$ modes.
Other models with which we compare have $\ell=2$ modes only.
When comparing two waveforms with different available modes, missing modes are simply
treated as being zero.

\subsection{The basic surrogate modeling approach} \label{subsec:sur_basics}

\begin{figure}[h]
  \includegraphics[width=.6\columnwidth]{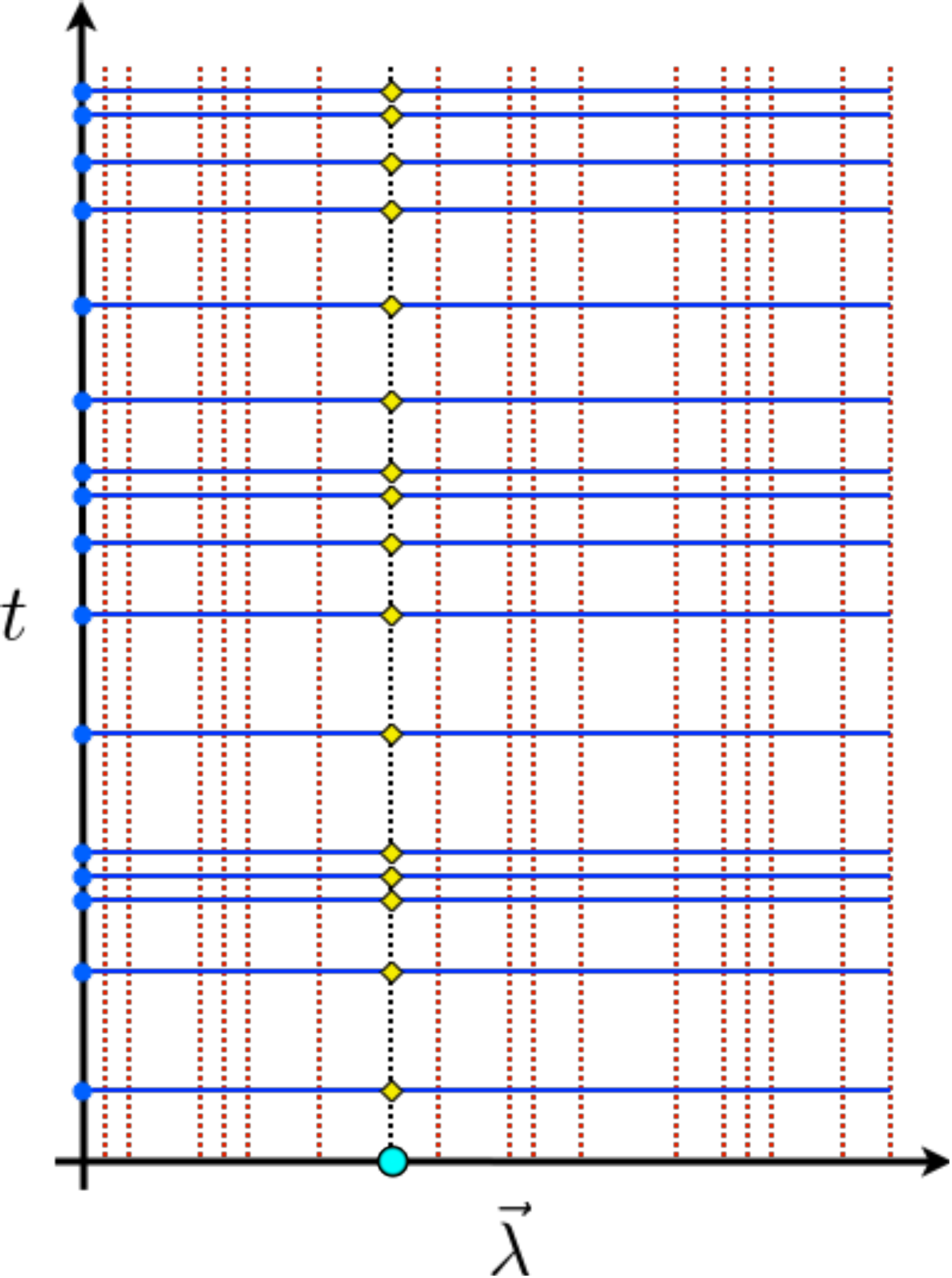}
  \caption{A schematic of the method for building a surrogate model for a
        function $X(t; \pmb{\lambda})$.
        The red dotted lines
        show $X(t)$
        evaluated at a selected set of greedy parameters
        $\pmb{\Lambda}_i$
        used to build a linear basis, and the blue dots show the associated
        empirical nodes in time from which
        $X_S(t; \pmb{\lambda})$
        can be reconstructed by interpolation with high accuracy.
        The blue lines indicate fits for
        $X(t; \pmb{\lambda})$ as a function of $\pmb{\lambda}$ at
        each of the empirical time nodes.
        The cyan dot shows a generic parameter $\pmb{\lambda}_0$
        that is not
        in the set of greedy parameters. To compute
        $X_S(t; \pmb{\lambda}_0)$, each fit is evaluated
        at $\pmb{\lambda}_0$ (the yellow diamonds), and then the
        empirical interpolant is used to construct
        $X_S(t; \pmb{\lambda}_0)$ at arbitrary times (the dotted black line).
        \label{fig:surrogate_alg}
    }
\end{figure}

\subsubsection{Problem statement}

Our surrogate modeling methods build on those outlined in~\cite{Field:2013cfa},
which we briefly describe here.
Consider a physical system parameterized by $\pmb{\lambda} \in \mathcal{T}$,
where $\mathcal{T}$ is a compact region in the space of possible
parameters.
We seek quick-to-evaluate time-dependent functions $X(t; \pmb{\lambda})$ that
describe this system for
times $t \in [t_\mathrm{min}, t_\mathrm{max}]$.
In our case, $\pmb{\lambda}$ will be the black hole masses and spins for
a single BBH system, and 
$\mathcal{T}$ will extend to some maximum spin magnitude
and maximum mass ratio for which we choose to compute NR waveforms.
The functions
$X(t; \pmb{\lambda})$ will be
obtained from decomposing $h^{\ell m}(t; \pmb{\lambda})$
as described in Sec.~\ref{sec:decomposition},
but here we discuss building a surrogate model for a single such function.

We already have a slow method of generating $X(t; \pmb{\lambda})$,
so we
seek a faster
\emph{surrogate model},
denoted as $X_S(t; \pmb{\lambda})$,
which approximates $X(t; \pmb{\lambda})$.
The surrogate model $X_S(t; \pmb{\lambda})$,
whose construction is summarized
in this section culminating in Eq.~\eqref{eq:surrogate_model_for_X}, 
is built to achieve small approximation errors
$\|X(\cdot; \pmb{\lambda}) - X_S(\cdot; \pmb{\lambda})\|$.
In our case, the slow method is performing a NR simulation, extracting
$h^{\ell, m}(t; \pmb{\lambda})$, and decomposing it 
to obtain
$X(t; \pmb{\lambda})$.
A solution $X(t; \pmb{\lambda})$ for a fixed $\pmb{\lambda}$ is represented
as a single (dotted red) vertical 
line in Fig.~\ref{fig:surrogate_alg}, which diagramatically
represents the surrogate model.


\subsubsection{Discovering representative binary configurations}

The first steps in building a surrogate model
are to determine a finite set of
{\em greedy parameters}
\[
G \equiv \{\pmb{\Lambda}_i \in \mathcal{T}\}_{i=1}^N \,.
\]
An NR simulation is then performed at each greedy parameter,
yielding the greedy solutions $\{X(t; \pmb{\Lambda}_i)\}_{i=1}^N$,
shown as vertical dotted red lines in Fig.~\ref{fig:surrogate_alg}.

One strategy
(described in more detail in \cite{Field:2013cfa})
to find the greedy parameters
begins by evaluating the slow method on a densely sampled
{\em training set}, $\mathcal{T}_\mathrm{TS} \subset \mathcal{T}$. 
This training set is 
input to a {\em greedy algorithm} (hence the name
greedy parameters) that works as follows.
First, the greedy algorithm is 
initialized by
arbitrarily selecting the first few greedy parameters 
which are sometimes called the algorithm's 
{\em seed}~\footnote{The final set of greedy parameters
selected by the greedy
algorithm will depend on that choice of seed. However, 
the number and distribution of greedy parameters 
is expected to be robust to the choice 
of seed~\cite{Field:2011mf,Caudill2012}.}.
The set of greedy parameters
is then extended iteratively by first
building an orthonormal linear
basis $B_n = \{e^i(t)\}_{i=1}^n$ spanning the $n$ current greedy solutions,
such that
\begin{equation}
X(t; \pmb{\Lambda}_j) = \sum_{i=1}^n c_i(\pmb{\Lambda}_j) e^i(t)\,.
\end{equation}
The aim of the greedy algorithm is to extend this basis such that
the approximation
\begin{equation}
X(t; \pmb{\lambda}) \approx \sum_{i=1}^n c_i(\pmb{\lambda}) e^i(t) \,, 
\quad \pmb{\lambda} \in \mathcal{T}_\mathrm{TS}
\label{eq:basis_expansion}
\end{equation}
is as accurate as possible and where the coefficient
$c_i(\pmb{\lambda})$ is the inner product of $X(t; \pmb{\lambda})$ with
$e^i(t)$.
Coefficients found in this way define an {\em orthogonal projection}
of the function $X(t; \pmb{\lambda})$ onto the span of the basis.
We compute the projection errors
\begin{equation}
E_n\left({\pmb{\lambda}}\right) = \|X(\cdot; \pmb{\lambda}) -
                    \sum_{i=1}^n c_i(\pmb{\lambda}) e^i(\cdot)\|
\label{eq:proj_errors}
\end{equation}
for each $\pmb{\lambda} \in \mathcal{T}_\mathrm{TS}$, and the
next greedy parameter $\pmb{\Lambda}_{n+1}$
is chosen to be the one yielding the largest projection error.
The next basis vector $e^{n+1}(t)$ is then obtained by orthonormalizing
$X(t; \pmb{\Lambda}_{n+1})$ against $B_n$,
and the basis set is extended as
$B_{n+1} = B_n \cup \{e^{n+1}(t)\}$.
The algorithm terminates once the basis 
achieves an accuracy requirement $E_N(\pmb{\lambda}) \leq \epsilon$,
for some predetermined error tolerance $\epsilon$,
over the whole training set.
With a dense enough training set and assuming $X$ varies
smoothly over $\mathcal{T}$, the projection errors outside of the
training set will be only mildly larger than $\epsilon$.

This method unfortunately requires evaluating the slow
method on each point in the (large) training set, so we make modifications
as described in Secs.~\ref{subsec:ts_sampling} and \ref{sec:greedyselection}.

\subsubsection{Temporal compression}

We have built a linear basis $B_N$ which can represent $X(t; \pmb{\lambda})$
for any $\pmb{\lambda} \in \mathcal{T}$ using Eq.~\ref{eq:basis_expansion},
up to some small projection error.
This reduces the problem of determining $X(t; \pmb{\lambda})$ to determining
the basis coefficients $\{c_i(\pmb{\lambda})\}_{i=1}^N$.
The most straightforward method of doing so would be to fit or interpolate
the basis coefficients $c_i$ over the parameter space $\mathcal{T}$
as is done in \cite{Purrer:2014,Purrer:2015tud}.
We have more intuition for the behavior over $\mathcal{T}$ of the solutions
$X(T; \cdot)$ evaluated at a fixed
time $T$ than we do for the basis coefficients.
We will therefore pursue an {\em empirical interpolation} approach,
described in
detail in \cite{Field:2013cfa}, which enables us to avoid fitting
the basis coefficients.

An empirical interpolant makes use of the orthogonal linear basis
$B_N = \{e^i(t)\}_{i=1}^N$ such that the errors given by
Eq.~\eqref{eq:proj_errors} are small, so
Eq.~\eqref{eq:basis_expansion} continues to provide 
a good approximation despite using a different
method to compute the coefficients.
During the construction of the empirical interpolant,
$N$ empirical time nodes $\{T_j\}_{j=1}^N$ will be used.
An algorithm to find these special time nodes will be described
later on.

We denote an $N$-node empirical interpolant of a function $f(t)$
by $I_N[f](t)$.
A conceptually helpful way to think of the empirical interpolant is that
$I_N[f](t)$ lies in the span of $B_N$,
passes through $f(T_j)$ at time $T_j$,
and is nearly as accurate as the orthogonal projection.
To construct the interpolant, we expand it in terms
of unknown coefficients $c_i$,
\begin{equation}
I_N[f](t) = \sum_{i=1}^N c_i e^i(t) \,.
\label{eq:f_basis_expansion}
\end{equation}
We then write a linear system of equations
\begin{equation}
\sum_{i=1}^N c_i e^i(T_j) = f(T_j)\,, \quad j=1,\dots, N
\label{eq:n_eq_n_unknowns}
\end{equation}
and we solve this system for all the coefficients
$c_i$.
A good choice of empirical time nodes will ensure that the matrix
$V_{ij} = e^i(T_j)$ is well-conditioned, 
thereby allowing an accurate solution
\begin{equation}
c_i = (V^{-1})_{ij} f(T_j)\,.
\end{equation}
We can then substitute the coefficients back
into Eq.~\eqref{eq:f_basis_expansion} to obtain
\begin{equation}
I_N[f](t) = \sum_{i=1}^N (V^{-1})_{ij} f(T_j) e^i(t)\,.
\end{equation}
If we then define
\begin{equation}
  b^j(t) = \sum_{i=1}^N (V^{-1})_{ij} e^i(t)\,,
  \label{eq:enjexpression}
\end{equation}
we obtain
\begin{equation}
I_N[f](t) = \sum_{j=1}^N f(T_j) b^j(t)\,.
\label{eq:eim_form}
\end{equation}
Here $b_N^j(t)$ is computed before
evaluating the surrogate, so evaluating the empirical
interpolant amounts to a matrix multiplication.

If $f(t)$ lies in the span of $B_N$, then $I_N[f](t) = f(t)$ for all times $t$.
Otherwise, there will be some interpolation error.
In practice, the
empirical time nodes are constructed iteratively using
bases $B_n$ for $n=1,\dots,N$. If $I_n$ is the $n$th iteration of
the interpolant, then the $n$th empirical time node $T_n$ is chosen
to be the time $t$ yielding the largest interpolation error when
interpolating $e^n(t)$ using the previous interpolant $I_{n-1}$.
The iteration begins with the initial interpolant chosen to
be $I_0[f](t) = 0$ for all $f$.

Note that since the empirical interpolant is linear
and $V$ is well-conditioned, if $f(t)$
has a deviation from the span of $B_N$ of order $\epsilon$,
then the empirical interpolation error will also be of order $\epsilon$.
Since our basis $B_N$ is constructed such that the projection errors
of $X(t; \pmb{\lambda})$ onto $B_N$ are small for all $\lambda \in
\mathcal{T}$, we can use the empirical interpolant $I_N[X](t)$
to obtain $X(t; \pmb{\lambda})$
for all times $t$ given the empirical node values
$\{X(T_j; \pmb{\lambda})\}_{j=1}^N$.
The remaining step is then to
approximate the $N$ functions
\begin{equation}
X_j(\pmb{\lambda}) = X(T_j; \pmb{\lambda})
\end{equation}
by fitting the available data
$\{X(T_j; \pmb{\Lambda}) : \pmb{\Lambda} \in G\}$ over the parameter space
$\mathcal{T}$.
We call these {\em parametric fits}, and denote the fitted approximation
for $X_j(\pmb{\lambda})$ by $X_{jS}(\pmb{\lambda})$.
The parametric fits are represented by the blue horizontal lines in
Fig.~\ref{fig:surrogate_alg}.
The explicit form of our {\it surrogate model for $X$} is then given by
\begin{equation}
X_S(t; \pmb{\lambda}) = \sum_{j=1}^N X_{jS}(\pmb{\lambda}) b^j(t)\,.
\label{eq:surrogate_model_for_X}
\end{equation}

\subsection{Modifications to the basic surrogate modeling approach}
\label{subsec:ts_sampling}

\begin{figure*}
    \centering
    \def\svgwidth{2.0 \columnwidth}
    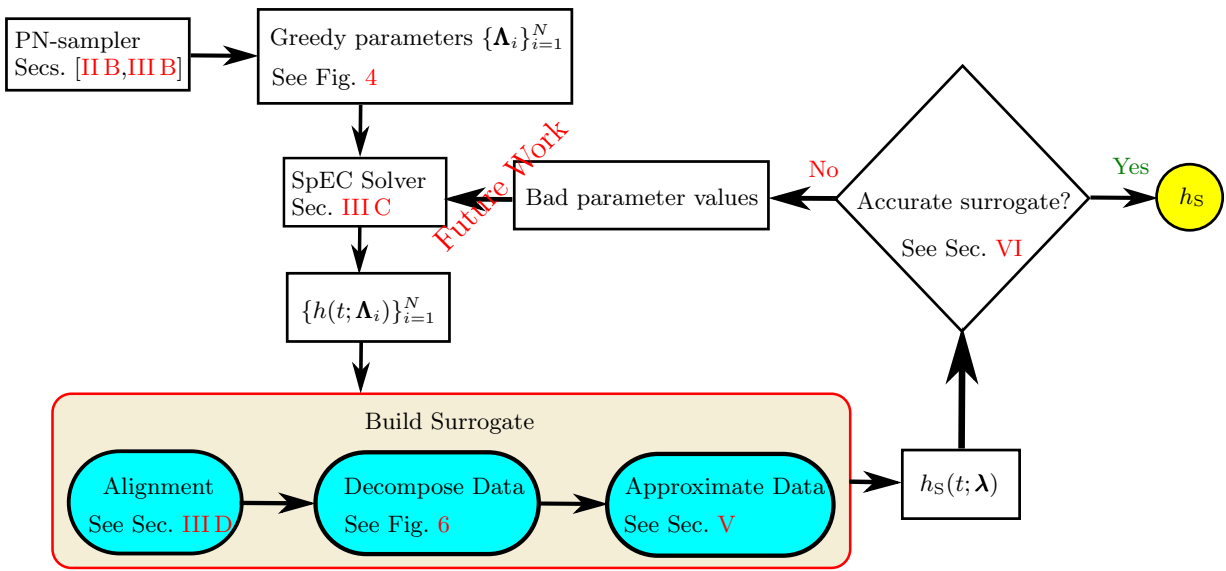
    \caption{Surrogate workflow. A greedy ``PN-sampler" selects the most informative parameter values $\{\pmb{\Lambda}_i\}_{i=1}^N$
                 for a fixed parametric and temporal range. For each
                 selected value $\pmb{\Lambda}_i$, SpEC generates a gravitational waveform. A surrogate model building
                 algorithm (cf.~Fig.~\ref{fig:surrogate_alg}) is applied to a
                 set of suitably aligned and decomposed (cf.~Fig.~\ref{fig:DecomposeData}) numerical relativity waveforms thereby producing a trial
                 surrogate. A handful of validation tests are performed to assess the surrogate's quality. If
                 the surrogate performs poorly for some parameter values,
                 one could produce additional numerical relativity waveforms
                 near those values, and rebuild a more accurate surrogate.
                 \label{fig:surrogate_work_alg}
}
\end{figure*}

A drawback of the algorithm presented in \S~\ref{subsec:sur_basics},
and of many previous surrogate modeling efforts, is the assumption
that the original slow model can be evaluated an arbitrary number
of times to build a dense training set.  Because of the significant computational
expense, this is not feasible for
waveforms found by numerically
solving the Einstein equations.
We can neither build NR surrogate models from dense
training data nor can we assess the surrogate's quality at
arbitrarily many randomly chosen validation points.
In previous work that used computationally inexpensive
waveform models~\cite{Field:2013cfa}, thousands of nonspinning 
waveforms comprised the training set, yet the final surrogate required only
a very small subset of greedy parameters $G$. 
If we could have predicted $G$ in advance
then dense training sets would not be required.

Since we cannot evaluate an arbitrarily large number of NR
waveforms, we instead first construct a temporary
{\em mock surrogate}
using a simpler waveform
model that is both fast to evaluate 
and is defined in the training region of interest. 
In this paper, for the purpose of discovering the most
relevant parameter values, we build a mock surrogate using the
precessing TaylorT4 post-Newtonian
(PN) waveform model as implemented in GWFrames~\cite{Boyle:2013nka,gwframes}.
We determine the PN greedy parameters $G^\mathrm{PN}$ using a
training set containing many thousands
of these PN waveforms, as described in Sec.~\ref{sec:greedyselection}.
If we then assume that
the distribution of parameters selected using PN waveforms
roughly mimics the distribution we would have
obtained had NR waveforms been available,
then $G^{\mathrm{PN}}$ should be a suitable set of greedy parameters
for building a NR surrogate.
This was found to work well for the non-spinning surrogates of
Ref.~\cite{Blackman:2014maa} and,
as judged by our validation studies, continues to remain
applicable to the precessing waveforms considered here.
Instead of PN, we could have used a different analytical
  waveform model~\cite{Damour:024009, Damour2009a, Taracchini:2013rva, Purrer:2015tud, Pan:2013rra, Bohe:2016gbl, Hannam:2013oca, Khan:2015jqa, Husa:2015}
  that contains merger and ringdown.  However, these other models either
  omit $\ell=3$ modes, omit precession, or yield waveforms that do
  not vary smoothly as a function of $\pmb{\lambda}$.  We find that these
  other considerations outweigh the inclusion of merger and ringdown.

This entire process just described is shown in
the first stage of the surrogate
workflow diagram~(Figure~\ref{fig:surrogate_work_alg})
as the ``PN-sampler".
Once the points $G^\mathrm{PN}$ have been selected, 
the corresponding NR waveforms are generated, and
the surrogate building 
proceeds as in Fig.~(\ref{fig:surrogate_work_alg}). 
We emphasize that \emph{no} PN waveforms are used to build the
resulting NRSur4d2s surrogate;
the PN model is used 
only to find the greedy parameters $G=G^\mathrm{PN}$.

While the PN waveforms are much cheaper to evaluate than NR waveforms,
building a dense training set remains
costly for high dimensional
parameter spaces.
In Ref.~\cite{Blackman:2014maa}, it was found that
an accurate basis can be achieved using small, sparse training sets
if each iteration $i$ of the greedy algorithm uses an independent
randomly-sampled training set $\mathcal{T}_\mathrm{R}^i$.
We extend this methodology by also including in our training sets a fixed
set of parameters $\mathcal{T}_\mathrm{B}$ on the boundary of
$\mathcal{T}$
(for example, the maximum mass ratio allowed
in $\mathcal{T}$).
This is motivated by the fact that the boundary of $\mathcal{T}$ carries
significant weight both when building a linear basis and when performing
parametric fits. At the $i$th greedy iteration, we then have
\begin{equation}
  \mathcal{T}^i_\mathrm{TS} = \mathcal{T}_\mathrm{B} \cup \mathcal{T}_\mathrm{R}^i
  \label{eq:TrainingSet}
\end{equation}
as our training set of parameters at which we evaluate PN waveforms.

Another issue with the standard greedy algorithm
is that it considers only a single function $X$.
For modeling waveforms, we will decompose each
waveform $h^{\ell m}(t; \pmb{\lambda})$ into many
such functions,
  which we call {\it waveform data pieces}
(cf. Sec.~\ref{sec:decomposition}).
Rather than generate a separate set of greedy parameters $G_X^{PN}$ for each
$X$, we construct a single set of greedy parameters $G^{PN}$ that can be used
for all waveform data pieces $X$.
We do so by replacing the projection errors for a single waveform data
piece given in Eq.~\eqref{eq:proj_errors} with a single error including
contributions from all waveform data pieces. This
will be described explicitly
in Sec.~\ref{sec:greedyselection} after the waveform decomposition
and error measures have been introduced.

The standard greedy algorithm guarantees
that the basis yields small projection errors given by
Eq.~\ref{eq:proj_errors}.
Therefore, if we have perfect parametric fits (so that
$X_{jS}(\pmb{\lambda}) = X_j(\pmb{\lambda})$ for all
$\pmb{\lambda} \in \mathcal{T}$) then the surrogate model $X_S$
given by Eq.~\ref{eq:surrogate_model_for_X} will
agree with $X$ in the sense that the $L^2$ norm
of $X_S(t; \pmb{\lambda})-X(t; \pmb{\lambda})$
will be small for all $\pmb{\lambda} \in \mathcal{T}$.
There is, however, no corresponding
guarantee that the
greedy points are sufficient for producing accurate parametric
fits $X_{jS}$.
In the one-dimensional models built in
Refs.~\cite{Field:2013cfa, Blackman:2015pia},
the parametric fits performed well using the samples produced
from the standard greedy algorithm.
As the dimensionality of the parameter space increases,
the number of greedy parameters required for an accurate basis
grows slowly~\cite{Herrmann:2012if}, but
the number of samples required
for accurate parametric fits can grow rapidly.
We therefore anticipate that the standard greedy algorithm
alone may lead to underresolved parametric fits in problems
with high dimensionality.

We overcome this problem
by first performing a greedy algorithm
to obtain greedy parameters $G^\mathrm{PN}_0$ that
ensure small basis projection errors,
and then performing a second greedy algorithm, seeded with $G^\mathrm{PN}_0$,
that produces the final set of PN greedy parameters $G^\mathrm{PN}$.
In each iteration of the second greedy algorithm, a
mock PN surrogate
is constructed from PN waveforms evaluated at the current set
of greedy parameters, including the parametric fits
at each empirical node.
To select the next greedy parameter in this second greedy algorithm,
for each $\pmb{\lambda} \in \mathcal{T}_{\rm{TS}}^i$
we compute an error between a PN waveform evaluated at $\pmb{\lambda}$
and the mock-PN surrogate evaluation at $\pmb{\lambda}$.
Since the basis is already
accurate and in general $\pmb{\lambda}$ will not have already been
selected as a greedy parameter, this procedure
selects points for which the parametric fits are
underresolved.

\subsection{Handling noise in the NR waveforms}

The presence of numerical
noise in the input NR waveforms
complicates the construction of surrogates.
The situation is simpler when building a
surrogate of a waveform model that is mostly noise-free, such as
post-Newtonian or EOB models that require the
solution of 
ordinary differential equations 
(which can be evaluated to almost arbitrary
accuracy) but not PDEs. For example, 
Ref.~\cite{Field:2013cfa} demonstrates in their Fig.~15 that 
EOB surrogates can be made to have arbitrarily small errors, and
Refs.~\cite{Purrer:2014,Purrer:2015tud} use interpolation across 
the parameter space without needing to avoid potential
pitfalls such as overfitting the noise.
We do not expect this to be the case for numerical relativity waveforms 
which are beset by numerous error sources, some of which cannot be made 
arbitrarily small with current computing technology. 

Systematic as well as numerical errors can influence the quality of
the NR waveform. 
For example, when attempting to model non-eccentric binaries, the
NR simulations will always have some small but non-zero orbital eccentricity.
In this paper
we will mostly focus on numerical truncation error.
This is typically the dominant source of error in SpEC
waveforms~\cite{Hemberger:2017},
and the other sources of error are expected to
be significantly smaller than truncation error, and smaller
than the surrogate error (see Fig. 3 of Ref.~\cite{Blackman:2015pia}).
The numerical error can be quantified through standard convergence
tests~\cite{Hemberger:2017}. Following Ref.~\cite{Blackman:2015pia}, 
we will (i) characterize SpEC waveform error across the parameter space
and, if necessary, remove
poorly-resolved waveforms (Sec.~\ref{subsec:omitBadData})
(ii) avoid overfitting the noise sources (App.~\ref{sec:fitappendix}), 
and (iii) set surrogate accuracy goals based on our answer
to the first question.
In future work it would be interesting to study the impact
of other noise sources. 

\subsection{Decomposing NR waveforms into simpler components}
\label{subsec:surrogate_techniques}

The detailed time dependence of an NR waveform is generally too
complicated to model directly with an acceptable degree of accuracy.
Instead, each NR waveform is decomposed into {\em waveform data pieces}
(cf. Sec.~\ref{sec:decomposition}), which are simpler,
more slowly-varying functions that can be modeled more easily.
A surrogate model is then built
for each waveform data piece (cf Sec.~\ref{sec:DataSurrogate}), and
then these models are recombined to produce a full surrogate waveform.
This process is shown in the ``Surrogate Build" step shown 
in Fig.~\ref{fig:surrogate_work_alg}.

Selecting the waveform data pieces is a critically important step.
For example, in nonspinning~\cite{Field:2013cfa,Blackman:2015pia} and 
spin-aligned~\cite{Purrer:2014,Purrer:2015tud} surrogate models,
the waveform data pieces are either the real and imaginary parts of the 
SWSH modes, $h^{\ell m}$, or the amplitude and phase
decompositions of these 
modes $A_{\ell m}$ and $\phi_{\ell m}$, where
$h^{\ell m} = A_{\ell m} \exp\left(-\mathrm{i}\phi_{\ell m}\right)$.
The idea is that it is easier to model every
$A_{\ell m}$ and $\phi_{\ell m}$, which are smooth and slowly-varying
functions of time, than it is to directly model the complicated waveform
$h(t,\theta,\phi;\pmb{\lambda})$, Eq.~(\ref{eq:strain_mode}).

Because of the complexity of precessing waveforms, 
we have needed to pursue a somewhat more complicated decomposition
scheme than in the nonprecessing case.
Fig.~\ref{fig:DecomposeData} summarizes the decomposition scheme used here.
Briefly, each waveform is 
transformed into a coordinate frame in which the binary is
not precessing
\cite{Schmidt2010, OShaughnessy2011, Boyle:2011gg, Boyle:2013nka, Boyle:2014};
specifically, we use the minimal-rotation
coprecessing frame of Boyle \cite{Boyle:2011gg}.
The waveform
modes in this frame have a simpler structure than their inertial frame
counterparts. Additional simplifications occur by applying further
transformations (described in detail in Sec.~\ref{sec:decomposition})
to the coprecessing-frame waveform modes.
The result of these steps is a set of waveform data pieces.  If 
$X(t, \pmb{\lambda})$ is a single waveform data piece,
then for that piece we build a surrogate
$X_{\rm S}(t,\pmb{\lambda}) \approx X(t;\pmb{\lambda})$. 
Here $X$ can stand for any of the decomposed waveform data pieces depicted 
as cyan ellipses in Fig.~\ref{fig:DecomposeData},
for example $A^{22}_{+}$, $\phi^{32}_{-}$,
$\varphi_p$, etc. The full NRSur4d2s surrogate waveform model is defined
by the individual data piece surrogates, $X_{\rm S}(t,\pmb{\lambda})$,
and the inverse transformations required to move back up the 
data decomposition diagram (Fig.~\ref{fig:DecomposeData}) and
reconstruct the waveform from all of the $X_{\rm S}(t,\pmb{\lambda})$.

\subsection{Tools for surrogate model validation}

Here we describe a useful framework for assessing the surrogate's 
predictive quality when only a limited number of waveforms are available.
This is a different setting from the EOB surrogates
of~\cite{Purrer:2014,Purrer:2015tud}
where \emph{out-of-sample} validation studies could be performed at arbitrarily
many parameter values.
The primary 
tool we shall use is {\em cross-validation}~\cite{Hastie2001}, which was also
used in~\cite{Blackman:2015pia}.
Cross-validation happens after the surrogate is built and 
determines whether or not more SpEC waveforms are needed to 
improve the accuracy of
the model (see Fig.~\ref{fig:surrogate_work_alg}).

We consider the case where our full dataset is composed of $N$ SpEC waveforms. 
From the full dataset, we select non-intersecting sets of trial and verification
waveforms with sizes $N_{\rm t}$ and $N_{\rm v}$,
such that $N_{\rm t} + N_{\rm v} \leq N$.
In the cross-validation step, 
a new {\em trial surrogate} is built solely from $N_{\rm t}$ trial waveforms. 
The remaining $N_{\rm v}$ verification waveforms serve as an {\em exact and independent} 
error measure of the {\em trial} surrogate's prediction.
The key assumption, which we 
believe to be true in practice, is that the surrogate built from all $N$ waveforms 
will have an accuracy similar to the trial surrogates, if not better.
Indeed, each step of the 
surrogate building algorithm will be more accurate so long as parametric 
overfitting is kept under control.
Hence, the trial surrogate's error should                                    
serve as a useful estimate
of the error associated with the full surrogate 
built from all $N$ waveforms.
We note, however, that when $N_{\rm v}$ is small or the surrogate
error is dominated by some systematic source of error,
the improved accuracy when including all $N$ waveforms may not be
enough to overcome the variance in the accuracy of the parametric fits
seen in Fig.~\ref{fig:pn_greedy}.
In that case the full surrogate error may in fact be slightly larger than a
trial surrogate error.

Two variants of cross-validation are considered. {\em Random} cross-validation 
proceeds by selecting the verification waveform set randomly. When $N_{\rm v} = 1$ 
this is known as the {\em leave-one-out} strategy. 
In Ref.~\cite{Blackman:2015pia}, all possible leave-one-out studies were performed.
In our case, $N$ is sufficiently large and surrogate-building is sufficiently
expensive that we opt to choose $N_{\rm v} = 10$.
We can perform many resamplings of the validation subset to infer
an error profile across the parameter space.

{\em Deterministic}
cross-validation proceeds by selecting the verification waveforms according to 
a rule. For example, the greedy bases are already ranked according to a 
``most important" criterion. We select the first $N_{\rm t}$ greedy waveforms
for our validation set. These should contribute most heavily to the surrogate's 
overall predictive ability, while the last $N_{\rm v}$ verification 
ones are quite dissimilar from the trial waveforms due to the greedy selection
process.
We fix $N_{\rm v}$ to have a consistent test of our trial surrogates, and
vary $N_{\rm t} \leq N - N_{\rm v}$ to estimate how the surrogate errors depend on $N$.

\subsection{Waveform error measurements} \label{subsec:err_def}

This subsection summarizes the most commonly used tools
to compare waveforms. A typical scenario is to quantify differences 
between waveforms, for example to compare 
a waveform model to NR waveforms or to estimate
the numerical truncation error associated with 
an NR waveform.

Let $h_1(t, \theta, \phi; \pmb{\lambda}_1)$ and 
$h_2(t, \theta, \phi; \pmb{\lambda}_2)$ denote waveforms from
two different models
(or two NR simulations with different numerical resolution)
potentially evaluated at different parameter 
values $\pmb{\lambda}_1$ and $\pmb{\lambda}_2$.
We assume the waveforms are already aligned according to the procedure
of Sec.~\ref{sec:alignment}.
Decomposing these waveforms 
into SWSHs we compute a time-dependent error 
\begin{equation}
\delta h(t) = \sqrt{ \sum_{\ell, m} \left| \delta h^{\ell m}(t) \right|^2} \,,
\end{equation}
from the individual mode differences
\begin{equation}
\delta h^{\ell m}(t) = h^{\ell m}_1(t; \pmb{\lambda}_1) - h^{\ell m}_2(t; \pmb{\lambda}_2) \,.
\label{eq:delta_h_of_t}
\end{equation}
We use the time-domain inner product
\begin{equation}
\langle a, b \rangle_{t} = \frac{1}{T}\int_{t_\mathrm{min}}^{t_\mathrm{max}}
        a(t) b^*(t) dt \,,
\end{equation}
between any complex functions of time $a$ and $b$,
where $T = t_\mathrm{max} - t_\mathrm{min}$ and $*$ denotes complex conjugation.
The associated norm $\| a \|^2 = \langle a, a\rangle_t$ can be used to compute
mean-squared errors,
and we compute the full time-domain waveform error
\begin{align}
\left(\delta h\right)^2 
& = \int_{S^2} \| h_1(t, \theta, \phi; \pmb{\lambda}_1) - 
                h_2(t, \theta, \phi; \pmb{\lambda}_2) \|^2 d\Omega\\
&= \sum_{\ell m} \| \delta h^{\ell, m} \|^2 \label{eq:deltahfromModes}\\
&= \frac{1}{T} \int_{t_\mathrm{min}}^{t_\mathrm{max}} \delta h(t)^2 dt
\end{align}
as a sum over individual mode errors $\| \delta h^{\ell, m} \|$.
We note that we do not perform any time or phase shifts to minimize
this error.
Since waveforms with different mass ratios and spins will have different
norms, the error we will use most often is defined as
\begin{equation}
\mathcal{E} [h_1, h_2] = \frac{1}{2}\frac{\delta h^2}{\|h_1\|^2}
\label{eq:cal_e}
\end{equation}
where $h_1$ is taken to be the more trusted waveform (usually the highest
resolution NR waveform).
The factor of $1/2$ is motivated in Appendix~\ref{app:TFErrors} and makes
$\mathcal{E}$ similar to a weighted average over the sphere of overlap
errors between $h_1(t, \theta, \phi; \pmb{\lambda}_1)$ and
$h_2(t, \theta, \phi; \pmb{\lambda}_2)$, where the overlap error
is $1 - \mathcal{O}$ with
\begin{equation}
\mathcal{O} = \frac{\langle h_1, h_2 \rangle}{
        \sqrt{\langle h_1, h_1 \rangle
          \langle h_2, h_2 \rangle}}\,.
\label{eq:Overlap}
\end{equation}
We note, however, that while the overlap error vanishes if $h_1$
and $h_2$ are identical up to a constant factor, $\mathcal{E}$ does not
and vanishes only when $h_1$ and $h_2$ are identical.
This is important as a different normalization will lead to a bias
when measuring the distance to the source of a gravitational wave.

Overlap errors are often computed in the frequency domain with a noise-weighted
inner product~\cite{Cutler:1994ys}
\begin{equation}
\label{eq:match}
\langle a, b \rangle_{f} = 4 \mathrm{Re} \int_{f_\mathrm{min}}^{f_\mathrm{max}}
        \frac{\tilde{a}(f) \tilde{b}^*(f)}{S_n(f)} df,
\end{equation}
where $S_n(f)$ is the noise power spectral density of a gravitational wave detector
and tildes are used to represent a Fourier transform.
We define the {\em mismatch} as the overlap error, 
  $1 - \mathcal{O}$, minimized over one or more
  extrinsic parameters such as an overall time shift.

\section{Populating the Set of NR Waveforms}
\label{sec:popul-train-set}

\subsection{Parameter space} \label{sec:paramspace}

Non-eccentric BBH systems are parametrized by the mass ratio $q = m_1/m_2 \geq 1$
as well as the two dimensionless BH spin vectors $\vec{\chi}_1$, $\vec{\chi}_2$.
The total mass $M = m_1 + m_2$ scales out of the problem, and can be used to
restore appropriate dimensions to times and distances.
Because the spin vectors precess and are therefore time-dependent,
to use them as parameters one must specify them at a particular time
or frequency.
We choose to specify parameters at a reference time of
$t_0=t_{\rm peak}-4500M$, where
  $t_{\rm peak}$ is the time at which the quadrature sum of the waveform modes,
\begin{equation}
\label{eq:peak}
A_\mathrm{tot}(t) = \sqrt{\sum_{\ell, m} |h^{\ell, m}(t)|^2},
\end{equation}
reaches
its maximum value.

\begin{figure}
  \includegraphics[width=\linewidth]{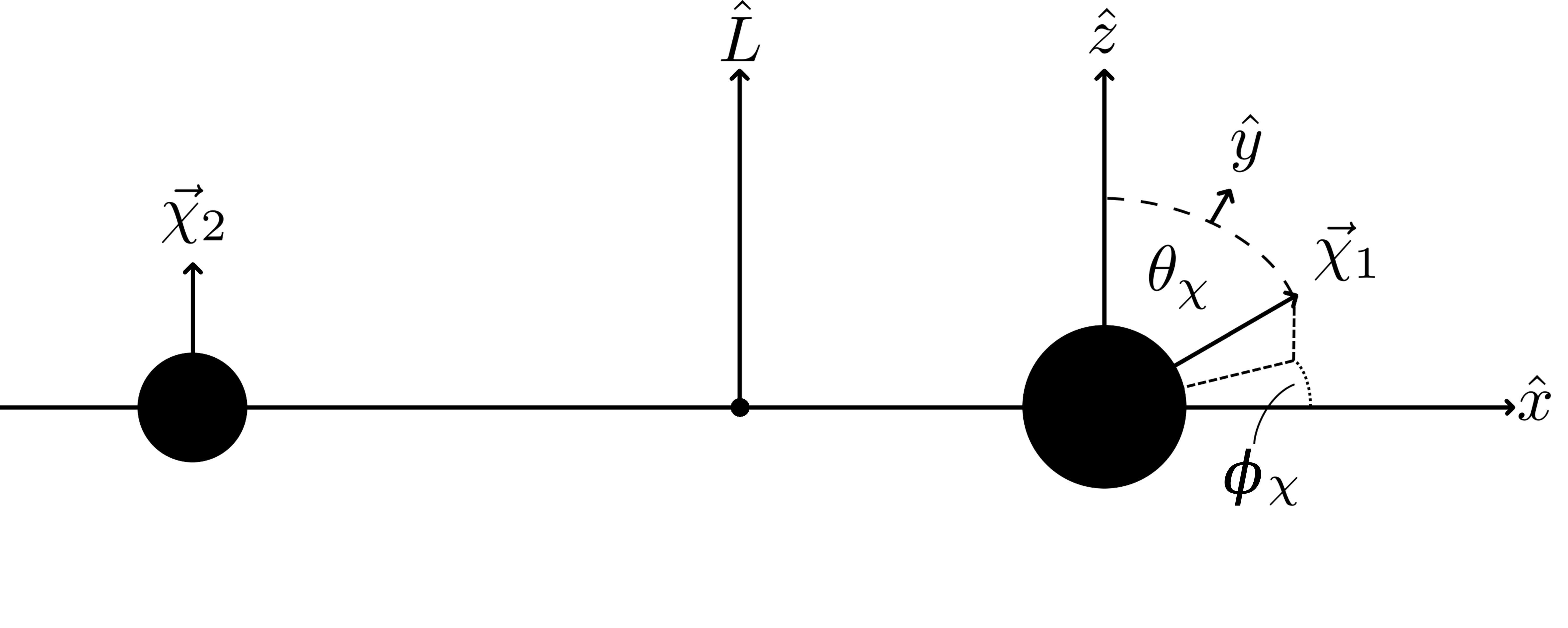}
  \caption{Diagram of the $4$ spin components in the $5d$ parameter
    subspace. We attempt to obtain $\phi_\chi = 0$ at $t_0=$
    $4500M$
    before peak amplitude,
    but in practice the NR simulations have
    arbitrary values of $\phi_\chi$.
    }
  \label{fig:diagram}
\end{figure}

We 
restrict to a $5d$ subspace of the parameter space where $\vec{\chi}_2$ is
aligned with the Newtonian orbital angular momentum $\hat{L}_N$ at the
reference time (in practice the NR simulations give us
small misalignments but ignore them;
see Sec. \ref{sec:parameterization}).
Let $\theta_\chi$ and $\phi_\chi$ be the polar and azimuthal angles
of $\vec{\chi}_1$ at the reference time.  Then our 5 parameters
are $q$, $|\vec{\chi}_1|$, $\chi_2^z$, $\theta_\chi$, and $\phi_\chi$
(see Fig.~\ref{fig:diagram}).
While NR simulations can be done for nearly extremal spins~\cite{Scheel2014}
and large mass ratios \cite{LoustoZlochower2010}, they are computationally
expensive and so we restrict to $|\vec{\chi}_1| \leq 0.8$,
$|\chi_2^z| \leq 0.8$ and $q \leq 2$.
These bounds were also motivated by the parameters of GW150914, which was
close to equal mass and did not show strong evidence of large spin magnitudes
\cite{TheLIGOScientific:2016wfe}.

To further simplify the surrogate, we attempted to reduce the
parameter subspace from 5d to 4d by restricting $\phi_\chi=0$.
While this can be done for analytic waveforms (PN, EOB, etc), it is
problematic for NR waveforms.  This is because it
is not possible to accurately predict the amount of
time between the start of an NR simulation and the peak of
$A_\mathrm{tot}(t)$, without having carried out the simulation.
Therefore, it is not possible to precisely
set initial conditions of the simulation so that
$\phi_\chi = 0$ at the reference time.  Therefore, our NR simulations
actually cover a $5d$ and not a $4d$ subspace of the
parameter space, and we must include $\phi_\chi$ as
one parameter.
Since we nevertheless
attempt to obtain $\phi_\chi=0$ when choosing the NR
initial data parameters, the actual distribution
  of $\phi_\chi$ is highly correlated with other parameters.
Since we do not have full coverage of this $5d$ parameter space,
we avoid including the extra dimension $\phi_\chi$ in the
NRSur4d2s surrogate
model by
using an analytic approximation for the $\phi_\chi$ dependence
  of the model, as described in Sec.~\ref{sec:phichi}.
The surrogate
model can then predict waveforms for parameters in the $5d$ subspace,
but the $\phi_\chi$ dimension is entirely described by the analytic
approximation.

\subsection{Selection of greedy parameters} \label{sec:greedyselection}

\begin{figure}
  \includegraphics[width=\linewidth]{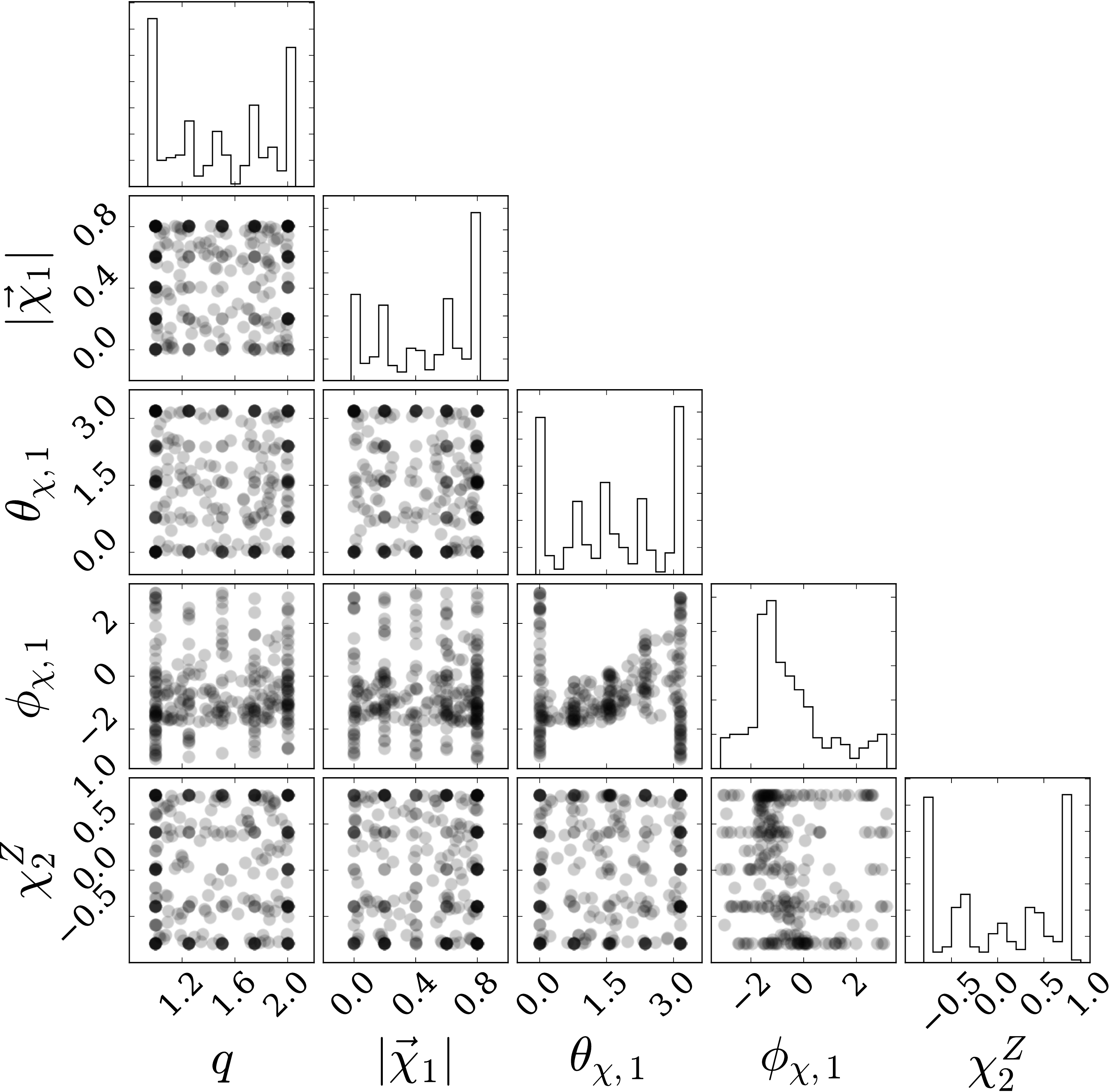}
  \caption{A ``triangle plot" showing all possible
two-dimensional projections and one-dimensional histograms of the
greedy parameters $G$ selected by the procedure of Sec.~\ref{sec:greedyselection}.
These are the parameters used for the numerical
 relativity simulations. Made using the Python package corner.py~\cite{corner}.
}
  \label{fig:training_points}
\end{figure}

We use $\phi_\chi = 0$ while determining the greedy parameters
$G = \{\pmb{\Lambda}_i\}$, and we use PN
waveforms to identify the most relevant and distinct points in parameter space
as outlined in sections
\ref{subsec:sur_basics} and \ref{subsec:ts_sampling}.
We first seed $G$ with the parameter space corner cases:
$q \in \{1, 2\}$, $|\vec{\chi}_A| \in \{0, 0.8\}$,
$\theta_\chi \in \{0, \pi\}$ and $\chi_2^z \in \{-0.8, 0.8\}$.
As described in Eq.~(\ref{eq:TrainingSet}), we
compute training sets $\mathcal{T}_\mathrm{TS}^i =
\mathcal{T}_\mathrm{B} \cup \mathcal{T}_\mathrm{R}^i$
consisting of a set of boundary parameters $\mathcal{T}_\mathrm{B}$
as well as a set of randomly sampled parameters $\mathcal{T}_\mathrm{R}^i$
that is
resampled at each greedy iteration $i$.
For $\mathcal{T}_\mathrm{B}$, we use a set of $216$ points where
$2$ components of $\pmb{\lambda}$ take on one of their extremal values
and the other $2$ are one of three intermediate (non-boundary) values.
This results in features that can be seen in
Fig.~\ref{fig:training_points}, where the $2$ boundary values and $3$
intermediate values occur frequently.
For example,
because some $2d$ projections of these special points
are selected multiple times, they appear as
darker points around the boundary of some of the subplots
in Fig.~\ref{fig:training_points}.  In addition, subplots involving
$\phi_\chi$ show an uneven distribution of stripes that
occur at these special points.
For $\mathcal{T}_\mathrm{R}^i$, we randomly sample each parameter component
uniformly in its range.

Next, we add parameters to $G$ using an initial
greedy algorithm
that
uses basis projection errors to select greedy parameters.
Given a point $\pmb\lambda$ as a candidate that might be added to $G$,
we compute a PN waveform $h$ corresponding to $\pmb\lambda$, we
decompose $h$ into waveform data pieces
(see Sec.~\ref{sec:decomposition}), and we project each waveform
data piece onto their respective bases.  Then we recombine the
projected waveform data pieces to produce a waveform $h_{\rm proj}$.
We then compute an error
$\mathcal{E}[h,h_{\rm proj}]$ using Eq.~(\ref{eq:cal_e}).
The point in $\mathcal{T}_\mathrm{TS}^i$
with the largest such error is the next point added to $G$.
This method is different than that of \cite{Blackman:2014maa}, in which
projection errors of each waveform data piece were computed separately,
and then these errors were combined in a weighted sum with coefficients
determined by hand. Our new method avoids the need to determine these
coefficients, and automatically ensures that the most significant waveform
data pieces are resolved accurately.
We use this initial greedy algorithm until
the error is $\mathcal{E} \leq 10^{-5}$.
At this point, the number of greedy points is approximately
$|G| = 30$.
Thus we have
built a linear basis for each waveform data piece.
For each iteration of
this initial greedy algorithm, we 
choose the number of randomly-sampled parameters to be
$|\mathcal{T}_\mathrm{R}^i| = 10 + 2|G^i|,$
where $|G^i|$ is the number of greedy parameters at the start of
the $i$th iteration.

\begin{figure}
  \includegraphics[width=\linewidth]{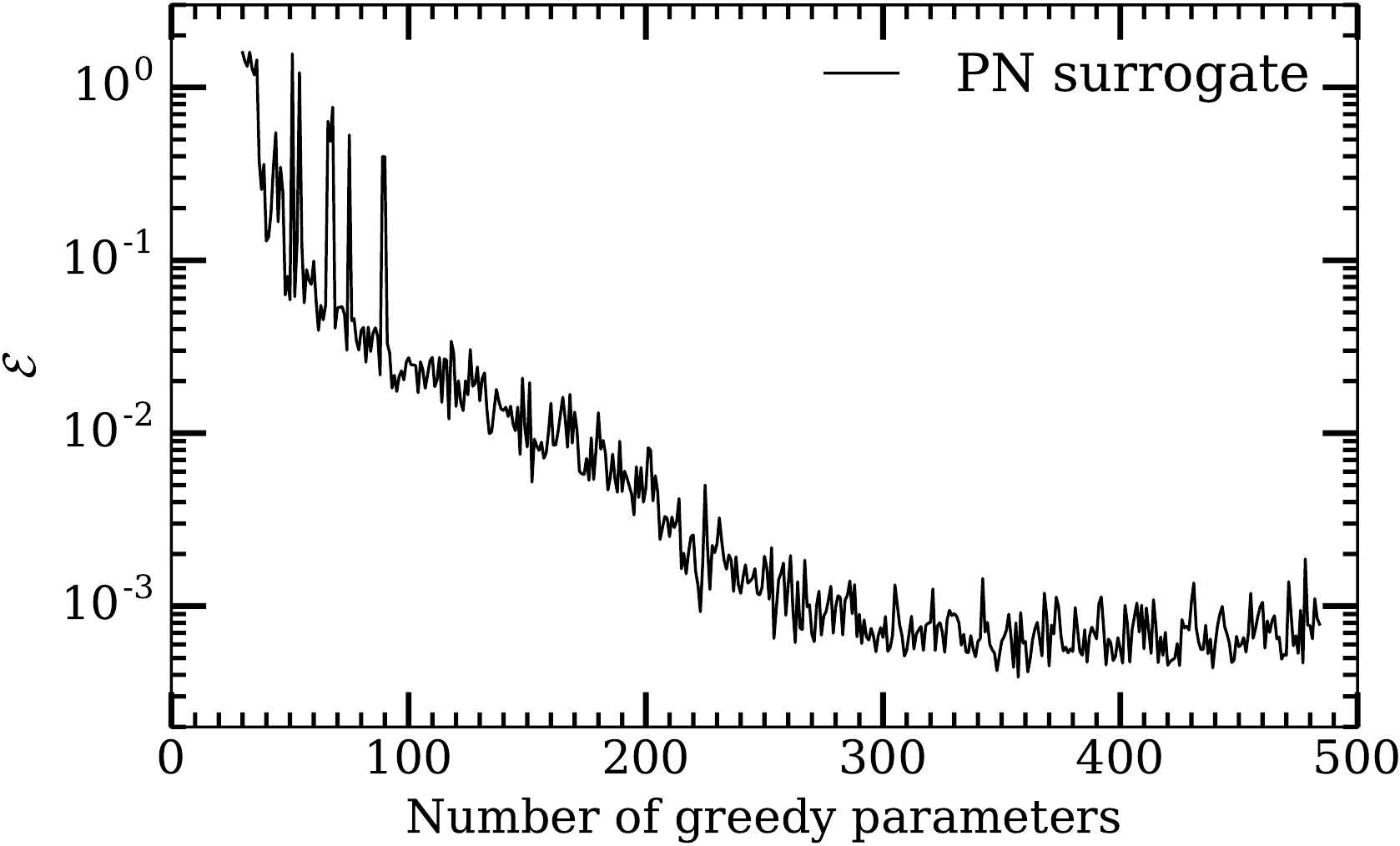}
  \caption{
        Maximum surrogate errors found during the second greedy algorithm
        (see Sec.~\ref{sec:greedyselection})
        for determining $\pmb{\Lambda}_i$ using trial PN surrogates.
        The noise is due to the random resampling, as well as the possibility
        of the parameter space fits becoming worse by adding a data point.
        The finite order of the fits leads to an error floor of
        $10^{-3}$, so we keep and perform NR simulations for
        only the first $300$ greedy parameters.
    }
  \label{fig:pn_greedy}
\end{figure}

Finally, we add parameters to $G$ using a second greedy algorithm
that uses surrogate errors to select greedy parameters.
At each iteration $i$, we construct a new trial PN waveform surrogate
(as described in Appendix~\ref{app:pn_surrogates}),
using the greedy parameters $G^i$,
and then for each point
$\pmb{\lambda} \in \mathcal{T}_\mathrm{TS}^i$,
we evaluate this surrogate and compare it to the corresponding PN
waveform by computing $\mathcal{E}$.
The parameter $\pmb{\lambda}$ that maximizes
this error is used as the next greedy parameter and
is added to $G$.
This error includes the errors in the parametric fits for each empirical node
of all waveform data pieces; the parametric fits are
shown as blue lines in Fig.~\ref{fig:surrogate_alg} and are
described in detail in Sec.~\ref{sec:param_space_fits} and appendices
\ref{sec:fitappendix} and \ref{app:pn_surrogates}.
For this step, we use $|\mathcal{T}_\mathrm{R}^i| = 6|G|$.
The maximum errors found in each iteration of this second greedy
algorithm are shown in Fig.~\ref{fig:pn_greedy} as a function of $|G|$.
The noise is due to the random resampling of the training set,
as well as the possibility of the
parametric fits becoming worse by adding a data point.
Because the parametric fits are restricted to a particular
order, the surrogate error in Fig.~\ref{fig:pn_greedy} does not
go below $10^{-3}$.  In principle one can reduce this error floor by
increasing the order of the fits, but here we simply keep only
the first $300$ greedy parameters.
We perform NR simulations for these $300$ parameters,
except for those parameters that can be obtained from other
parameters by symmetry, for example by exchanging the black hole labels.
These symmetry considerations reduce the number of simulations to  $276$.

\subsection{Numerical Relativity Simulations} \label{sec:simulations}

To build our time-domain model,
we use the $276$ NR waveforms computed
by the SXS collaboration
with the Spectral Einstein Code (SpEC)
described in Ref. \cite{Hemberger:2017}.
Each NR simulation is performed at three different numerical
resolutions, labeled `Lev1', `Lev2', and `Lev3', in order of increasing
resolution; Lev$i$ has an adaptive-mesh-refinement (AMR) error tolerance
that is a factor of 4 smaller than Lev$i-1$.
For each resolution, the waveform is extracted at multiple finite radii
from the source, and then the waveform is extrapolated to future null
infinity~\cite{Boyle-Mroue:2008}. The extrapolation is done using an $N$-th
order polynomial in $1/r$, where $r$ is a radial coordinate.
To estimate errors in extrapolation, we perform extrapolation with
several values of $N$~\cite{Boyle-Mroue:2008}. Similarly,
to estimate numerical truncation error, we compare simulations that
are identical except for resolution~\cite{Hemberger:2017}.
However, for building surrogates, we
always use the highest available resolution (Lev3) simulations,
and use the $N=2$ extrapolated waveforms. 
The simulations 
begin at a time of $\sim 5000M$ before merger
where $M = m_1 + m_2$ and $m_i$
are the Christodoulou masses of each black hole.
We ignore the small eccentricities present in the simulations,
which have a median of $0.00029$ and a maximum of $0.00085$ for the
highest resolution simulations.
The masses we use are those measured
after the initial burst of junk radiation~\cite{Aylott:2009ya} leaves the computational
domain.

The BH spin vectors are measured on the apparent horizons of the BHs
during the evolution of the NR simulation.
The spin directions are therefore
gauge-dependent
The potential concern is that when the surrogate model is evaluated,
the spin directions must be provided with the gauge used to build the model,
so that the spin directions obtained in gravitational wave parameter
estimation can be interpreted correctly. However,
it
has been found that the time-dependent 
spin and orbital angular momentum vectors in the damped harmonic
gauge used by SpEC agree very
well with the corresponding vectors in PN theory~\cite{Ossokine:2015vda}.
Therefore, this is of no
more concern than the interpretation of spin directions with PN-based
gravitational wave models.

For the purposes of surrogate modeling, we need to associate
  each gravitational waveform with a {\em single} value of the
  parameter vector $\pmb{\lambda}$, even though some of the parameters
  (in particular the spin directions) are time-dependent.  To do this,
  we measure the parameters at some fiducial time.  To define this
  time, we (arbitrarily) equate the time coordinate of the simulation
  with the time coordinate of the waveform at future null infinity,
  offset so that the beginning of the simulation and the beginning of
  the NR waveform correspond to the same coordinate $t$.  We then set
  $t=0$ at the peak amplitude of the waveform, and we measure
  $\pmb{\lambda}$ at a fiducial value of $t_0=-4500M$.  We emphasize
  that there is no unique way to map coordinates in the near zone to
  coordinates at infinity. However, choosing a different map changes
  nothing in the surrogate model other than the time at which
  $\pmb{\lambda}$ is measured.  Because the spin directions change
  only on the precession timescale and not the orbital timescale, any
  other choice that measures $\pmb{\lambda}$ at a time near the
  beginning of the simulation should yield similar results.

As described above, we selected the first $300$ points in parameter
space chosen by the PN greedy algorithms, and we reduced this number
to $276$ points after removing configurations that were equivalent
because of symmetries.  We therefore performed $276$ NR simulations.
However, the total number of NR waveforms represented by these $276$
simulations is greater than $276$ if we use symmetry to restore
additional configurations.  For example, for equal mass cases with
$\theta_\chi \in \{0, \pi\}$, exchanging the two black holes yields
another configuration in the parameter subspace.  For each of these
  cases, we produce the additional configuration by relabeling the
  black holes and rotating the coordinates by 180 degrees in the orbital
  plane; this results in a total of $288$ NR waveforms.  In addition,
configurations with $|\vec{\chi}_1| = 0$ are invariant under changes
in $\theta_\chi$, so we might add additional such configurations that
differ only in $\theta_\chi$.  In principle, we could add an arbitrary
number of such configurations, but it is unclear how many to add.
Also, $|\vec{\chi}_1|$ is never exactly zero for NR simulations,
so we have an unambiguous choice of $\theta_\chi$ for each simulation.
We therefore choose not to
restore these additional configurations, so we are left with a total of
$288$ NR
waveforms.

\subsection{Waveform alignment} \label{sec:alignment}

Our surrogate model is built assuming that the
waveform has peak amplitude at $t=0$, and that the parameters
$\pmb{\lambda}$ (mass ratio and spin vectors) are measured at some fixed
time $t=t_0$, which we choose to be $t_0=-4500M$.  Furthermore,
our surrogate model assumes a coordinate system in the source frame
  such that at $t=t_0$, black hole 1
  lies along the positive $\hat{x}$ axis, black hole 2
  lies along the negative $\hat{x}$ axis, and the instantaneous
  Newtonian orbital angular momentum lies along the positive $\hat{z}$ axis.

Ideally, all of the input NR waveforms used in the surrogate should also
  have peak amplitude at $t=0$, and each simulation's black holes
  should have the same orientation vector $\hat{n}$ at $t=t_0$,
  where $\hat{n}$ is a unit vector pointing from the large black hole
  to the small black hole.  However, when setting up
  an NR simulation, the time between the beginning of the simulation until
  merger is {\it a priori} unknown, and depends on the mass ratio and
  the black hole spins.  Furthermore, the orientation $\hat{n}$ of the black holes,
  and the mass and spin parameters, are chosen at the
  beginning of the simulation, which (because the merger time is {\it a priori}
  unknown) is not at a fixed time before merger.
  Therefore, for each of our 276 NR waveforms
  the peak amplitude occurs at a different time, and the orientation
  of the black holes with respect to the coordinates does not agree
  at any given time relative to the time of peak amplitude.
  Therefore, it is necessary to {\it align} all the NR waveforms by
  time-shifting them so that the maximum amplitude occurs at $t=0$,
  rotating the coordinates so that the black holes are oriented
  in the same way at $t=t_0$, and then remeasuring the mass and spin
  parameters at $t=t_0$.

To align the waveforms, we shift them
in time such that the peak of the total waveform amplitude
as given in Eq.~\ref{eq:peak}
occurs at $t=0$.
We then use a cubic spline
to interpolate the real and imaginary parts of the waveform
onto a uniformly-spaced time series with $dt = 0.1M$.
Next, we rotate
the waveforms to align the orientation of the
  binary at $t_0 = -4500M$ in two steps:
first we perform an approximate rotation
using the black hole trajectories,
and then we perform a small correction using only the waveform.
For the initial approximate
  rotation, we use the horizon trajectory to
align the Newtonian orbital angular momentum with $\hat{z}$
and rotate about $\hat{z}$ such that black hole $1$
lies along
the positive
$\hat{x}$ axis.
We then use the waveform modes to perform an additional rotation,
aligning the principal eigenvector of the angular momentum
operator~\cite{Boyle:2011gg}
with $\hat{z}$ and equating the
phases of $h^{2, 2}$ and $h^{2, -2}$ at $t=t_0$.
The first coarse alignment was used since the second alignment is
ambiguous - we can change the sign of the principal eigenvector
and/or rotate by an additional $\pi$ about $\hat{z}$,
which we resolve by choosing the smallest of the rotations,
since the waveform is already nearly aligned.
We perform identical rotations on the spin directions and then
measure them at $t_0.$

\subsection{Post-alignment parameterization} \label{sec:parameterization}

While the initial orbital parameters were chosen using PN
approximations such that
$\vec{\chi}_2(t_0) \propto \hat{z}$ after this alignment,
in practice we obtain
small misalignments leading to orthogonal components of $\vec{\chi}_2$ less
than $0.016$ in magnitude.
We ignore these spin components, leading to a $5d$ parameter space:
\begin{itemize}
    \item $q = \frac{m_1}{m_2} \in [0.9999, 2.0005]$
    \item $|\vec{\chi}_1| \in [0, 0.801]$
    \item $\theta_\chi \equiv \cos^{-1}\left(\frac{
            \chi_1^z(t_0)}{|\vec{\chi_1}|}\right) \in [0, \pi]$
    \item $\phi_\chi \equiv \mathrm{arctan2}\left(
            \chi_1^y(t_0),\, \chi_1^x(t_0) \right) \in (-\pi, \pi]$
    \item $\chi_2^z(t_0) \in [-0.8, 0.800006]$
\end{itemize}
as shown in Fig.~\ref{fig:diagram}.
We will often omit the time dependence of the last parameter and simply write
$\chi_2^z.$

\section{Waveform Decomposition} \label{sec:decomposition}

\begin{figure*}
    \centering
    \def\svgwidth{2.0 \columnwidth}
    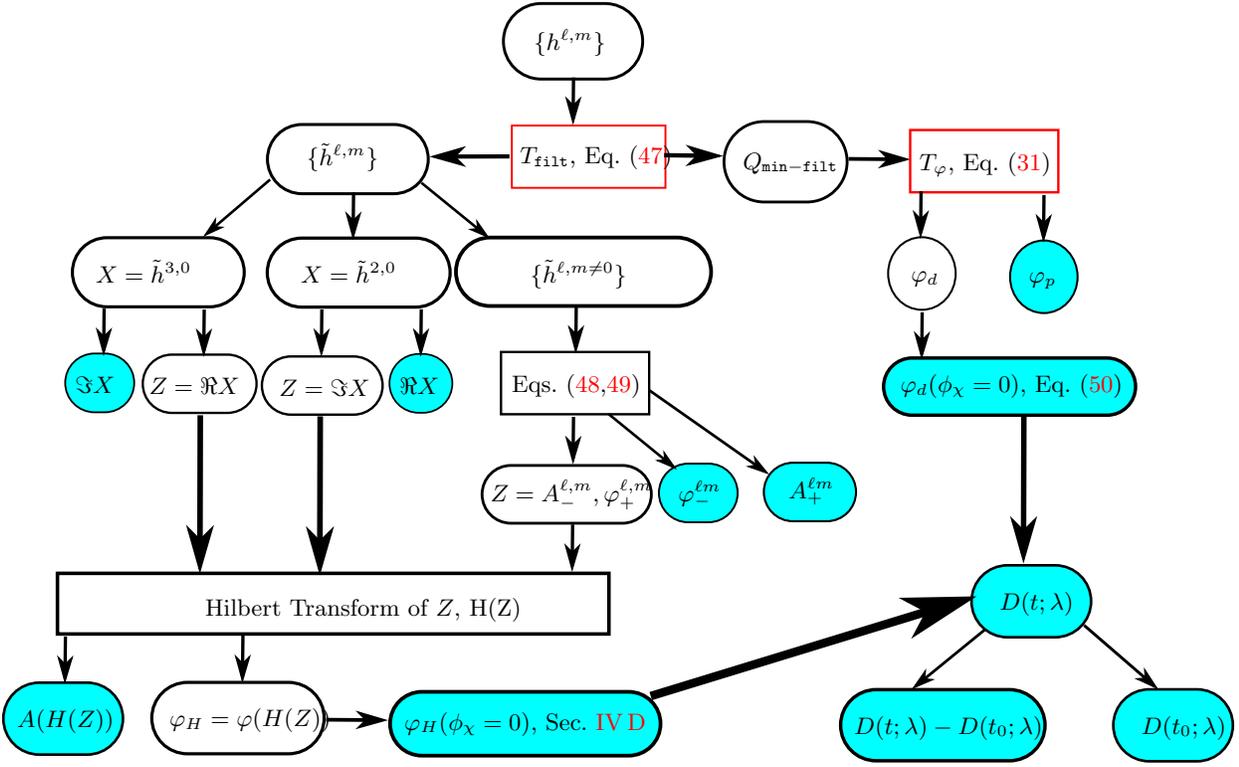
    \caption{Waveform decomposition schematic.
A series of decompositions are applied to a set of NR waveform modes $\{h^{\ell m}\}$
yielding easier-to-approximate waveform data pieces (shown as cyan ellipses) for which we
ultimately fit.  Two types of objects are shown: 
timeseries data as an ellipse and operators/maps as rectangles. A red outlining border
identifies an object which uses a modeling approximation which will {\em not} go away with additional
NR waveforms.
These {\em decomposition errors} are quantified and shown to be smaller
than other sources of error in Sec.~\ref{sec:Assessment}.
An additional source of error that will not converge away with more 
NR waveforms results from the assumption that each data piece
transform in a simplistic way with changes of $\phi_\chi$.
\label{fig:DecomposeData}
}
\end{figure*}

This section describes how each input NR waveform is decomposed
into a set of ``waveform data pieces'', which are simple functions that can
be modeled easily and can be recombined to produce the original waveform.
This decomposition was outlined briefly in \S~\ref{subsec:surrogate_techniques},
  and a
flowchart of this process is shown in Fig.~\ref{fig:DecomposeData}.

\begin{figure}
  \includegraphics[width=\linewidth]{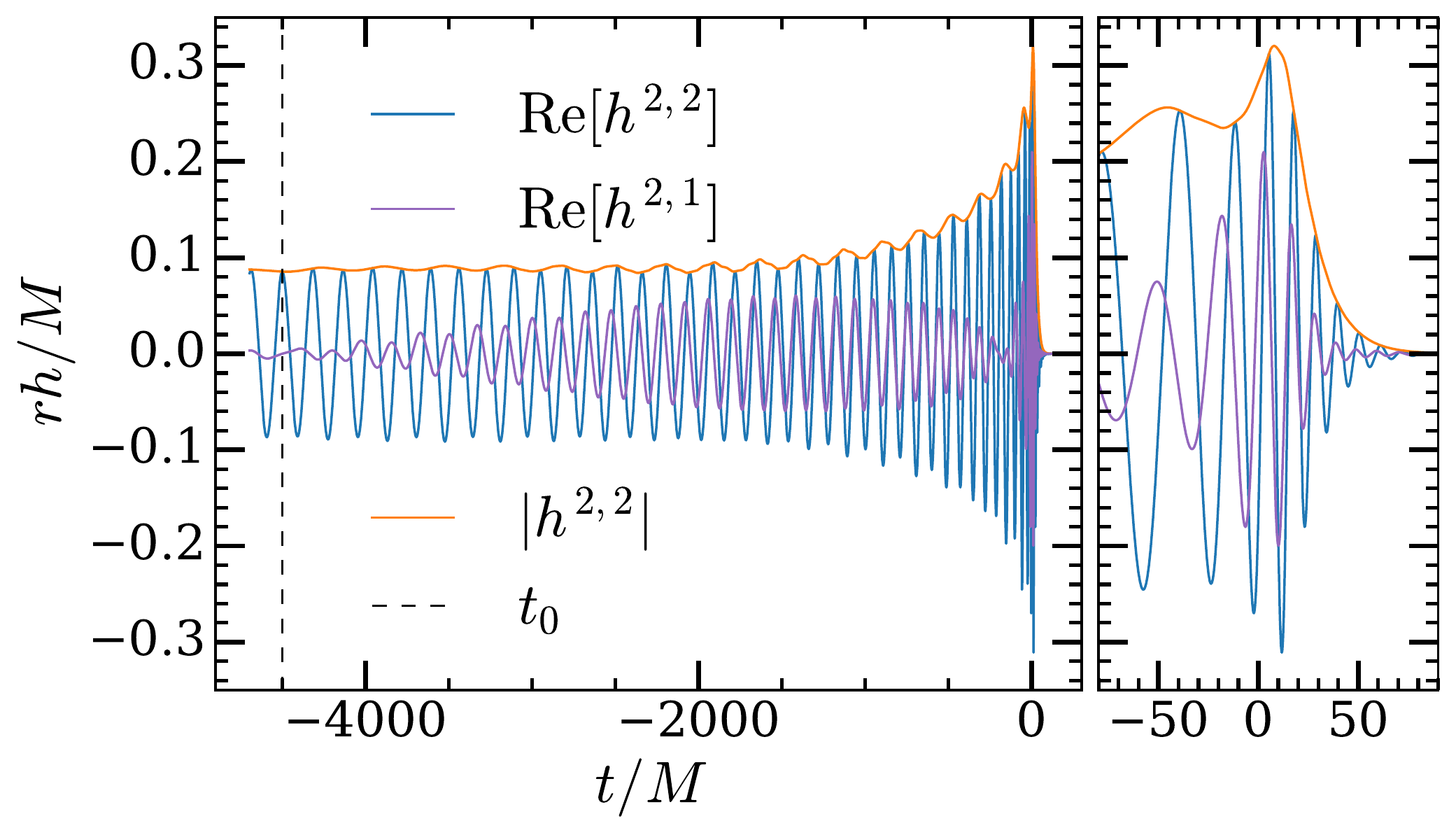}
  \caption{Waveform modes in the inertial frame for SXS:BBH:0338 with
        $q=2$, $|\vec{\chi}_1| = 0.8$, $\theta_\chi = 1.505$,
        $\phi_\chi = -1.041$ and $\chi_2^z = 0.8$. For precessing
        systems, all $\ell=2$ modes contain significant power in the
        inertial frame. The NR waveform is aligned to have the
        canonical orientation at $t=t_0$.
    }
  \label{fig:decomp_inertial}
\end{figure}

We write each input waveform as a set of modes
$H = \{h^{\ell, m}(t)\}$, with $t \in [t_\mathrm{min}, t_\mathrm{max}]$.
Here $t_\mathrm{min}$ and $t_\mathrm{max}$ are chosen to be
the same for all waveforms, and are selected in the following way:
Recall that each waveform is time-shifted so that the maximum
amplitude occurs at $t=0$; this means that each time-shifted finite-length NR
waveform $H_i$ has a different beginning time $t^{\rm begin}_i$ and a different
ending time $t^{\rm end}_i$. We choose
$t_\mathrm{min}=\max_i(t^{\rm begin}_i)+150M$ and
$t_\mathrm{max}=\min_i(t^{\rm end}_i)$.
The value $150M$ is chosen to remove the worst
  of the ``junk radiation''~\cite{Aylott:2009ya} that results from the failure
  of NR initial data to precisely describe a quasiequilibrium inspiral.
Although the surrogate output will cover only the smaller
time interval $[t_0, t_f=70M]$, we use waveforms over the
larger time interval $[t_\mathrm{min}, t_\mathrm{max}]$ in order
to mitigate edge effects that can occur in
later steps in the decomposition process
(filtering and Hilbert transforms, described below).
Selected modes of $H$ are shown in Fig.~\ref{fig:decomp_inertial}.

\subsection{Transforming to a coprecessing frame}
\label{sec:xformCoPrecessing}

\begin{figure}
  \includegraphics[width=\linewidth]{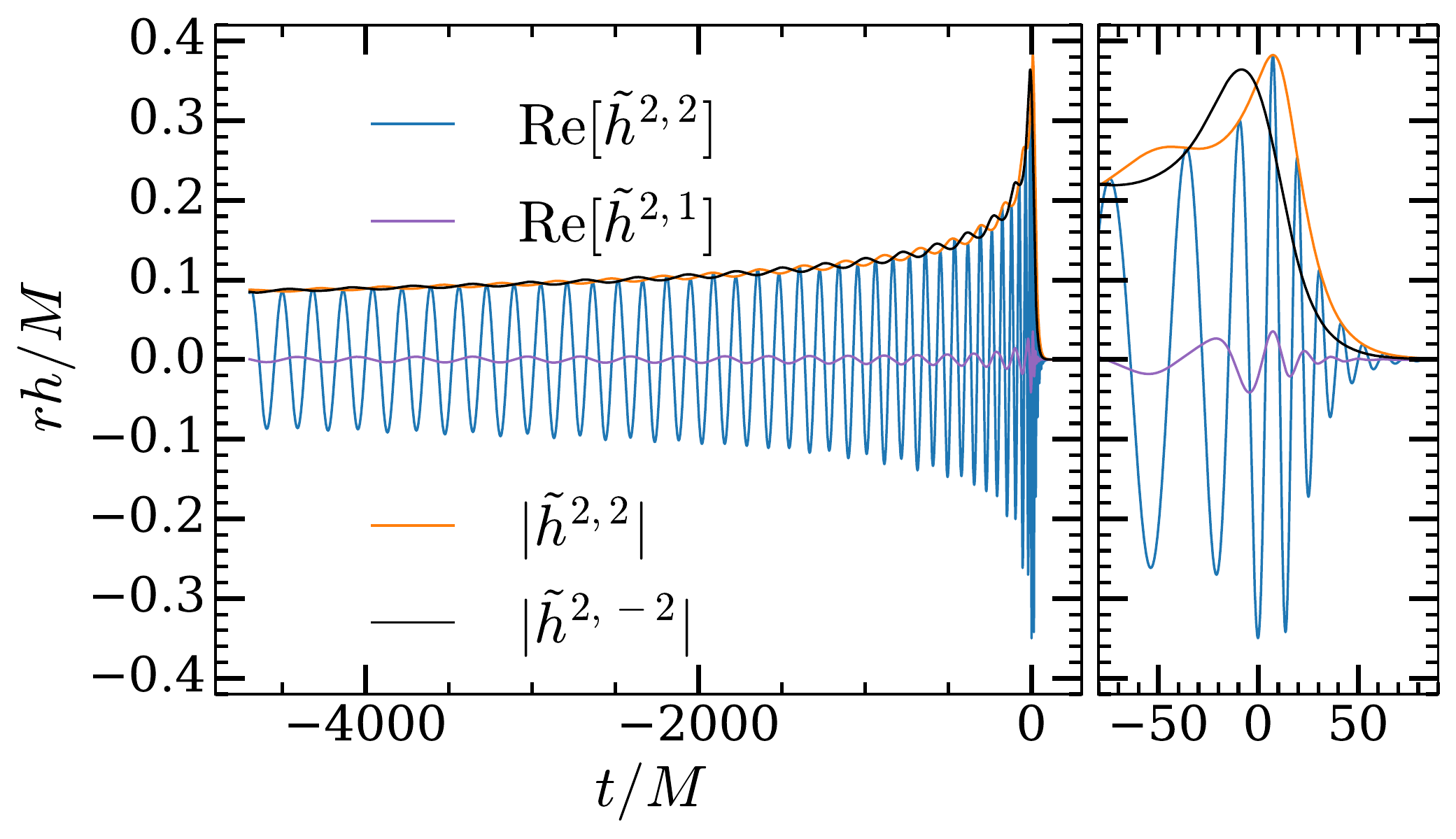}
  \caption{Waveform modes in the coprecessing frame for SXS:BBH:0338.
    The mode power hierarchy is now
      the same as for a non-precessing waveform, with the $(2, \pm2)$
      modes dominating, but small
      effects of precession
      are still present in the mode amplitudes and phases.
      The amplitudes of the $(2, \pm2)$ modes have small
      nearly opposite oscillations.
    }
  \label{fig:decomp_copr}
\end{figure}

The first step in the waveform decomposition is transforming
to a rotating coordinate frame in
which the binary is not precessing.  Thus the original waveform is described
by a (much simpler) waveform in this coprecessing frame, plus functions
that describe the time-dependent rotation.
We transform~\footnote{Throughout this work we use 
GWFrames~\cite{Boyle:2013nka,gwframes} to enact our transformations.} 
$H$ to the minimally rotating coprecessing frame of Ref.
\cite{Boyle:2011gg}, and thereby
obtain the waveform modes
$\tilde{H} = \{\tilde{h}^{\ell, m}(t)\}$ in this frame,
  as well as a time-dependent unit quaternion $q(t)$ that describes
  the rotation of the frame.
Throughout this section we will use a tilde, i.e., $\tilde{h}^{\ell, m}(t)$,
to denote a time-domain waveform mode in the coprecessing frame, as
opposed to the Fourier transform of a waveform mode.
Selected modes of  $\tilde{H}$ are shown in
Fig.~\ref{fig:decomp_copr}.
We
denote this transformation by 
\begin{equation} 
T_C : H \rightarrow (\tilde{H}, q),
\end{equation}
where the `C' stands for the coprecessing frame.
If we also define a different transformation
\begin{equation}
  T_Q : (H', q) \rightarrow H
  \label{eq:T_Q}
\end{equation}
that takes an arbitrary waveform $H'(t)$ and rotates it
  by an arbitrary unit quaternion $q(t)$, then 
  $T_Q$ is the left inverse of $T_C$, that is, $T_Q(T_C(H)) = H$.
However, an arbitrary waveform $H'(t)$ and
  an arbitrary unit quaternion $q(t)$ do not necessarily represent
  the decomposition of any inertial-frame
  waveform $H$ into a coprecessing frame.  Therefore,
  for arbitrary $H'(t)$ and $q(t)$ we have in general
  $T_C(T_Q(H', q) ) \neq (H',q)$. This property will
    be important in \S~\ref{sec:filtCoPrecessing} below.

\begin{figure}
  \includegraphics[width=\linewidth]{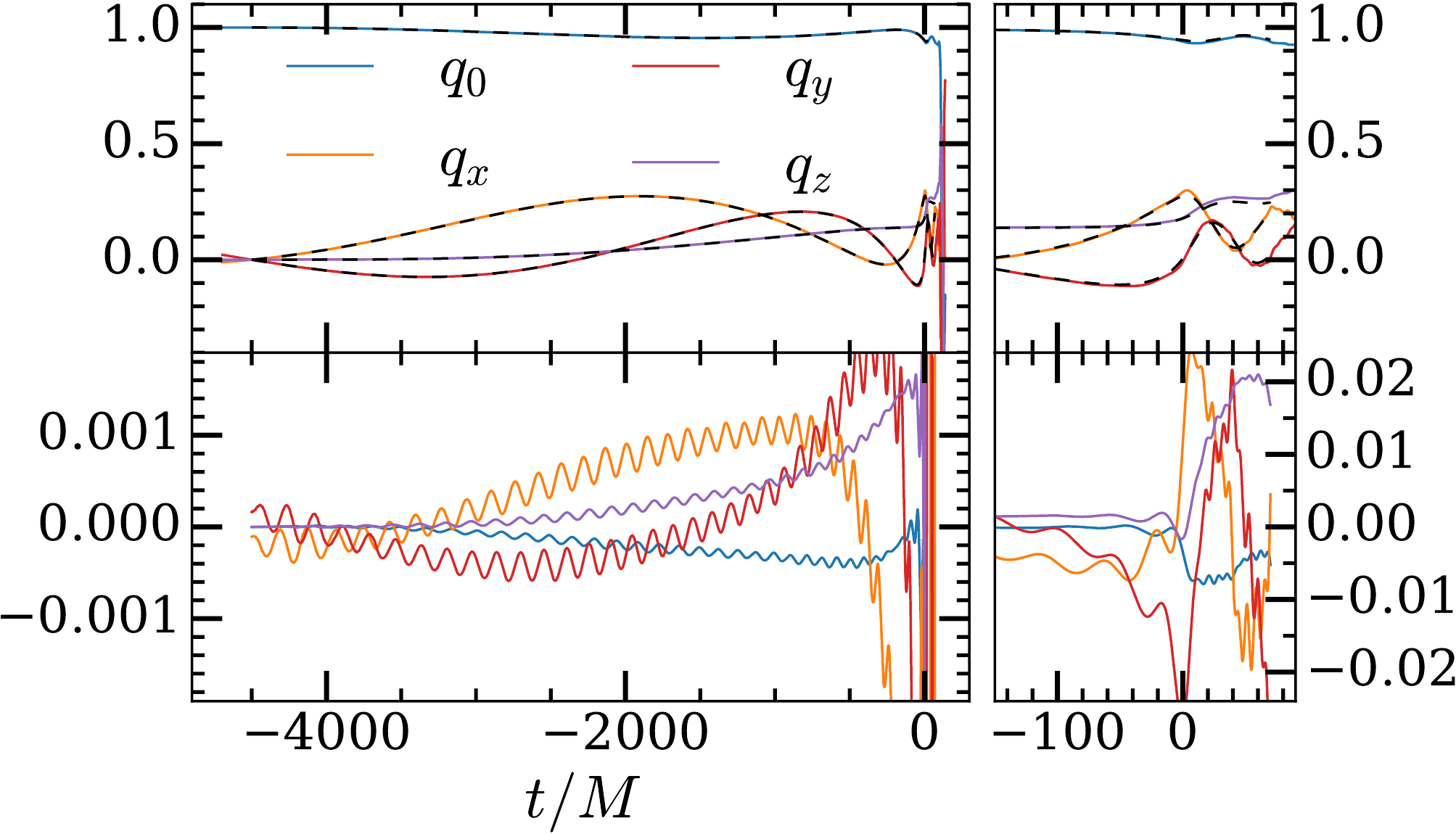}
  \caption{
     Top: Quaternion $q$ representing the time-dependent rotation
        from the coprecessing frame to the inertial frame (solid lines)
        and the filtered quaternion $q_\mathrm{min-filt}$ (dashed lines)
        for SXS:BBH:0338. Bottom: Differences between the filtered and unfiltered
        quaternions. This difference results in an error when reconstructing
        the waveform in the inertial frame, contributing to a ``decomposition"
        error in the surrogate model.
    }
  \label{fig:decomp_quat}
\end{figure}

The unit quaternion $q(t)$ has
$4$ components shown as
solid lines in Fig.~\ref{fig:decomp_quat}.
However, the minimally-rotating coprecessing frame 
  constrains $q(t)$ so as
  to minimize the magnitude of the frame's instantaneous angular velocity
  (the ``minimal rotation condition'') \cite{Boyle:2011gg}.
  This condition, combined with
  the unit norm, imply that $q(t)$ has only $2$ independent
  components.

Therefore, we will further decompose $q(t)$ into these
  two independent components, so that we have only two functions to
  model in order to describe the rotation.  To do this, consider first
 the relative instantaneous rotation of the frame
\begin{equation}
dq(t) = q^{-1}(t) q(t + dt) = 1 + 2 \vec{\omega}(t) dt +
        \mathcal{O}(dt^2).
\end{equation}
The minimal rotation condition says that $\omega_z = \mathcal{O}(dt^2)$, while
$\omega_x$ and $\omega_y$ are $\mathcal{O}(1)$, so in the limit $dt\to 0$
we find that $\vec{\omega}(t)$ has only two independent components.
The precession angular frequency $\omega_p(t) = |\vec{\omega}(t)|$
describes the velocity of the path on the unit sphere traced out by
the $z$-axis of the coprecessing frame.

We approximate $dq(t)$ using
finite differences:
\begin{equation}
\delta q(t) = q^{-1}(t) q(t + \delta t) = s(t) + \delta t\,\vec{u}(t),
\label{eq:FiniteDifferenceQ}
\end{equation}
where the scalar component $s(t)$ is $1 + \mathcal{O}(\delta t^2)$.
Thus, for a given $\delta t$, Eq.~(\ref{eq:FiniteDifferenceQ})
  defines $\vec{u}(t)$ in terms of $q(t)$, and furthermore, $\vec{u}(t)$
  approaches $\frac{1}{2}\vec{\omega}(t)$ as
  $\delta t\to 0$. We find that if we use $\delta t=0.1M$, the $\vec{u}(t)$
  we obtain is sufficiently close to this limit that the error we make is
  negligible compared to other errors; this error is included in the
  {\em decomposition error} discussed in \S~\ref{sec:Assessment}.
  Finally, instead of using $\vec{u}(t)$ directly as independent components
  of $q(t)$, we define
$\tilde{\omega}_p = \frac{1}{2}|\vec{u}(t)|$
and

\begin{align}
\varphi_p(t) &= \delta t\sum_{\tau < t} \tilde{\omega}_p(\tau) \\
\varphi_d(t) &= \mathrm{arg} \big( u_x(t) + iu_y(t) \big)\,.
\end{align}
The length of the path on the unit sphere traced out by the
$z$-axis of the coprecessing frame is given
by $\varphi_p(t)$.
In a frame instantaneously aligned with the coprecessing frame,
$\varphi_d$ is the phase of the projection of 
$\vec{u}(t)$
into the $xy$-plane.

We have thus decomposed the quaternion $q(t)$ into
two functions $\varphi_p(t)$ and $\varphi_d(t)$.  These are the
two functions we will model in constructing the surrogate.
We denote this transformation by
\begin{equation} \label{eq:T_varphi}
T_\varphi : q \rightarrow (\varphi_p, \varphi_d).
\end{equation}

To perform the inverse transformation, that is, to
compute $q(t)$ from  $\varphi_p$
and $\varphi_d$, we compute
\begin{align}
\tilde{\omega}_p(t) &= \frac{\varphi_p(t + dt) - \varphi_p(t)}{\delta t} \\
u_x(t) &= 2\tilde{\omega}_p(t) \mathrm{cos}(\varphi_d(t)) \label{eq:DeltaQReconstructionx}
\\
u_y(t) &= 2\tilde{\omega}_p(t) \mathrm{sin}(\varphi_d(t)) \\
u_z(t) &= 0 \label{eq:DeltaQReconstructionz} \\
s(t) &= \sqrt{1 - \left(2 \tilde{\omega}_p(t) \delta t\right)^2}
\label{eq:DeltaQReconstructions}
\\
\delta q(t) &= s(t) + \vec{u}(t)\delta t \, . \label{eq:DeltaQReconstruction}
\end{align}
We include the $(\delta t)^2$ term in Eq.~(\ref{eq:DeltaQReconstructions})
so that the reconstructed $q(t)$ has unit norm.
Because we assume $\delta q_z=0$, the $\delta q$ we compute in
Eq.~(\ref{eq:DeltaQReconstruction})
is not exactly the $\delta q$ we started with in Eq.~(\ref{eq:FiniteDifferenceQ}); however, the error we make is only $\mathcal{O}(\delta t^3)$.
Given $q(t)$ and $\delta q$, we can then compute $q(t + \delta t)$
using
\begin{equation}
\label{eq:reconstructquat}
q(t + \delta t) = q(t) \delta q(t),
\end{equation}
which results in an $\mathcal{O}(\delta t^2)$ error in
$q(t+\delta t)$.
Because we have $q(t_0) = 1$ at the alignment time $t_0$,
we can use the recurrence relation Eq.~(\ref{eq:reconstructquat})
to construct $q(t)$ at
all times, given $\varphi_p(t)$ and $\varphi_d(t)$.

\subsection{A ``filtered" coprecessing frame} \label{sec:filtCoPrecessing}

\begin{figure}
  \includegraphics[width=\linewidth]{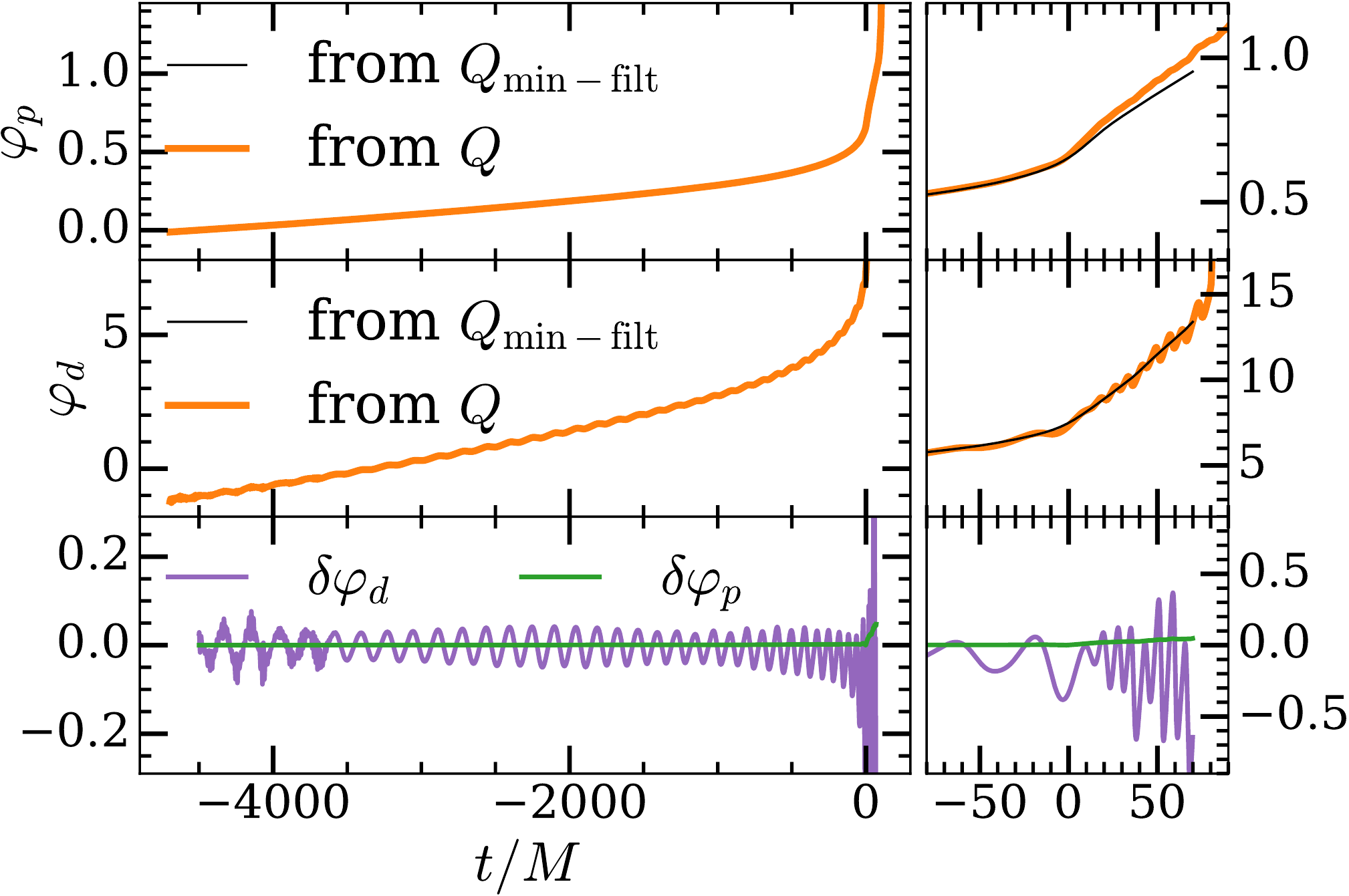}
  \caption{Phases $\varphi_p$ (top) and $\varphi_d$ (middle)
        for SXS:BBH:0338. These phases represent
        the total amount of precession and the instantaneous direction
        of precession respectively.
        Shown are phases computed
        from the unfiltered coprecessing quaternion
        (thick orange lines) and the filtered quaternion
        (thin black lines).
        The orbital timescale oscillation in $\varphi_d$ is suppressed after
        filtering.
        Bottom: Differences between the filtered and unfiltered
        phases.
    }
  \label{fig:decomp_phases}
\end{figure}

The quaternion $q(t)$ representing the coprecessing
frame oscillates
mostly on the slow precession timescale, which makes it
  easier to model. However, it also has
small 
oscillations on the much faster orbital timescale,
as shown by the purple curve
in the bottom plot of Fig.~\ref{fig:decomp_phases}.
These oscillations are due to the nutation of the
rotation axis of the coprecessing frame, relative to the inertial frame.
These small oscillations can make it more difficult
to fit $\varphi_d$ across parameter space.
Since the effect of the nutation on the inertial frame waveform is small,
we filter out the nutation in the coprecessing frame.
We use a Gaussian filter with a width of $\pi$ radians of the orbital phase,
which is computed from the angular velocity
of the waveform as described in \cite{Boyle:2013a}.
Near the edges of the domain, we truncate the filter on both sides to keep
the filter centered.
Specifically, if the (monotonic) orbital phase is given by
$\varphi_\mathrm{orb}(t)$, then we can invert the relationship to find
$t(\varphi_\mathrm{orb})$.
For a given time $\tau$ with corresponding orbital phase
$\varphi^* = \varphi_\mathrm{orb}(\tau)$ we then compute
\begin{align}
\varphi_\mathrm{min} &= \varphi_\mathrm{orb}(t_\mathrm{min}) \\
\varphi_\mathrm{max} &= \varphi_\mathrm{orb}(t_\mathrm{max}) \\
\Delta \varphi &= \min(4\pi,
                       |\varphi^* - \varphi_\mathrm{min}|,
                       |\varphi^* - \varphi_\mathrm{max}|) \\
\varphi_\pm &= \varphi^* \pm \Delta \varphi \\
G(\varphi) &= \mathrm{exp}\left[
        -\left(\frac{\varphi - \varphi^*}{\pi}\right)^2\right]\\
q_\mathrm{init-filt}(\tau) &= \frac{
        \int_{\varphi_-}^{\varphi_+} q(t(\varphi))G(\varphi) d\varphi}{
        \int_{\varphi_-}^{\varphi_+} G(\varphi) d\varphi}\\
q_\mathrm{filt}(\tau) &= \frac{q_\mathrm{init-filt}(\tau)}
                             {|q_\mathrm{init-filt}(\tau)|}\,.
\end{align}
This filtered frame corresponding to
$q_\mathrm{filt}$ is no longer minimally rotating, but
we can compute
\begin{equation}
\tilde{H}', q_\mathrm{min-filt} = T_C( T_Q(\tilde{H}, q_\mathrm{filt}))
\end{equation}
and use the frame corresponding to $q_\mathrm{min-filt}$,
which is minimally rotating and has
much less nutation than the frame corresponding to $q$.
The components of the filtered quaternion
$q_\mathrm{min-filt}$ are shown in Fig.~\ref{fig:decomp_quat} as dashed lines.
We use $\tilde{H}$, and not $\tilde{H}'$, as the filtered coprecessing
waveform, because $\tilde{H}'$ is not as slowly-varying as
$\tilde{H}$ and is therefore slightly more difficult to fit.
We have verified that
the error in the final model caused by choosing $\tilde{H}$ instead of
$\tilde{H}'$ is small compared to other errors. Note that even if we
choose $\tilde{H}'$, introducing a filter produces some information
loss, and therefore results in some error in the final surrogate model.
This {\em decomposition error}
is discussed in Sec.~\ref{sec:Assessment} and is plotted in
Figs.~\ref{fig:check5d_inertial} and~\ref{fig:error_hist}.
We thus denote the filtering transformation by
\begin{equation} \label{eq:T_filt}
T_\mathrm{filt} :   (\tilde{H}, q) \rightarrow
                    (\tilde{H}, q_\mathrm{min-filt}).
\end{equation}
Applying $T_\varphi$ to $q_\mathrm{min-filt}$ results in less oscillatory
behavior in $\varphi_d$ than when $T_\varphi$ is applied to $q$,
as seen in
Fig.~\ref{fig:decomp_phases}.
When evaluating the surrogate and reconstructing the inertial frame
waveform, we do not attempt to invert $T_\mathrm{filt}$ which
contributes to the decomposition errors
shown in Fig.~\ref{fig:error_hist}.

\subsection{Decomposition of coprecessing-frame waveforms}
\label{sec:CoPrecessing}

Once we have computed waveform modes $\tilde{h}^{\ell, m}(t)$ in the
  coprecessing frame, we decompose each of these modes (except for the
  $m=0$ modes, which are discussed separately below) into an amplitude
  and a phase.  However,
  these amplitudes and phases are difficult to model because they
contain oscillations on the
  orbital timescale.
These oscillations are due to asymmetries of waveforms from precessing
systems and cannot be completely removed with a different choice of
frame \cite{Boyle:2014}.
Fig.~\ref{fig:decomp_copr} shows an example of these oscillations.
To better model the amplitudes and phases of $\tilde{h}^{\ell, m}(t)$,
  we seek to further decompose them into simpler slowly-varying functions.
To do this, first note 
that the amplitudes of
  $\tilde{h}^{2,2}(t)$ and
$\tilde{h}^{2,-2}(t)$ shown in Fig.~\ref{fig:decomp_copr}
oscillate in opposite directions. The same is true for
  the phases, although it is not apparent in the figure, and it is
  also true for some (but not all) higher-order modes.
This motivates the use of symmetric and antisymmetric amplitudes and phases
\begin{align}
A_\pm^{\ell, m}(t) &= \frac{1}{2}\left( |\tilde{h}^{\ell, m}(t)|
                    \pm  |\tilde{h}^{\ell, -m}(t) |\right) \label{Eq:AmpParts} \\
\varphi_\pm^{\ell, m}(t) &= \frac{1}{2}\left(
    \varphi(\tilde{h}^{\ell, m}(t))
    \pm \varphi(\tilde{h}^{\ell, -m}(t))\right)  \label{Eq:PhaseParts}
\end{align}
for $m > 0$ where $\varphi(x(t)) = \mathrm{arg}(x(t))$.
The symmetric amplitude $A_+^{2,\pm 2}(t)$ and the
antisymmetric phase $\varphi_-^{2, \pm 2}(t)$ contain almost no oscillations
and are slowly-varying, so we use these as waveform data pieces.
However, the antisymmetric
amplitude $A_-^{2,\pm 2}(t)$ and the symmetric phase $\varphi_+^{2,
\pm 2}(t)$ of the $(2, \pm 2)$ mode are small oscillatory real
functions, so to model them we taper each of these functions in the
  intervals $[t_\mathrm{min}, t_0]$ and $[t_f, t_\mathrm{max}]$ with a
  Planck window~\cite{McKechan:2010kp} and take a Hilbert transform, thereby
  producing an amplitude and phase for each of these
  functions; these amplitudes and phases are slowly-varying, so we use
  these as our waveform data pieces.

For subdominant modes, 
we treat $\varphi_+^{\ell, m}$ differently than
for the $(2,\pm 2)$ modes.
We model $\varphi_+^{\ell, m}$ directly instead
of using a Hilbert transform, because for these modes
the Hilbert transform does not improve the model's accuracy.
Fortunately, errors in $\varphi_+^{\ell, m}$
for $\ell>2$ contribute very little to the overall error of the
final model waveform,
as seen in Table~\ref{tab:component_errs} below.

An additional difficulty is that subdominant modes can vanish
at certain points in parameter space, and this makes
phases ill-defined.
Consider a system with $q=1$, $|\vec{\chi}_1| = 0$, and some $\chi_2^z$.
For $\chi_2^z = 0$, the $(2, 1)$ mode vanishes.
For small $\chi_2^z \neq 0$, switching 
  the sign of $\chi_2^z$ will switch the sign of
the $(2, 1)$ mode, meaning that
the phase of the $(2, 1)$ mode has a
discontinuity of $\pi$ as $\chi_2^z$ passes through $0$.
We wish to avoid such discontinuities when building surrogate models.
In this particular example, the discontinuity can be avoided
by defining the amplitude of the
$(2, 1)$ mode to be negative and the phase to be increased by
$\pi$ when $\chi_2^z \leq 0$.

Now consider the general case with arbitrary $\vec{\chi}_1$.  At the
alignment time $t_0$, the orbits of all NR waveforms are aligned.
Because of this, at time $t_0$ the phase of a given $(\ell, m)$ mode
with $m>0$ and even will be approximately equal for all NR waveforms,
i.e. for all choices of parameters.  Similarly, at time $t_0$ the
phase of a given $(\ell, m)$ mode with $m$ odd will either be
approximately equal or will differ by approximately $\pi$ for all
choices of parameters.  
Therefore at $t_0$, the phases of each non-vanishing $(\ell, m)$ mode,
for all choices of parameters, are clustered around either one or two
values, depending on the mode.  Furthermore, when the phases of a
given $(\ell, m)$ mode are clustered around two values instead of one,
the clusters are separated by $\pi$ and the phases of the
corresponding $(\ell, -m)$ mode are also clustered around two values
and not one.  For modes $(\ell, m)$ with phases that are are clustered
around one value, there is no discontinuity in phase as a function of
parameters, and nothing more needs to be done.  But for modes
$(\ell,m)$ with phases clustered around two values, we remove the
discontinuity.  To do this, we arbitrarily choose one of the two
values as the reference phase $\varphi^{\ell, m}_0$, and then compute
the initial phase deviations $\delta \varphi^{\ell, m} =
|\varphi(\tilde{h}^{\ell, m}(t_0)) - \varphi^{\ell, m}_0|$.  Whenever
$\delta \varphi^{\ell, m} + \delta\varphi^{\ell, -m} > \pi$ we take
the amplitudes of the $(\ell, \pm m)$ modes to be negative and
increase the phases of these modes by $\pi$.  This causes the initial
phase of either the $+m$ or $-m$ mode to be $>\pi$, so we subtract
$2\pi$ from that phase.
These transformations preserve the complex waveform mode $\tilde{h}^{\ell, m}$
but transform $A_\pm^{\ell, m} \rightarrow -A_\pm^{\ell, m}$ and
$\varphi_-^{\ell, m} \rightarrow \varphi_-^{\ell, m} + \pi$, leaving
$\varphi_+^{\ell, m}$ unmodified.

\begin{figure}
  \includegraphics[width=\linewidth]{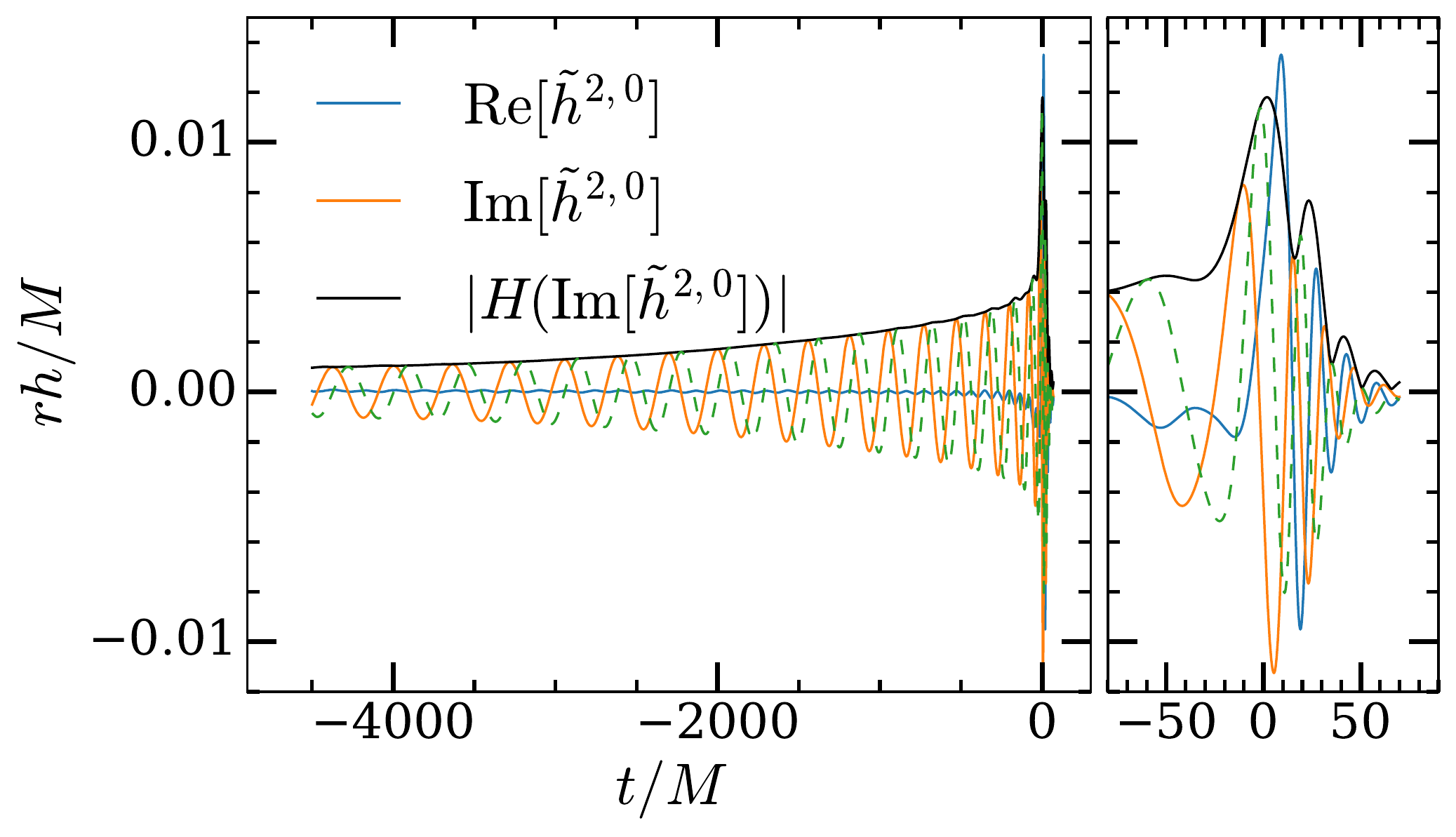}
  \caption{For the real-valued oscillatory components $X$ such as
        $\mathrm{Im}[\tilde{h}^{2, 0}]$, we perform a Hilbert transform
        to obtain a complex signal $H(X)$ and extract an amplitude and phase.
        The dashed green line shows the imaginary part of $H(X)$.
    }
  \label{fig:decomp_2_0}
\end{figure}

Now we discuss modes $\tilde{h}^{\ell,0}(t)$, with $m=0$.
As seen in Fig.~\ref{fig:decomp_2_0}, the
$(2, 0)$ mode has a non-oscillatory real part during the
inspiral, while the imaginary part is small but oscillatory.
The $(3, 0)$ mode is similar, with the roles of the real and imaginary
parts reversed.
Therefore, we do not decompose $\tilde{h}^{\ell,0}(t)$ according
to Eq.~(\ref{Eq:AmpParts}) and~(\ref{Eq:PhaseParts}).  Instead, 
we model the non-oscillatory component directly, and
we take a Hilbert transform
of the oscillatory component to obtain an amplitude and phase,
after tapering
that component
in the intervals $[t_\mathrm{min}, t_0]$ and
$[t_f, t_\mathrm{max}]$.

The decomposition of the 
  NR waveforms is summarized in
Fig.~\ref{fig:DecomposeData}.
The NR waveforms begin at the top of the diagram and are processed
going downwards. Each blue endpoint represents one of the
  slowly-varying waveform data pieces that we fit as a function of
  parameters $\pmb{\lambda}$ at each of the empirical time nodes.
To evaluate the surrogate, the fits and empirical interpolants are
evaluated for each of the blue endpoints,
and the waveform is reconstructed by
going upwards in the diagram and undoing each decomposition,
eventually yielding $h^{\ell, m}(t)$.

\subsection{Removing the dependence on $\phi_\chi$}
\label{sec:phichi}

As discussed in \S~\ref{sec:paramspace}, we attempt to start all
  NR simulations so that at the reference time $t=t_0$ we have
  $\phi_\chi=0$, where $\phi_\chi$ is the azimuthal angle of the
  spin of the larger black hole, as shown in Fig.~\ref{fig:diagram}.
  However, in practice we obtain NR simulations with nonzero values
  of $\phi_\chi$ at $t=t_0$.  In this section we describe how we
  analytically approximate the dependence of the waveform on $\phi_\chi$.
  The surrogate model is then built assuming $\phi_\chi=0$, so that
  when the surrogate model predicts waveforms with $\phi_\chi \ne 0$,
  the $\phi_\chi$ dependence is described fully by this analytical
  approximation.
For an orbit-averaged PN waveform of any order
that is decomposed into waveform data pieces as described
above, it turns out that one can show from
  the equations (e.g. as written in \cite{Ossokine:2015vda}) that
none of the waveform data pieces
depend on the parameter $\phi_\chi$ except for the phase
$\varphi_d(t)$. This phase has a particularly simple dependence:
\begin{equation} \label{eq:Remove_phi_chi}
\varphi_d(t; \pmb{\lambda}, \phi_\chi) = \varphi_d(t; \pmb{\lambda}, 0)
                                            + \phi_\chi,
\end{equation}
where $\pmb{\lambda}$ describes all of the parameters except
$\phi_\chi$.   So we will make the approximation
that Eq.~(\ref{eq:Remove_phi_chi}) applies not only to orbit-averaged
PN waveforms,
but also to NR waveforms lying within the 5d parameter space.
In addition, we
find empirically for NR waveforms
that 
the phases of the
Hilbert transforms of $A_-^{\ell, m}$ and $\varphi_+^{\ell, m}$
also obey Eq.~(\ref{eq:Remove_phi_chi}), but with the opposite sign
on the last term.

Therefore, given a point $\pmb{\lambda}$ in 5d parameter space,
  we first decompose $h_\mathrm{NR}(t;\pmb{\lambda})$ into waveform data
  pieces, and we then subtract
    $\phi_\chi$
from $\varphi_d$ and add $\phi_\chi$ to
the phases of the Hilbert
transforms of $A_-^{\ell, m}$ and $\varphi_+^{\ell, m}$.
We then consider the waveform data pieces as functions of only the
4 parameters
($q$, $|\vec{\chi}_1|$, $\chi_2^z$, and $\theta_\chi$), and we build a 4d
model of each of these waveform data pieces.
When evaluating the surrogate model waveform at a point $\pmb{\lambda}$
in the full 5d
parameter space, 
  we first evaluate the 4d surrogate model expressions for
  the waveform data pieces at
  the parameters ($q$, $|\vec{\chi}_1|$, $\chi_2^z$, and $\theta_\chi$),
  we add $\phi_\chi$ to $\varphi_d$, and we subtract
  $\phi_\chi$ from the phases of the
  Hilbert transforms of $A_-^{\ell, m}$ and $\varphi_+^{\ell, m}$.
  Then we combine the waveform data pieces to yield the model waveform
  $h_\mathrm{sur}(t;\pmb{\lambda})$.

\begin{figure}
  \includegraphics[width=\linewidth]{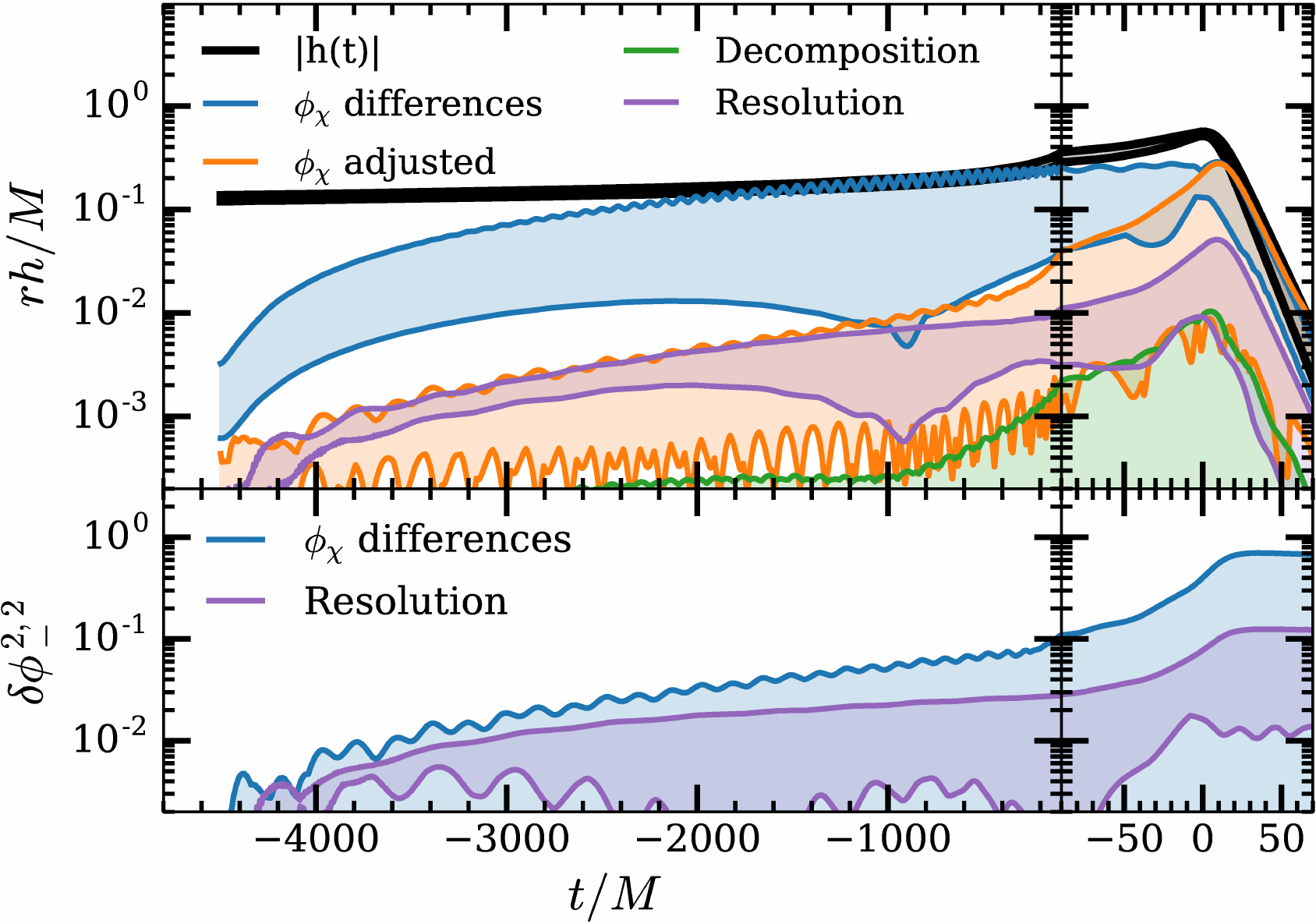}
  \caption{
    Top: waveform differences
        $\delta h(t)$ investigating the removal of the $\phi_\chi$
        dependence on the waveform.
        Each colored band includes waveforms compared to
        SXS:BBH:0346 and SXS:BBH:0346
        for several different
        values of $\phi_\chi$.
        Before making any adjustment, the errors ($\phi_\chi$ differences)
        are large.
        After adjusting, the errors ($\phi_\chi$ adjusted) are comparable
        to resolution errors during the inspiral but grow large at merger.
        The decomposition errors are negligible.
    Bottom: differences in $\varphi_-^{2, 2}$.
        Our analytic approximation to remove the effect of $\phi_\chi$
        on the waveform does not affect $\varphi_-^{2, 2}$, but here we
        see that the orbital phase at merger can vary by nearly a radian
        for different values of $\phi_\chi$, which is the most significant
        contribution to the $\phi_\chi$ adjusted errors in the top figure.
    }
  \label{fig:check5d_inertial}
\end{figure}

To verify how well this procedure removes the dependence on $\phi_\chi$,
we performed additional SpEC simulations with parameters identical to cases
SXS:BBH:0346 and SXS:BBH:0346
but with different values of $\phi_\chi$.
We then analytically remove the $\phi_\chi$ dependence from all
  these waveforms, as described above, thereby generating
  $\phi_\chi=0$ versions of these waveforms, which we compare with each
  other. The agreement (or lack thereof) of these $\phi_\chi=0$ waveforms
  is a measure of the effectiveness of our analytical procedure
  for removing the $\phi_\chi$ dependence.
We find that while the dependence on $\phi_\chi$ is removed well during
the inspiral, $\varphi_-^{2, 2}(t)$ varies by nearly a radian during the
merger as we vary $\phi_\chi$, which leads to errors significantly larger
than the SpEC resolution errors as shown in
Fig.~\ref{fig:check5d_inertial}.
Incidentally, we note that for two waveforms for which
$\phi_\chi$ originally differs by $\pi$,
the corresponding $\phi_\chi=0$ waveforms are nearly identical.
Before removing $\phi_\chi$, the largest difference in the waveforms
used in this test is
$\mathcal{E} = 0.0285$, while after removing $\phi_\chi$, the
largest difference
is $\mathcal{E} = 0.00684$.
While our $\phi_\chi$-removal procedure
successfully accounts for most of the effect of
$\phi_\chi$, the error associated with this procedure
is larger than the median surrogate error
(see Fig.~\ref{fig:error_hist} and Sec.~\ref{sec:Assessment})
and indicates this approximation could be the
dominant source of error in the surrogate model.

\subsection{Handling undefined phases}
\label{subsec:omitBadData}
Our waveform
decomposition scheme results in many phases, which become
undefined when their corresponding amplitudes vanish.
For example, $\varphi_d$ is undefined for non-precessing systems,
as are the phases of the Hilbert transforms of nutating quantities.
Additionally, the amplitudes of subdominant modes in the coprecessing
frame can briefly become $0$, making the corresponding
$\varphi_\pm^{\ell, m}$ quantities undefined.
Since the NR waveforms contain numerical noise, in practice the
phases become poorly resolved when the corresponding amplitude becomes
comparable to the noise level.

When decomposing each NR waveform into waveform data pieces,
if one of the amplitudes $A(t)$ falls below some threshold at any time $t$
before the merger, then the corresponding phase $\varphi(t)$ is omitted
from the model for that NR waveform.
This means that when building empirical interpolants or
fitting across parameter space at empirical nodes, we use
fewer than our entire set of 288
waveforms to fit that particular $\varphi(t)$.
The thresholds are described in Table
\ref{tab:thresholds}.

\begin{table}
\begin{tabular}{c | c | c | c | c}
Data        & Quantity used    &   Tol &   N pass  & N reject \\
\hline
$\varphi_d$ & $|\vec{\chi}_1|\mathrm{sin}(\theta_\chi)$ &   $10^{-3}$ & $192$ & $96$ \\
$\varphi[H[A_-^{2, 2}]]$ & $|H[A_-^{2, 2}]]|$ & $10^{-6}$ & $192$ & $96$ \\
$\varphi[H[\varphi_+^{2, 2}]]$ & $|H[\varphi_+^{2, 2}]]|$ & $10^{-4}$ & $169$ & $119$ \\
$\varphi_-^{2, 1}$ & $A_+^{2, 1}$ & $10^{-4}$ & $260$ & $28$ \\
$\varphi_+^{2, 1}$ & $A_+^{2, 1}$ & $\infty$ & $0$ & $288$ \\
$\varphi[H[A_-^{2, 1}]]$ & $|H[A_-^{2, 1}]]|$ & $3 \times 10^{-6}$ & $97$ & $191$ \\
$\varphi[H[\mathrm{Im}[\tilde{h}^{2, 0}]]]$ &
        $|H[\mathrm{Im}[\tilde{h}^{2, 0}]]|$ & $2 \times 10^{-6}$ & $190$ & $98$ \\
$\varphi_\pm^{3, 3}$ & $A_+^{3, 3}$ & $10^{-3}$ & $210$ & $78$ \\
$\varphi[H[A_-^{3, 3}]]$ & $|H[A_-^{3, 3}]]|$ & $3 \times 10^{-6}$ & $166$ & $122$ \\
$\varphi[H[A_-^{3, 2}]]$ & $|H[A_-^{3, 2}]]|$ & $10^{-6}$ & $140$ & $148$ \\
$\varphi_\pm^{3, 1}$ & $A_+^{3, 1}$ & $10^{-4}$ & $137$ & $151$ \\
$\varphi[H[A_-^{3, 1}]]$ & $|H[A_-^{3, 1}]]|$ & $2 \times 10^{-6}$ & $135$ & $153$ \\
$\varphi[H[\mathrm{Im}[\tilde{h}^{3, 0}]]]$ &
        $|H[\mathrm{Im}[\tilde{h}^{3, 0}]]|$ & $2 \times 10^{-6}$ & $86$ & $202$ \\
\end{tabular}
\caption{Tolerances used to omit poorly resolved phases.
        Other than the tolerance for $\varphi_d$, which is based on
        the amount of in-plane spin, the tolerances are based on the
        minimum value of some amplitude before $t=0$. If a tolerance
        is not listed for a particular phase parameter,
        for example $\varphi_\pm^{\ell, 2}$,
        then that phase parameter is always
        included in the surrogate.  The columns $N$ pass and $N$ reject describe
        the number of waveforms for which a phase is included in the
        surrogate, and the number for which it is not.
        Note that we have a total of 288 waveforms but only
        276 NR simulations, because a few of the NR simulations allow us
        to compute waveforms for more than one set of parameters because of
        symmetry considerations (cf. \S~\ref{sec:simulations}).
        }
\label{tab:thresholds} 
\end{table}

\section{Building a surrogate model from decomposed waveforms}
\label{sec:DataSurrogate}

We have decomposed each NR waveform into many functions
$X(t; \pmb{\lambda})$ that are smoothly varying as a function of
parameters $\pmb{\lambda}$.
Here, $X$ represents one of the many decomposed waveform data pieces
such as $\varphi_p$ or $A^{2, 2}_+$.  Note that while
different waveform data pieces
$X$ will have different
linear basis sizes, empirical time nodes, empirical interpolants, and
parameter space fits,
we will not always label the explicit $X$ dependence
of these quantities.
For each $X$ we have several NR solutions with different parameters
$\{X(t; \pmb{\lambda}) \, : \, \pmb{\lambda} \in G^X\}$
where $G^X \subset G = \{\pmb{\Lambda}_i\}_{i=1}^N$.
We note that the only reason we might not have $G^X = G$
is due to omitting cases with
undefined phases discussed in Sec.~\ref{subsec:omitBadData}.
The next step is to model each of those functions $X$
with its own surrogate model
$X_S$ by building an empirical interpolant and fitting the empirical
nodes across the parameter space $\mathcal{T}$.
The surrogate model for the waveform $h_S^{\ell, m}(t; \pmb{\lambda})$
will then evaluate $X_S(t; \pmb{\lambda})$ for each waveform data piece,
from which the inertial frame
waveform modes $\{h_S^{\ell, m}(t; \pmb{\lambda})\}$ will be reconstructed.
These stages are discussed below.

\subsection{Empirical Interpolation}
\label{sec:empir-interp}
For each waveform data piece $X$, we build an empirical interpolant
using the available solutions
$\{X(t; \pmb{\lambda}) \, : \, \pmb{\lambda} \in G^X\}$.
Here we address modifications to the standard empirical interpolation
method discussed in
Sec.~\ref{subsec:sur_basics}.

We require an orthonormal basis $B$ spanning the space
of solutions $\{X(t; \pmb{\lambda})\,:\,\lambda \in \mathcal{T}\}$.
While the standard method is to use a reduced basis that was previously constructed
when determining the greedy parameters $G$, in our case we used PN
waveforms to find the greedy parameters and have not yet built a basis for
NR solutions of $X$.
Greedy and singular value decomposition (SVD) algorithms
have been used within the gravitational wave surrogate modelling
community~\cite{Blackman:2015pia,Field:2013cfa,Purrer:2014,Purrer:2015tud},
and will both provide
an accurate basis provided any $X(t, \pmb{\lambda})$ can be
accurately approximated in the span of
$\{X(t, \pmb{\lambda})\,:\,\pmb{\lambda} \in G^X\}$.
A short discussion,
including advantages and disadvantages of SVD and greedy
algorithms in the context of surrogate waveform modeling,
is given in Appendix~\ref{sec:RB_SVD}. Despite using a greedy sampling
strategy to identify the set of greedy parameters,
we use a SVD basis for the NR solutions, primarily for
its ability to average out uncorrelated noise sources
(see Appendix~\ref{sec:RB_SVD}).

We truncate the orthonormal basis and use the first $n$
singular values and vectors such that all projection errors are below the tolerances
given in Table \ref{tab:basistol}.
We note that $n$ will be different for different waveform data pieces.
We then proceed according to Sec.~\ref{subsec:sur_basics}, finding empirical
time nodes $\{T_j\}_{j=1}^{n}$ and building an empirical interpolant
$I_{n}$.
If we are given $X_S$ at the empirical nodes $T_j$, we can now determine
\begin{equation}
X_S(t; \pmb{\lambda}) = I_{n}[X_S](t)
\end{equation}
for all times $t \in [t_\mathrm{min}, t_\mathrm{max}]$.

\begin{table}
\begin{tabular}{c | c || c | c || c | c || c | c}
Data & Tol & Data & Tol & Data & Tol & Data & Tol \\
\hline
$\varphi_p$ & $0.005$ & $\varphi_+^{3, 3}$ & $10.0$ &
        $\varphi[H[A_-^{2, 2}]]$ & $0.3$ & $\varphi_-^{2, 2}$ & $0.15$ \\
$\varphi_d$ & $0.03$ & $\varphi_+^{3, 2}$ & $10.0$ &
        $\varphi[H[A_-^{2, 1}]]$ & $1.0$ & $\varphi_-^{2, 1}$ & $1.0$ \\
$\varphi[H[\tilde{h}^{2, 0}]]$ & $0.5$ & $\varphi_+^{3, 1}$ & $10.0$ &
        $\varphi[H[A_-^{3, 3}]]$ & $10.0$ & $\varphi_-^{3, 3}$ & $0.3$ \\
$\varphi[H[\tilde{h}^{3, 0}]]$ & $0.5$ & $A_+^{2, 2}$ & $0.001$ &
        $\varphi[H[A_-^{3, 2}]]$ & $10.0$ & $\varphi_-^{3, 2}$ & $0.3$ \\
$|H[\varphi_+^{2, 2}]|$ & $0.15$ & $A_+^{2, 1}$ & $0.001$ &
        $\varphi[H[A_-^{3, 1}]]$ & $1.0$ & $\varphi_-^{3, 1}$ & $10$ \\
$\varphi[H[\varphi_+^{2, 2}]]$ & $10.0$ & $A_+^{3, 2}$ & $0.0003$ &
        & & & \\
\end{tabular}
\caption{Projection error RMS tolerances for each basis.
        Unlisted quantities have a default tolerance of $0.003$ for amplitudes
        and $0.03$ for phases.}
\label{tab:basistol}
\end{table}

\subsection{Parametric Fits}
\label{sec:param_space_fits}

The next step is to model the dependence on $\pmb{\lambda}$
of the waveform data pieces at the empirical nodes
\begin{equation}
X_j(\pmb{\lambda}) = X(T_j; \pmb{\lambda})\,.
\end{equation}
We build an approximate model
for $X_j$ denoted by $X_{jS}$
by fitting it to the available data
$\{X_j(\pmb{\lambda})\,:\,\pmb{\lambda} \in G^X\}$.
We do so using a forward-stepwise least-squares fit~\cite{Hocking1976}
described in Appendix~\ref{sec:fitappendix}, using products of
univariate basis functions in $q$, $|\vec{\chi}_1|$, $\theta_\chi$
and $\chi_2^z$ as the fit features.
For each fit, the number of fit coefficients is determined through
a cross validation study using $50$ trials, each of which
uses $N_v = 5$ randomly chosen validation points.
The number of fit coefficients used is the one minimizing the
sum in quadrature over the error in each trial, which is the
maximum fit residual for the validation points.

\subsection{Complete Surrogate Waveform Model in inertial coordinates}

Given parameters
$\pmb{\lambda}_5 = (q, |\vec{\chi}_1|, \theta_\chi, \phi_\chi, \chi_2^z)$,
we extract $\pmb{\lambda} = (q, |\vec{\chi}_1|, \theta_\chi, \chi_2^z)$ and
evaluate the fits and empirical interpolants of each waveform data piece
$X$, obtaining
\begin{equation}
X_S(t; \pmb{\lambda}) = \sum_{j=1}^{n} X_{jS}(\pmb{\lambda})
                                                b^{j}(t)\,.
\end{equation}
We then obtain the inertial frame waveform $h_S^{\ell, m}(t)$
by combining the waveform
data pieces and flowing upwards in Fig.~\ref{fig:DecomposeData}.
Explicitly,
\begin{align}
Q(t) &= T_\varphi^{-1}\left(\varphi_{d}(t; \pmb{\lambda}) +
                \phi_\chi, \varphi_{p}(t; \pmb{\lambda})\right) \\
\varphi_\mathcal{I}^{2, 0}(t) &= \varphi(H(\mathcal{I}\tilde{h}^{2, 0}))(
                                 t; \pmb{\lambda}) - \phi_\chi \\
\mathcal{I}\tilde{h}^{2, 0}(t) &= A(H(\mathcal{I}\tilde{h}^{2, 0}))(
                t; \pmb{\lambda})\mathrm{cos}(\varphi_\mathcal{I}^{2, 0}(t)) \\
\tilde{h}^{2, 0}(t) &= \mathcal{R}\tilde{h}^{2, 0}(t; \pmb{\lambda}) +
                    i\mathcal{I}\tilde{h}^{2, 0}(t)\\
\varphi_\mathcal{R}^{3, 0}(t) &= \varphi(H(\mathcal{R}\tilde{h}^{3, 0}))(
                                 t; \pmb{\lambda}) - \phi_\chi \\
\mathcal{R}\tilde{h}^{3, 0}(t) &= A(H(\mathcal{R}\tilde{h}^{3, 0}))(
                t; \pmb{\lambda})\mathrm{cos}(\varphi_\mathcal{R}^{3, 0}(t)) \\
\tilde{h}^{3, 0}(t) &= \mathcal{R}\tilde{h}^{3, 0}(t) +
                    i\mathcal{I}\tilde{h}^{3, 0}(t; \pmb{\lambda})\\
\varphi^{\ell, \pm m}(t) &= \varphi_{+}^{\ell, m}(t; \pmb{\lambda}) \pm
                             \varphi_{-}^{\ell, m}(t; \pmb{\lambda}),\,\,\, m>0 \\
A^{\ell, \pm m}(t) &= A_{+}^{\ell, m}(t; \pmb{\lambda}) \pm
                       A_{-}^{\ell, m}(t; \pmb{\lambda}),\,\,\, m>0 \\
\tilde{h}^{\ell, \pm m}(t) &= A^{\ell, \pm m}(t)\mathrm{cos}(
            \varphi^{\ell, \pm m}(t)),\,\,\, m>0 \\
\{h^{\ell, m}(t)\} &= T_Q(\{\tilde{h}^{\ell, m}(t)\}, Q(t))
\end{align}
where we have included the dependence on $\pmb{\lambda}$ explicitly
for surrogate evaluations of waveform data pieces $X_S$.
The full NRSur4d2s surrogate evaluation producing all $2 \leq \ell \leq 4$ modes
for an array of times between $t_\mathrm{min}$ and $t_\mathrm{max}$
with spacing $\delta t = 0.1$ takes $\sim 1s$ on a single modern processor.
Roughly half of this time is spent computing the transformation $T_Q$
from the coprecessing frame to the inertial frame, Eq.~(\ref{eq:T_Q}).
\section{Assessing the Model Errors} \label{sec:Assessment}

\subsection{Time Domain Errors}
\label{sec:time-domain-errors}
To determine how well the output of the NRSur4d2s
surrogate matches a NR waveform
with the same parameters, we compute
\begin{equation}
  \mathcal{E}[h_\mathrm{NR},h_\mathrm{Sur}]
  = \frac{1}{2}\frac{\delta h}{\|h_\mathrm{NR}\|^2},
  \label{eq:TimeDomainError}
\end{equation}
where $h_\mathrm{NR}$ and $h_\mathrm{Sur}$ are the
  NR and surrogate waveforms,
and $\delta h$ is given by Eq.~(\ref{eq:deltahfromModes}). This
quantifies the surrogate error as a whole at one point in parameter space.
For NR waveforms that were used to build the surrogate, we
  call Eq.~(\ref{eq:TimeDomainError}) the {\em training error}.  For NR
  waveforms that were not used to build the surrogate, but are used to
  test the accuracy of the surrogate model
  versus NR, we call Eq.~(\ref{eq:TimeDomainError}) the
  {\em validation error}.
Because we decompose each waveform into a set of slowly-varying functions
that are modeled independently (i.e., the waveform data pieces of
\S~\ref{sec:decomposition}), it is
useful to consider the contribution to the surrogate error
that arises from modeling a single waveform data piece.
If $X$ denotes the waveform data piece in question, then we compute this
error contribution by decomposing the NR waveform $h_\mathrm{NR}$
into waveform data pieces,
we replace the NR version of $X$ with the surrogate model for $X$ while
leaving all waveform data pieces other than $X$ untouched, and we
recombine the waveform data pieces, thus producing a waveform we call
$h_\mathrm{X}$.
The error contribution from $X$ is then $\mathcal{E}_X \equiv
\mathcal{E}[h_\mathrm{NR},h_\mathrm{X}]$. Values of $\mathcal{E}_X$ for
various waveform data pieces $X$ are listed in Table~\ref{tab:component_errs}.
Note that if we decompose $h_\mathrm{NR}$ into waveform data pieces
and then recompose the waveform data pieces, we
do not recover $h_\mathrm{NR}$ exactly, but instead we get a different
waveform $h_\mathrm{\varnothing}$ because there is error associated
with the decomposition.  This error, $\mathcal{E}_\varnothing \equiv
\mathcal{E}[h_\mathrm{NR},h_\mathrm{\varnothing}]$, is also shown in
Table~\ref{tab:component_errs}.

\begin{figure}
  \includegraphics[width=\linewidth]{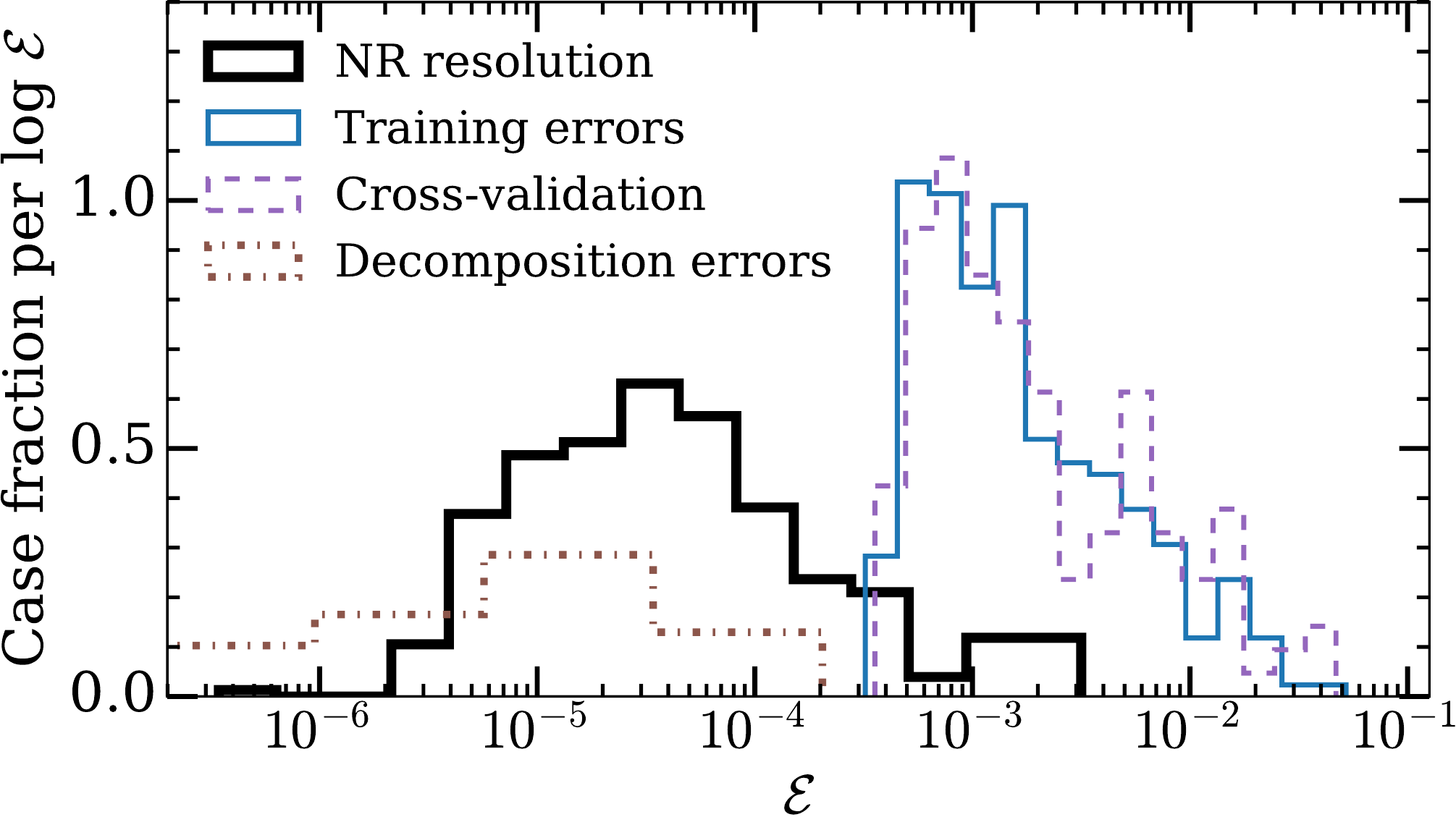}
  \caption{
    Histograms of time
    domain waveform errors $\mathcal{E}$ relevant to the surrogate.
    Equal areas under the curves correspond to equal numbers of cases,
    and the curves are normalized such that the total area under each curve
    when integrated over $\mathrm{log}_{10}(\mathcal{E})$
    is $1$.
    Solid black: The resolution error comparing the highest and second highest
        resolution NR waveforms.
    Dotted brown: The error intrinsic to the surrogate's waveform decomposition.
        Filtering out nutation in the quaternions and neglecting the
        small but non-zero $\delta q_z$ due to discrete time sampling
        leads to errors in the reconstructed waveforms.
        These errors are nearly zero for non-precessing
        cases, and even for precessing cases they are smaller than
        the resolution errors.
        Thin solid blue: The errors when the full
        surrogate attempts to reproduce the
        set of waveforms from which it was built.
    Dashed purple: The errors when trial surrogates attempt to reproduce
        NR waveforms that were omitted during the surrogate construction.
    }
  \label{fig:error_hist}
\end{figure}

\begin{table}
\begin{tabular}{| c | c | c | c || c | c | c | c |}
\hline
$X$ & $\mathcal{E}_X^0$ & $\mathcal{E}_X^\mathrm{max}$ & $\mathcal{E}_X^\mathrm{median}$ &
$X$ & $\mathcal{E}_X^0$ & $\mathcal{E}_X^\mathrm{max}$ & $\mathcal{E}_X^\mathrm{median}$ \\
\hline

$\varnothing$       & $0.0006$ & $0.0006$ & $0.0003$ &
        $q$                 & $0.2450$ & $0.0089$ & $0.0004$ \\
$h$                 & $0.5$     & $0.0521$ & $0.0014$ &
        $\varphi_p$         & $0.2450$ & $0.0095$ & $0.0004$ \\
$\tilde{h}$         & $0.5$     & $0.0478$ & $0.0013$ &
        $\varphi_d$         & $0.4171$ & $0.0008$ & $0.0003$ \\
$\tilde{h}^{2, 0}$  & $0.0006$ & $0.0006$ & $0.0003$ &
        $\tilde{h}^{2, \pm 2}$  & $0.4999$ & $0.0461$ & $0.0011$ \\
$\tilde{h}^{2, \pm 1}$  & $0.0044$ & $0.0016$ & $0.0004$ &
        $A_+^{2, 2}$        & $0.4999$ & $0.0007$ & $0.0003$ \\
$\tilde{h}^{3, 0}$  & $0.0006$ & $0.0006$ & $0.0003$ &
        $A_-^{2, 2}$        & $0.0018$ & $0.0010$ & $0.0003$ \\
$\tilde{h}^{3, \pm 1}$  & $0.0006$ & $0.0006$ & $0.0003$ &
        $\varphi_+^{2, 2}$  & $0.0027$ & $0.0049$ & $0.0004$ \\
$\tilde{h}^{3, \pm 2}$  & $0.0008$ & $0.0007$ & $0.0003$ &
        $\varphi_-^{2, 2}$  & $0.9959$ & $0.0446$ & $0.0009$ \\
$\tilde{h}^{3, \pm 3}$  & $0.0043$ & $0.0020$ & $0.0004$ &
        & & & \\

\hline
\end{tabular}
\caption{Maximum and median errors when attempting to reproduce the
    set of NR waveforms
    when a single waveform data piece
    is replaced $X$ with its surrogate
    evaluation $X_S$ and the waveforms are reconstructed.
    This can be compared with $\mathcal{E}_X^0$, which is the maximum
    error when replacing $X$ with $0$ (or the identity
    quaternion when $X=q$) instead of with $X_S$.
    When $X=\varnothing$ we replace no waveform data piece,
    but there is still decomposition error due to the lack of $\ell > 3$
    modes in the surrogate waveforms, filtering, and neglecting $q_z$.
    Note that the errors for $\tilde{h}^{\ell, \pm m}$ include
    replacing both the $(\ell, m)$ and $(\ell, -m)$ coprecessing modes.
    Some components $X$ (such as $X=\tilde{h}^{3, 0}$) have
    $\mathcal{E}_X^0 \sim \mathcal{E}_\varnothing^\mathrm{max}$, indicating
    the error associated with replacing $X$ with $0$ is similar to or
    smaller than the decomposition errors.
    $\varphi_-^{2, 2}$ is the biggest source of error in the surrogate,
    although $\varphi_p$ also contributes significantly.
}
\label{tab:component_errs}
\end{table}

\begin{figure}
  \includegraphics[width=\linewidth]{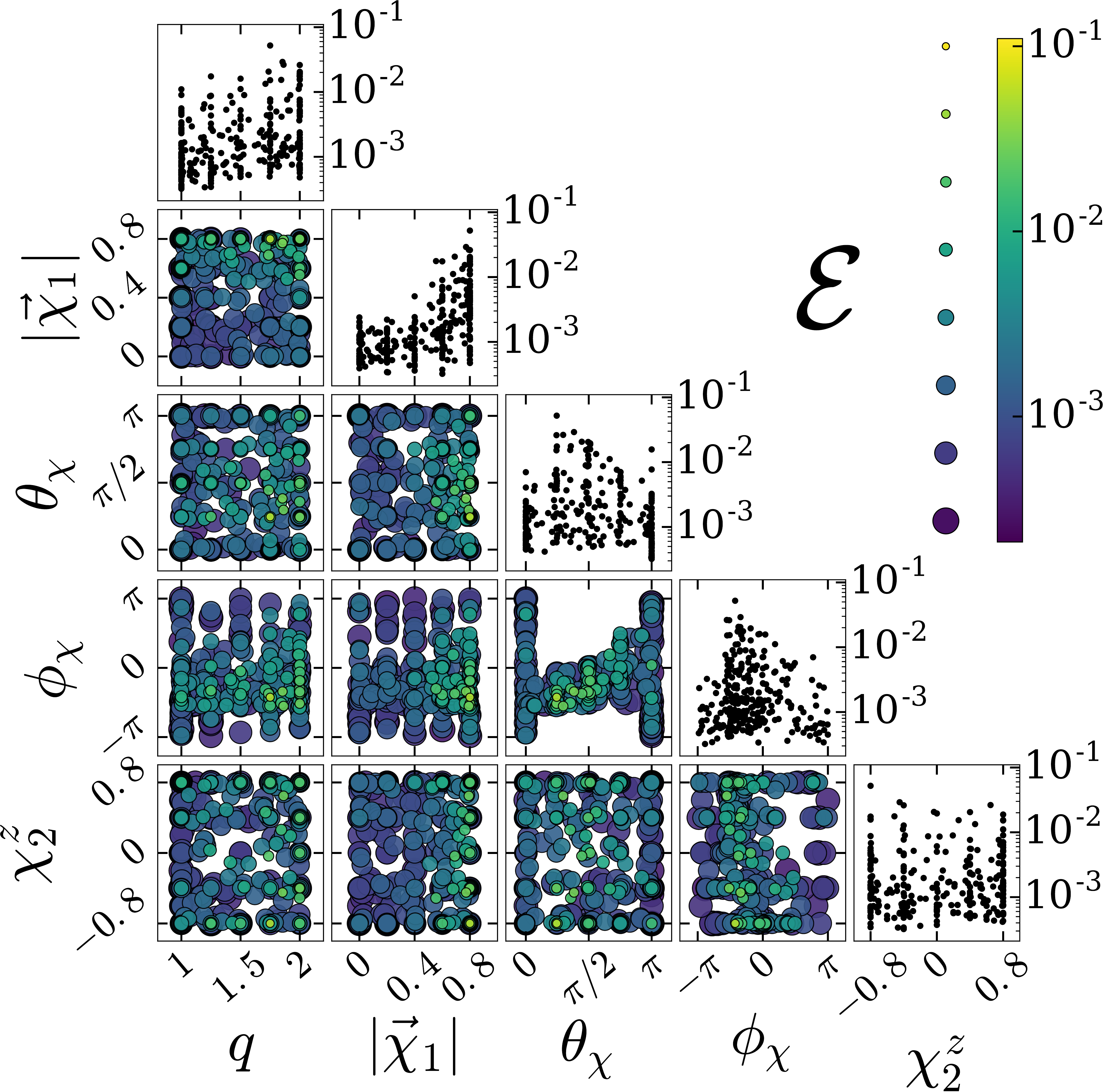}
  \caption{
    Parameter dependence of 
        the error $\mathcal{E}[h,h_S]$ when reproducing the set of NR waveforms
        with the surrogate.
    Diagonal: For each parameter plot, the black dots label the
    (parameter value, $\mathcal{E}[h,h_S]$) pairs.
    Off-diagonal: For each pair of parameters, we show the $2d$
        projection of parameters
        as in Fig.~\ref{fig:training_points}
        while varying the color and size of the point based on the
        error $\mathcal{E}[h,h_S]$.
        Points are placed in order of increasing error, to ensure
        the small yellow points with large errors are visible.
        Larger spin magnitudes, especially for precessing spin
        configurations, correlate with larger errors.
    }
  \label{fig:err_vs_params}
\end{figure}

A first test is to verify that the NRSur4d2s surrogate can reproduce the
set of NR waveforms from which it was built.
The errors for those parameters are shown as the solid blue curve in
Fig.~\ref{fig:error_hist}.
These 
errors are significantly larger than the NR resolution errors
(cyan curve), which compare the highest and second highest NR
resolutions.
This indicates either that 
  including additional NR waveforms when building the surrogate model
would reduce the training error,
or that 
the error is dominated by approximations made when building the model,
such as the analytic treatment of $\phi_\chi$.
The median training
error is $0.00136$,
and in Sec.~\ref{sec:phichi} we found
that our approximation for the waveform's dependence on
$\phi_\chi$ resulted in errors up to $0.00684$, indicating
the model errors could be dominated by the error in this
approximation.
While the maximum training
error is $0.05212$, we only investigated
the dependence on $\phi_\chi$ for three cases and only for
a few values of $\phi_\chi$.
The parametric dependence of the training errors is illustrated in
Fig.~\ref{fig:err_vs_params}.
Perhaps unsurprisingly, the largest errors occur at larger mass ratios and
spin magnitudes, and for precessing spin directions.

To test the interpolation accuracy of the surrogate, we perform a
cross-validation study.  For each of $10$ trials,
we randomly select $N_{\rm v}=10$ waveforms which we call
{\em validation waveforms}, and we build a trial surrogate using the
remaining $N_{\rm t} = N - N_{\rm v}$ waveforms.  The trial surrogate
is evaluated at the $N_{\rm v}$ validation parameters, and the results
are compared to the validation NR waveforms.  These validation errors
are shown as the purple dashed curve in
Fig.~\ref{fig:error_hist}.  The validation errors are quite similar to
the training errors, indicating we are not overfitting the data.

\begin{figure}
  \includegraphics[width=\linewidth]{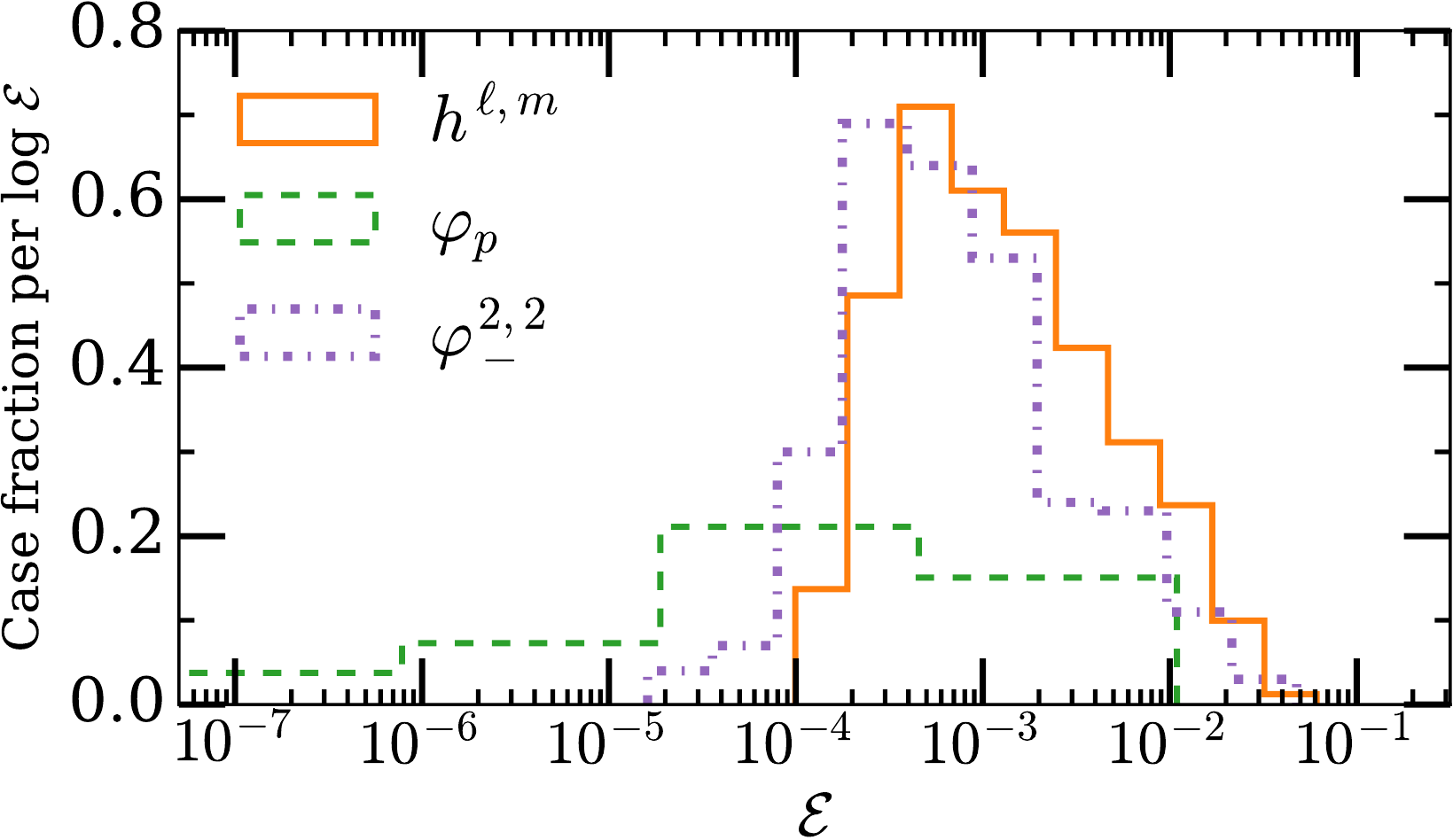}
  \caption{
    Errors $\mathcal{E}_X$ showing the error contribution of a single
    surrogate component $X$.
    }
  \label{fig:component_hist}
\end{figure}

\begin{figure}
  \includegraphics[width=\linewidth]{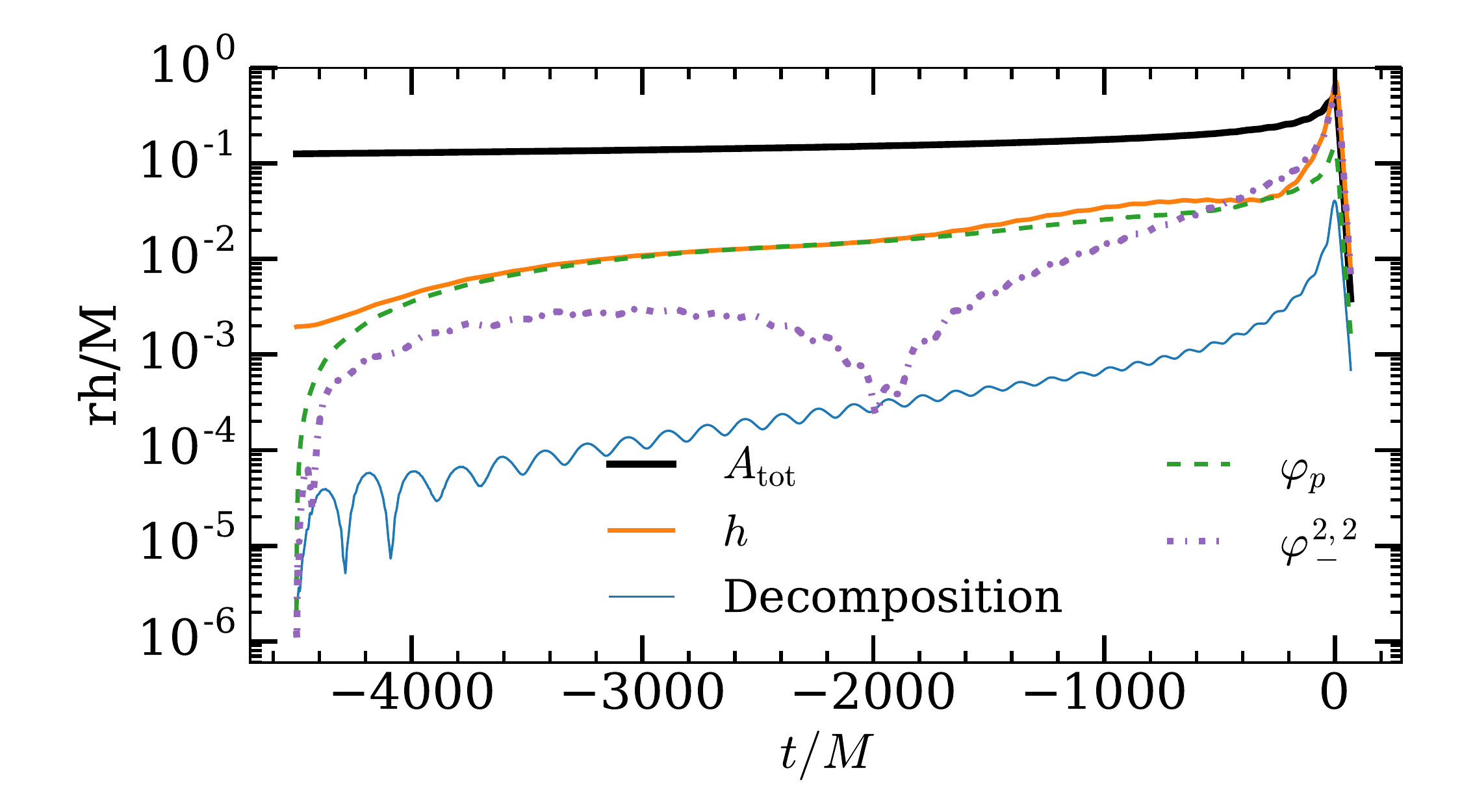}
  \caption{
    Error contributions
    $\delta h(t)$ of those waveform data pieces $X$
    that have the largest
    error $\mathcal{E}[h,h_X]$ for a selected simulation: ID $79$.
    To compute the error, the NR waveform
    is decomposed into the surrogate components, and component $X$
    is replaced with its surrogate evaluation. The waveform is then
    reconstructed, and $\delta h(t)$ is computed from
    Eq.~(\ref{eq:delta_h_of_t}). The solid black curve
    is given by Eq.~(\ref{eq:peak}). The dashed curve
    is the error in $\varphi_p$, which is the dominant error in modeling
    the precession, and the dominant error source during the
    inspiral. The dotted curve is the error in a quantity similar to twice
    the
    orbital phase, and becomes the dominant error source during the
    merger and ringdown. The contribution from errors in the other
    waveform data pieces is smaller, as shown in
    Table~\ref{tab:component_errs}.
    }
  \label{fig:component_err}
\end{figure}

The maximum and median values
of the training errors $\mathcal{E}_X$ 
are listed in Table \ref{tab:component_errs}.
The decomposition errors $\mathcal{E}_\varnothing$, also shown as the
dotted brown curve
in Fig.~\ref{fig:error_hist}, are similar or smaller to the
NR resolution errors and are therefore negligible.
All component errors $\mathcal{E}_X$ include the decomposition errors
by construction, and we see that $X=\tilde{h}^{\ell, m}$ leads to
negligible errors except for the $(2, 2)$, $(2, 1)$ and $(3, 3)$ modes.
The $(2, 2)$ mode is the dominant contribution to the error,
and its error is dominated by the error in $\varphi^{2, 2}_-$.
The precession phase $\varphi_p$ is the dominant precession error,
and is the next most significant contribution to the total error in $h$.
Fig.~\ref{fig:component_hist} shows histograms of the dominant sources of
error, and Fig.~\ref{fig:component_err} shows the time-dependent errors
of these components for the case with the largest training
error.

We have constructed the surrogate models and computed $\mathcal{E}$
assuming zero orbital eccentricity.  However, it is not possible to
construct NR simulations with exactly zero eccentricity, and
the simulations used to build
the surrogate have eccentricities of up to $0.00085$.
To estimate the effect that the eccentricity of the NR waveforms
has on our surrogate, we repeated two of our NR simulations
changing nothing except the eccentricity.
The errors we found are listed in Table \ref{tab:ecc_errors}.
The largest eccentricities in these additional simulations are
several times larger than the maximum eccentricity in the NR simulations
used to build the surrogate, yet the resulting waveform errors are smaller
than the surrogate errors and comparable to the NR resolution errors.
This suggests that the small eccentricities present in the NR waveforms
used to build the surrogate are negligible compared to the NR resolution
errors.

\begin{table}
\begin{tabular}{ | c | c | c | }
\hline
Reference Case & Ecc & $\mathcal{E}$ \\
\hline
SXS:BBH:0534 & $0.000375$ & $0.000007$ \\
SXS:BBH:0534 & $0.002272$ & $0.000162$ \\
SXS:BBH:0546 & $0.000316$ & $0.000004$ \\
SXS:BBH:0546 & $0.000381$ & $0.000005$ \\
SXS:BBH:0546 & $0.002389$ & $0.000106$ \\
\hline
\end{tabular}
\caption{Errors $\mathcal{E}[h_0, h_{\rm{ecc}}]$ where
$h_0$ is the waveform from a reference case used to build the surrogate and
$h_{\rm{ecc}}$ is a waveform from a NR simulation
with nearly identical parameters
but with a larger eccentricity.
For SXS:BBH:0534, $h_0$ has an eccentricity of $0.000027$, and
for SXS:BBH:0534, $h_0$ has an eccentricity of $0.000055$.}
\label{tab:ecc_errors}
\end{table}

\subsection{Frequency-domain comparisons}

In this section we compute mismatches in the frequency domain
  between surrogate waveforms and NR waveforms.  To ascertain the
  significance of these mismatches, we also compute mismatches between
  two NR waveforms with the same parameters but different resolutions.
  For comparison, we also compute mismatches between NR waveforms
  and the phenomenological
  inspiral-merger-ringdown waveform model IMRPhenomPv2 
  (which follows the procedure outlined in~\cite{Hannam:2013oca}
  with IMRPhenomD~\cite{Khan:2015jqa} as the aligned-spin model)
  and between the
  effective-one-body model SEOBNRv3
  \cite{Pan:2013rra}, both of which include the effects of precession.

We minimize the frequency domain mismatches over time
and polarization angle shifts analytically as described in
Appendix~\ref{app:mismatches}, and also minimize them
over orbital phase shifts numerically.
When we compare two
waveforms, we choose one waveform as the {\em reference} waveform
with fixed parameters, and optimize over the parameters of the other
waveform.  When comparing two NR waveforms, the reference waveform
is the one with the highest resolution; when comparing NR with some
model waveform, the NR waveform is chosen as the reference.

The SEOBNRv3 and IMRPhenomPv2
waveforms are generated with the
lalsimulation package \cite{LAL}.
Each SEOBNRv3 waveform is generated in the time domain;
the spin directions are specified at the start of the waveform,
which is determined by specifying a minimum frequency.
We ensure the spin directions are consistent with those of the NR
waveforms by varying the minimum frequency in order to obtain a waveform
with a peak amplitude occurring $4500M$ after its initial time.
The IMRPhenomPv2 waveforms are generated in the frequency domain,
and the spin directions are specified at a reference frequency
$f_\mathrm{ref}$ that can
be freely chosen. For IMRPhenomPv2 it is not straightforward
to determine $f_\mathrm{ref}$ such that the spin directions are specified
at a time of $4500M$ before the peak amplitude.  Therefore, we
instead choose $f_\mathrm{ref}$ differently: we minimize the mismatches
by varying $f_\mathrm{ref}$, with an initial guess of twice the initial
orbital frequency of the NR waveform.

To
transform the time domain waveforms into the frequency domain, we
first taper them using
Planck windows\cite{McKechan:2010kp}, rolling on for 
$t \in [t_0,t_0+1000M]$ and rolling off
for $t \in [50M, 70M]$ where $t_0=-4500M$ is the time at which
  the parameters are measured, and t=0 is the time of peak waveform
amplitude.
We then pad them with zeros and compute the frequency domain
waveforms via the fast Fourier transform (FFT).
For the reference NR waveform, 
we obtain $30$ random samples of
the direction of gravitational wave propagation
$(\theta, \phi)$ from a distribution uniform in $\cos\theta$ and in $\phi$,
and we
uniformly sample the polarization angle $\psi$ between $[0, \pi]$ to obtain
\begin{equation}
h_\psi(t) = h_+(t)\mathrm{cos}(2\psi) + h_\times(t)\mathrm{sin}(2\psi).
\end{equation}
For the non-reference waveform, we use the same parameters except
we add an additional initial azimuthal rotation angle $\phi$,
a polarization angle $\psi$, and a time offset, and we optimize over
these three new parameters to yield a minimum mismatch.
Because the waveform models do not intrinsically depend on the total mass,
we first use a flat noise curve to evaluate the overlap integrals;
this provides a raw comparison between models.
We evaluate Eq.~\ref{eq:match}
with $f_\mathrm{min}$ being twice the orbital frequency of the NR
waveform at $t=-3500M$.

\begin{figure}
  \includegraphics[width=\linewidth]{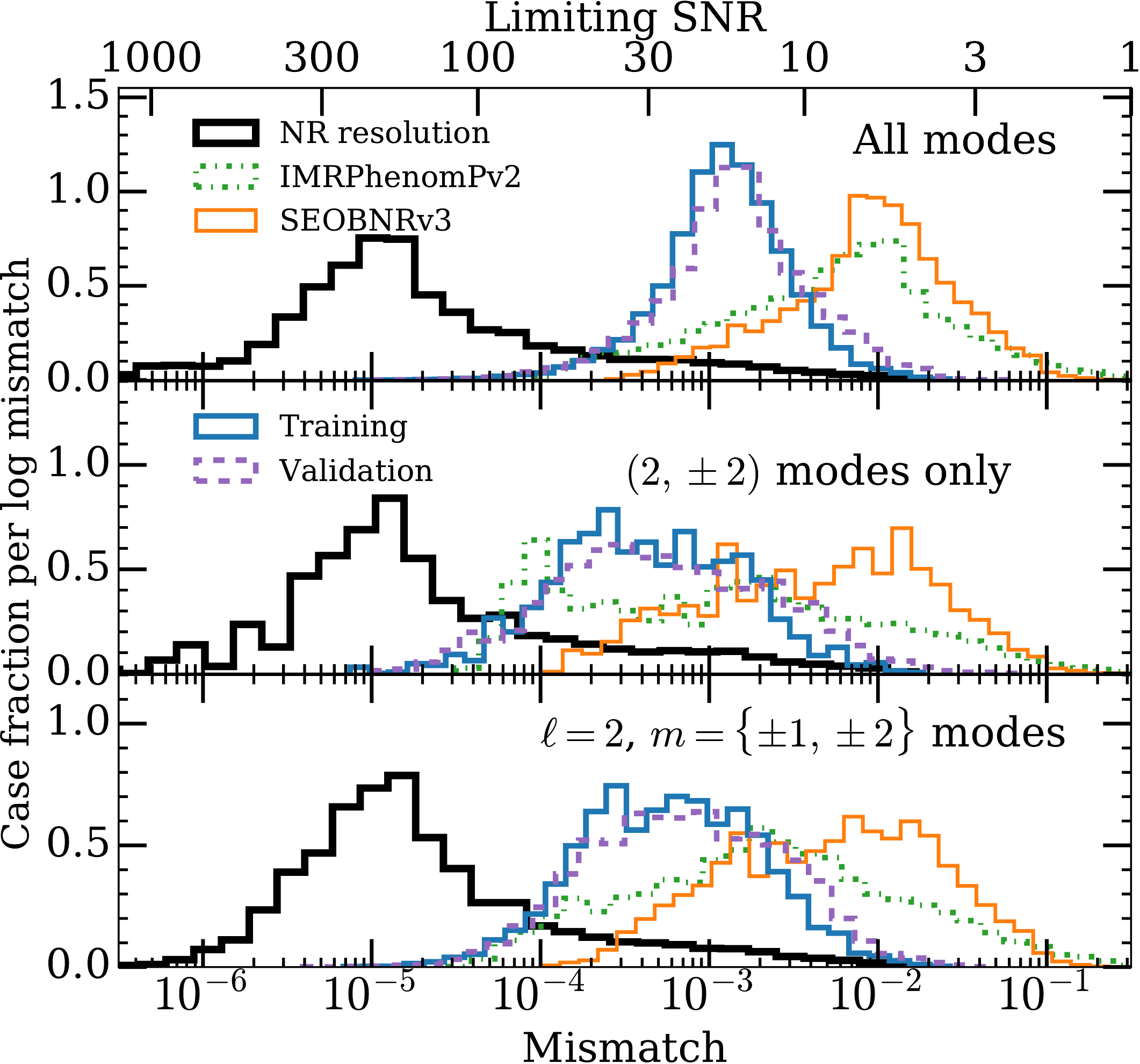}
  \caption{
    Mismatches, computed using a flat noise curve, versus
    the highest resolution NR waveforms.
    Histograms are normalized to show the error fraction per log-mismatch,
    such that the area under each curve is the same.
    A sufficient but not necessary condition for a mismatch
        to have a negligible effect is that the signal-to-noise ratio
        (SNR) lies below the limiting SNR
        $\rho_* = 1/\sqrt{2\mathrm{Mismatch}}$ given on the top
        axis~\cite{Lindblom2008}.
    Top: All modes available to each waveform model are included,
        and the NR waveforms use all $\ell \leq 5$ modes.
    Middle: All {\em coprecessing-frame} modes other than $(2,\pm 2)$
        are set to zero in all waveforms.
    Bottom: All {\em coprecessing-frame} modes other than $(2, \pm 1)$ and
        $(2, \pm 2)$ are set to zero in all waveforms.
    These restricted mode studies are done to compare more directly with
        IMRPhenomPv2 and SEOBNRv3, which retain the coprecessing-frame modes
        of the middle and bottom panels respectively.
    }
  \label{fig:mismatch_hist}
\end{figure}

The mismatches using a flat noise curve are shown in
the top panel of Figure \ref{fig:mismatch_hist}.
We find that both the IMRPhenomPv2 (green dot-dashed curve) and SEOBNRv3
(solid curve) models have median mismatches of $\sim 10^{-2}$ with the NR
waveforms.
The mismatches between our surrogate model and the NR waveforms
are given by the ``Training'' (solid blue) and ``Validation''
(dashed purple) curves and have median mismatches of $\sim 10^{-3}$
with the NR waveforms; see \S~\ref{sec:time-domain-errors}
for a discussion of training and validation errors.
Finally, NR waveforms of different resolution have median mismatches
(solid black curve)
of $\sim10^{-5}$.
In the middle and bottom panels, we repeat this study while restricting
which \emph{coprecessing-frame} modes are used.
IMRPhenomPv2 contains only the $(2, \pm 2)$ modes, while
SEOBNRv3 also contains the $(2, \pm 1)$ modes. Obtaining larger mismatches
in the top panel when comparing against all NR modes indicates these
waveform models would benefit from additional modes.
We find that our surrogate performs roughly an order
of magnitude better than the other waveform models
in its range of validity,
but still
has mismatches two orders of magnitude larger than the intrinsic
resolution error of the NR waveforms. This suggests
that the surrogate
could be improved with additional waveforms and/or improved
model choices. However, we also note that neither
IMRPhenomPv2 nor SEOBNRv3 have been calibrated to 
precessing NR simulations.

\begin{figure}
  \includegraphics[width=\linewidth]{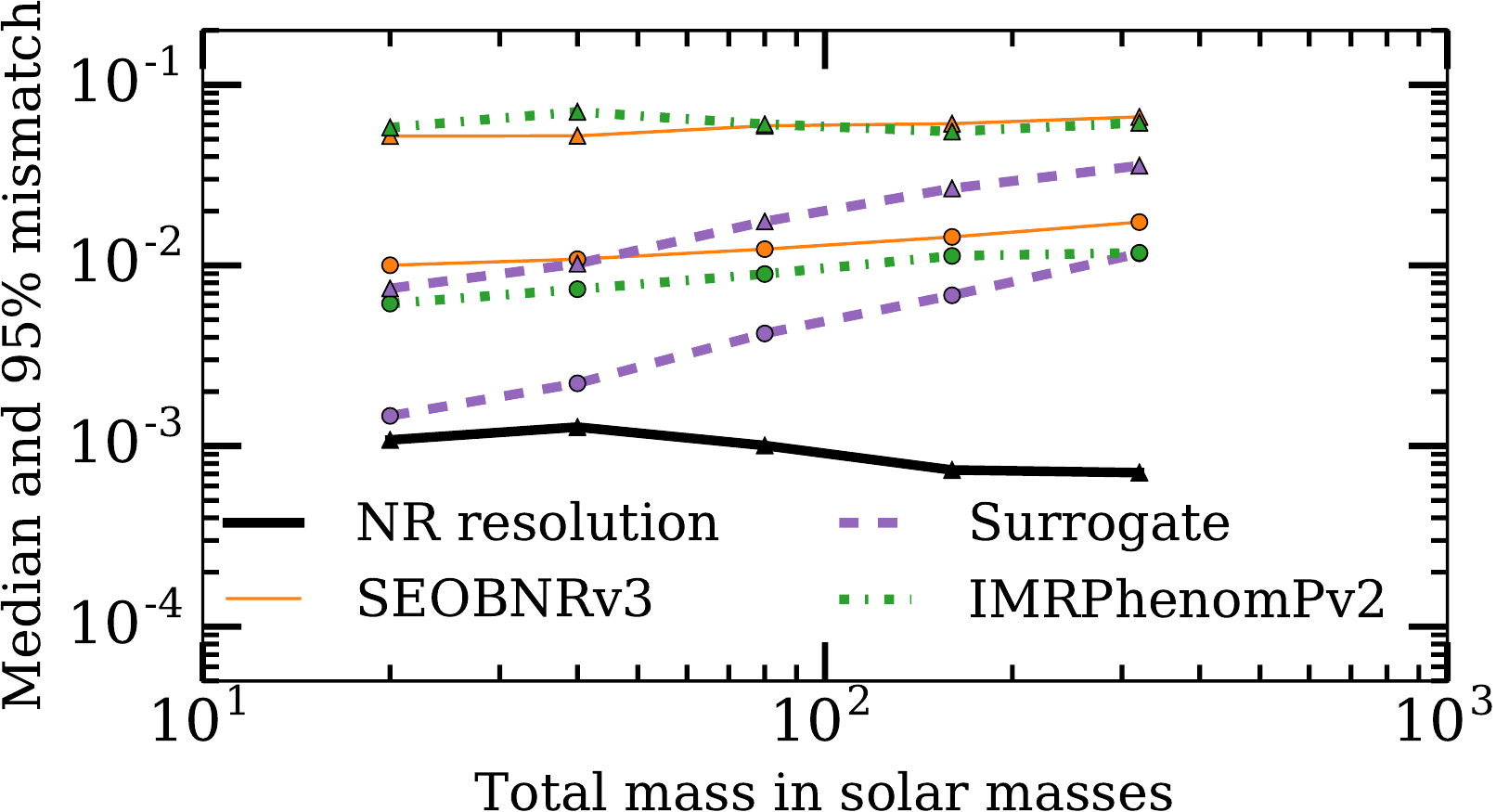}
  \caption{Median (lower curves, circles) and 95th percentile
        (upper curves, triangles) mismatches
        for various total masses $M$ using the advanced LIGO design sensitivity.
        The median NR resolution mismatches are all below $2 \times 10^{-5}$.
        The ``Surrogate" mismatches shown here are ``Validation" errors
        described in \S~\ref{sec:time-domain-errors}.}
  \label{fig:mismatch_vs_mass}
\end{figure}

Since a realistic noise curve will affect mismatches, we also compute
mismatches for total masses $M$ between $20M_\odot$ and $320M_\odot$
using the advanced LIGO design sensitivity~\cite{Shoemaker2009}.
In Fig.~\ref{fig:mismatch_vs_mass},
the lower and upper curves for each waveform model
denote the median mismatch and $95$th percentile
mismatch.
We note that for $M < 114_\odot$, some NR and surrogate waveforms
begin at $f_\mathrm{min} > 10\,\mathrm{Hz}$ and the noise-weighted inner
products will not cover the whole
advanced LIGO design sensitivity band.
The surrogate model errors increase with total mass, indicating a larger
amount of error in the merger phase and less error in the inspiral phase.
Note that our largest systematic source of error,
  the approximate treatment of the waveform's dependence on
  the angle $\phi_\chi$, is much larger during the merger than during
  the inspiral,
  as discussed in \S~\ref{sec:phichi}
  and plotted in Fig.~\ref{fig:check5d_inertial}.  This 
  error source arises from our attempt to model a 5d parameter space
  with a 4d surrogate model, so it will not be relevant for a full
  7d surrogate model.
Even with this error, our surrogate model performs better
than the other waveform models up to $320M_\odot$ within the surrogate
parameter space.

\begin{figure}
  \includegraphics[width=\linewidth]{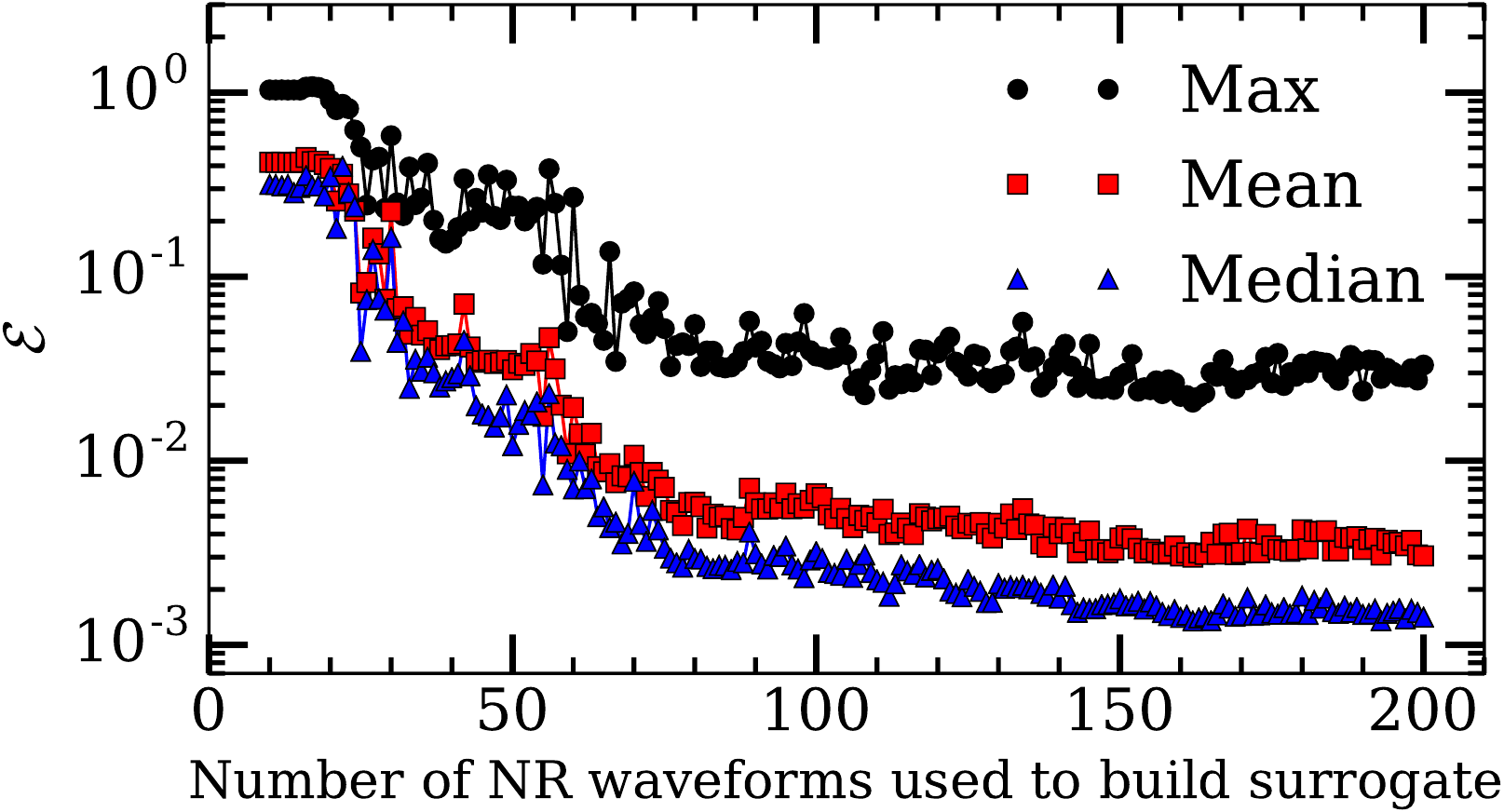}
  \caption{
    Max, mean and medians of the distributions of $\mathcal{E}$
    when building a
    surrogate using the first $N$ waveforms
    and a validation set consisting
    the remaining $200-N$
    waveforms.
    }
  \label{fig:errs_vs_size}
\end{figure}

To determine if the discrepancy between the surrogate errors and NR resolution
errors is due to an insufficient number of
NR waveforms in the surrogate, we study
how the errors depend on the number of waveforms used to build the surrogate.
We construct trial
surrogates using the first $N_\mathrm{train}$ NR waveforms
for $N_\mathrm{train} \in [30, 200]$;
  for validating the surrogate, we use the $N - 200$ waveforms that are
  not used to build any of these trial surrogates.
By using the same $N - 200$ validation waveforms for all choices of
$N_\mathrm{train}$, we ensure that any changes in the error distribution
resulting from changes in $N_\mathrm{train}$ are due to changes in the surrogate
model and not in the set of validation waveforms.
The validation errors, shown in Fig.~\ref{fig:errs_vs_size},
decrease quite slowly with additional waveforms
when $N_\mathrm{train} > 100$, suggesting that the number of NR waveforms
would have to increase dramatically to have a noticeable affect on the
predictive ability of the surrogate.

\subsection{Representing arbitrary spin directions}

One of the limitations
of the NRSur4d2s surrogate model
is that it only produces
waveforms for binaries with a restricted spin direction on the smaller black
hole.  However, it is possible to make
use of effective spin parameters to create a parameter
mapping
\begin{equation}
f: (q, \vec{\chi}_1, \vec{\chi}_2) \rightarrow \vec{x}_\mathrm{model}
\end{equation}
from the $7d$ space of binaries with arbitrary spin directions
to a lower-dimensional parameter subspace
\cite{Ajith2009, Santamaria:2010yb, Schmidt:2014iyl}.
The use of a model with such a parameter space mapping in gravitational wave
source parameter estimation
leads to equivalence classes
\begin{equation}
\{ (q, \vec{\chi}_1 , \vec{\chi}_2) : f(q, \vec{\chi}_1, \vec{\chi}_2) = \vec{x}_\mathrm{model}\}
\end{equation}
where multiple values of the 7d parameters map to the same
lower-dimensional parameter vector $\vec{x}_\mathrm{model}$.
For parameter estimation, all members of the equivalence class
have the same likelihood, so distinguishing parameters within one
equivalence class can be done only using knowledge of the prior.

Here we investigate several possible
mappings from the full $7d$ parameter space to the $5d$ subspace covered
by the NRSur4d2s surrogate model,
and we investigate the accuracy of these mappings
using $3$ SpEC simulations
with parameters outside the $5d$ subspace.
In our case, $\vec{x}_\mathrm{model}$ is
the vector $(q, \vec{\chi}_1,\chi_2^z)$
at $t=t_0$. To construct a parameter space mapping
from $(q, \vec{\chi}_1,\vec{\chi}_2)$ to $\vec{x}_\mathrm{model}$, we
use the values of $\vec{\chi}_1$ and $\vec{\chi}_2$ at $t=t_0$ to form an effective
spin $\vec{\chi}_\mathrm{eff}$, and then construct $\vec{x}_\mathrm{model}$ using
$\vec{\chi}_\mathrm{eff}$ instead of $\vec{\chi}_1$.
This preserves the values of $q$ and $\chi_2^z$, while reducing the other $5$ spin
components to $3$.

The most simple mapping would be to ignore the $x$ and $y$ components of $\vec{\chi}_2$
at $t=t_0$ and take
\begin{equation}
    \vec{\chi}_\mathrm{eff}^\mathrm{Drop} = \vec{\chi}_1.
    \label{eq:param_Drop}
\end{equation}
A second possibility would be to use a similar parameter mapping as is used in
IMRPhenomP~\cite{Hannam:2013oca} with an effective precessing spin
$\chi_\mathrm{p}$~\cite{Schmidt:2014iyl} and take
\begin{align}
    B_1 &= \left(2 + \frac{3}{2q}\right) \left(\frac{q}{1 + q}\right)^2, \\
    B_2 &= \left(2 + \frac{3q}{2}\right) \left(\frac{1}{1 + q}\right)^2, \\
    i^* &= \argmax_{i=1,2} B_i\|\vec{\chi}_i^\perp\|, \\
    \vec{\chi}_\mathrm{eff}^\mathrm{\chi_\mathrm{p}} &= \frac{B_{i^*}}{B_1}
            \vec{\chi}_{i^*}^\perp + \chi_1^z \hat{z}\,,
    \label{eq:param_chi_p}
\end{align}
where $\vec{\chi}_i^\perp$ is the part of $\vec{\chi}_i$ orthogonal to the
Newtonian orbital
angular momentum, which is $(\chi_i^x, \chi_i^y, 0)$ at $t=t_0$.
This mapping uses the in-plane spin components of whichever spin contributes
the most to precession at leading PN order, scaled appropriately 
and placed on the heavier black hole.
This mapping is particularly effective when the in-plane spins of the smaller
BH are negligible, i.e., for high mass ratios, and for long duration GWs.
However,
it has also been shown to prove sufficient for binaries similar to
GW150914 \cite{TheLIGOScientific:2016wfe, Abbott:2016izl}.

In our case, we have a couple precession cycles at most, and we might
consider adding the effects of the in-plane components of the two spins.
A further motivation to add the spins is that for nearly equal masses,
the precession rates of the two spins will be nearly equal
\cite{Apostolatos1994, Buonanno2004}.
When adding the dimensionless spins, we can either do so directly
\begin{equation}
    \vec{\chi}_\mathrm{eff}^\mathrm{Add} = \vec{\chi}_1 + \frac{1}{q^2} \vec{\chi}_2^\perp
    \label{eq:param_Add}
\end{equation}
or again using the leading order PN contribution to precession
\begin{equation}
    \vec{\chi}_\mathrm{eff}^\mathrm{PN} = \vec{\chi}_1 + \frac{B_2}{B_1} \vec{\chi}_2^\perp.
    \label{eq:param_PN}
\end{equation}

\begin{table}
  \begin{tabular}{| c || c | c | c |}
    \hline
    SSX:BBH:ID    &   $q$     &   $\vec{\chi}_1$  &   $\vec{\chi}_2$ \\
    \hline

    0607  &   $1.5$   &   $(0.067, -0.199, 0.212)$    &   $(0.139, -0.374, 0.202)$ \\
    0608   &   $1.7$   &   $(0.053, -0.085, 0.001)$    &   $(0.494, 0.337, 0.113)$ \\
    0609   &   $1.9$   &   $(0.094, -0.145, 0.099)$    &   $(-0.398, 0.576, 0.001)$ \\
    \hline

    \end{tabular}
  \caption{Parameters for $3$ additional SpEC simulations with unrestricted spin
            directions. The spins are measured at $t=t_0$.}
  \label{tab:7dparams}
\end{table}

\begin{figure}
  \includegraphics[width=\linewidth,clip]{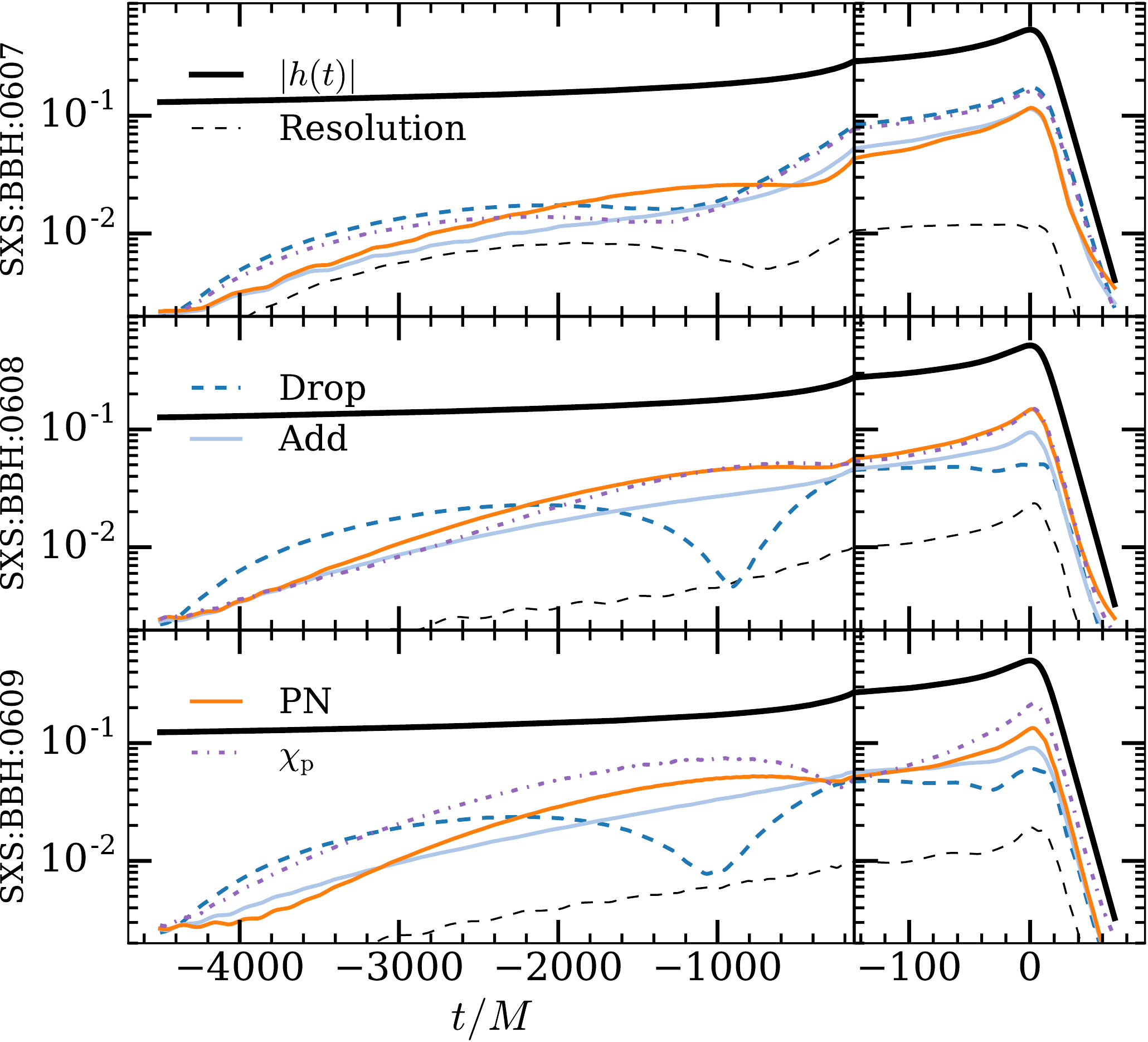}
  \caption{
    Comparing surrogate evaluations to three NR waveforms (top, middle and bottom
    plots) with spins outside the $5d$ parameter subspace.
    For each case and each of the $4$ parameter mappings, the surrogate
    model waveform error is shown.
    In all cases, the `PN' mapping performs well at the very start of the
    waveform and the `Drop' mapping performs poorly, but there is no clear
    overall best mapping. Surrogate modeling errors contribute to the
    difficulty in assessing the quality of the mappings.
  }
  \label{fig:check7d}
\end{figure}

We do a brief investigation of the quality of these parameter space mappings
using three additional SpEC simulations.
The waveforms are aligned as described in Sec.~\ref{sec:alignment}, and their
parameters at $t=t_0$ are measured and listed in Table \ref{tab:7dparams}.
For each case and each parameter space mapping,
we compute the mapped parameters and compare the surrogate evaluation
with the mapped parameters to the NR waveform.
The time-dependent waveform errors are shown in Fig.~\ref{fig:check7d}
and $\mathcal{E}$ values as well as mismatches are given in Table \ref{tab:check7d_errs}.
$\sim 0.01$, which is larger than the median surrogate errors
but well within the possible range of surrogate errors,
so we cannot rule out that these errors are dominated by surrogate error.
The ``Drop" parameter space mapping performs reasonably well
since the cases investigated are far enough away from equal mass that the
spin of the smaller black hole has a small effect on the waveform.

\begin{table}
  \begin{tabular}{| c | c  c  c  c  c  c  c|}
    \hline
    Map    &  \multicolumn{3}{c}{$\mathcal{E}$} & \multicolumn{3}{c}{Median Mismatch} & \\
    & 0607  & 0608  &  0609       &   0607   &   0608   &   0609  & \\
    \hline
    Drop (Eq.~\ref{eq:param_Drop})    & \textbf{0.016} & 0.007 & 0.008  &   \textbf{0.0054} & 0.0026 & 0.0031 & \\ 
    Add (Eq.~\ref{eq:param_Add})    & 0.007 & 0.009 & \textbf{0.013}  &   0.0046 & 0.0051 & \textbf{0.0076} & \\
    PN (Eq.~\ref{eq:param_PN})     & 0.008 & 0.019 & \textbf{0.021}  &   0.0041 & 0.0075 & \textbf{0.0109} & \\
    $\chi_\mathrm{p}$ (Eq.~\ref{eq:param_chi_p}) & 0.014 & 0.018 & \textbf{0.044}  &   0.0050 & 0.0074 & \textbf{0.0161} & \\
    \hline

    \end{tabular}
  \caption{Errors between the $3$ NR waveforms and the surrogate evaluation for
            a given parameter space mapping. Mismatches are optimized over time,
            polarization angle and orbital phase shifts.
            For each mapping, the largest error is in bold.
        }
  \label{tab:check7d_errs}
\end{table}

\section{Building the Frequency Domain Surrogate}
\label{sec:freqSurrogates}
Evaluating the NRSur4d2s surrogate takes $\sim 1s$ on a single modern
processor. Evaluating all coprecessing modes takes $\sim 0.21s$,
evaluating the frame quaternions $q(t)$ takes $\sim 0.38s$ and is dominated
by evaluating Eq.~\ref{eq:reconstructquat} sequentially for all times,
and rotating the modes into the inertial frame with the transformation
$T_Q$ takes $\sim 0.41s$.
Gravitational wave parameter estimation is typically done using
Markov-chain Monte Carlo \cite{Veitch:2015} and can require $\mathcal{O}(10^8)$
waveform evaluations; this motivates us to build a faster surrogate model.
We also wish the faster surrogate model to be in the frequency domain, where
most parameter estimation is currently done.
Accelerated frequency-domain surrogates have been built
in $3d$~\cite{Purrer:2015tud,Purrer:2014}
using cubic tensor-spline interpolation of the waveform amplitudes and phases
at some sparsely sampled frequency points.

To build the frequency-domain NRSur4d2s\_FDROM surrogate, we first
choose a uniformly spaced grid of $N = N_q \times ... \times N_{\chi_2^z}$
points in our $5d$ parameter space and
evaluate the NRSur4d2s
surrogate model at each point on the grid.
We taper the waveforms with Planck windows~\cite{McKechan:2010kp}, rolling
on for $t \in [-4500M, -3500M]$ and rolling off for $t \in [50M, 70M]$.
We then pad the waveform modes with zeros and perform
a fast Fourier transform to obtain the frequency
domain modes $\tilde{h}^{\ell, m}(f)$.
We then downsample the frequency domain waveforms to a non-uniformly spaced
set of frequencies, which are chosen to be the same for all waveforms
and to be uniformly spaced in gravitational-wave phase for an equal-mass zero-spin binary.
This significantly reduces the cost
of evaluating the model,
with a negligible loss in accuracy.
For each mode $\tilde{h}^{\ell, m}(f)$,
we build an empirical interpolant in frequency
using all $N$ waveforms,
and we keep the first $100$ basis vectors.
At each empirical frequency node,
we fit the real and imaginary parts of each mode across parameter space
using a cubic tensor-product spline; we use ``not-a-knot'' boundary conditions
that have a constant third derivative across the first and last
knots~\cite{deBoor1978}.
Finding the spline coefficients
involves solving a sparse linear system of size
$(N_q + 2) \times ... \times (N_{\chi_2^z} + 2)$, for which we used
Suitesparse \cite{Davis:2004UMFPACK, Davis:2004} and/or
SuperLU DIST \cite{superlu_dist, superlu_ug99}.
The advantage of using a spline is that the evaluation cost is nearly
independent of the grid size $N$,
and requires only
$4^{d=5}$ coefficients and basis
functions to be evaluated.

Implementing the NRSur4d2s\_FDROM
surrogate model in both C
and Python,
we find it takes $50 \mathrm{ms}$ to
evaluate a single waveform in either case.
Empirical interpolation accounts for roughly $10\%$
of the cost, and the remaining $90\%$
  comes from to the $2400$ spline evaluations.
Assembling the waveform at a desired sky direction
from the modes and interpolating
onto the desired frequencies have negligible cost.

\begin{table}
\begin{tabular}{| c || c | c | c | c | c |}
\hline
Grid label  &
$N_q$       &
$N_{|\chi_1|}$ &
$N_{\theta_\chi}$ &
$N_{\phi_\chi}$ &
$N_{\chi_2^z}$ \\
\hline
5   &   5   &   4   &   7   &   4   &   6   \\
6   &   6   &   4   &   8   &   4   &   7   \\
7   &   7   &   5   &   9   &   4   &   8   \\
8   &   8   &   6   &   11  &   4   &   9   \\
9   &   9   &   6   &   13  &   5   &   11  \\
10  &   10  &   7   &   14  &   6   &   12  \\
11  &   11  &   8   &   15  &   7   &   14  \\
12  &   12  &   9   &   17  &   8   &   16  \\
13  &   13  &   10  &   19  &   9   &   19  \\
\hline
\end{tabular}
\caption{Grid sizes for tensor-spline interpolation in the
    frequency-domain surrogate.
  The size in each dimension is chosen such that
   surrogates for $1d$ slices in all dimensions have
   comparable interpolation errors.
}
\label{tab:gridSizes}
\end{table}

\begin{figure}
  \includegraphics[width=\linewidth]{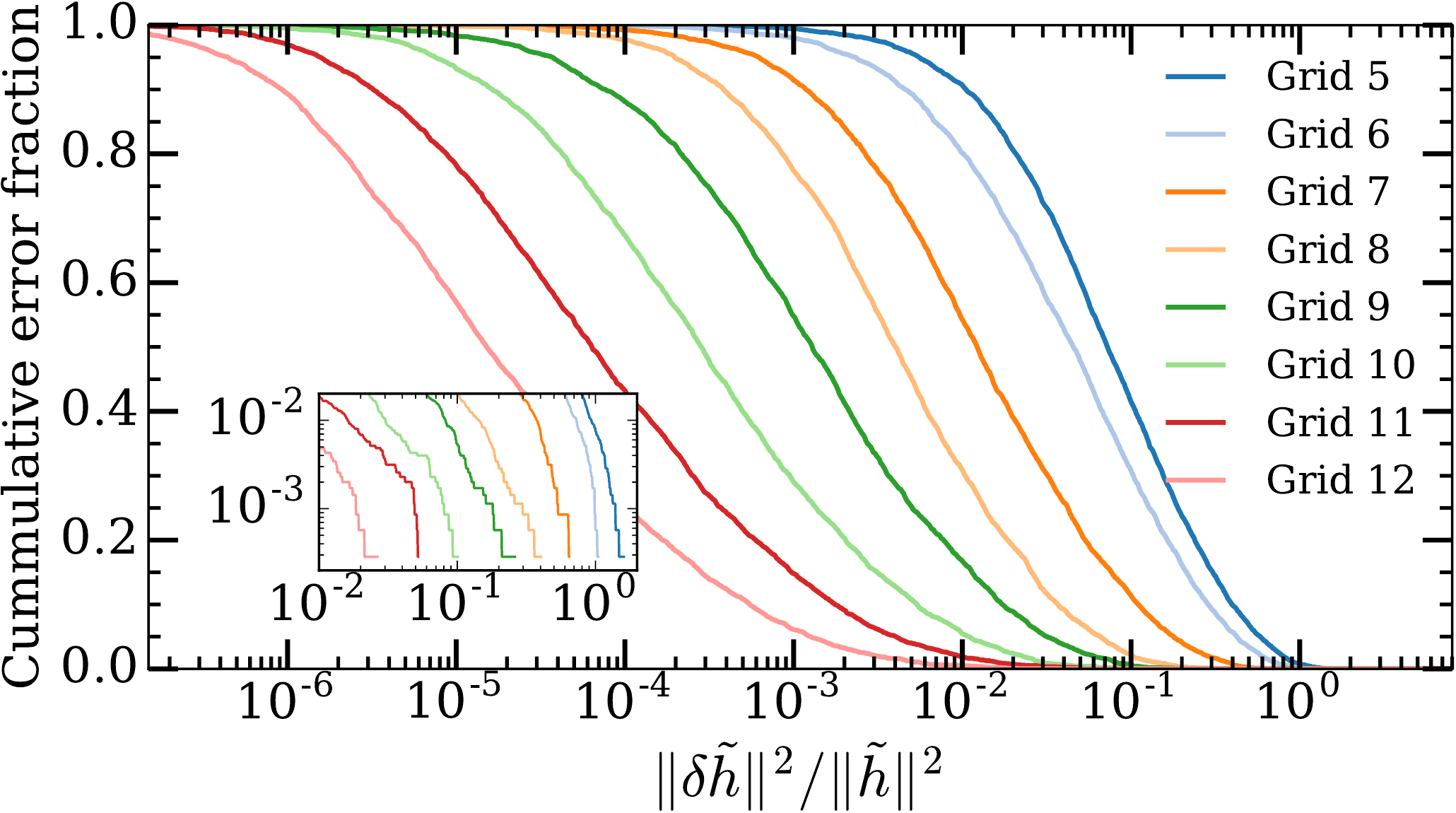}
  \caption{
    Cumulative error distributions of the frequency domain
    NRSur4d2s\_FDROM surrogate
    waveforms compared to the time domain NRSur4d2s
    surrogate waveforms transformed
    to the frequency domain, evaluated for randomly chosen uniformly
    distributed parameters.
    The curves indicate the fraction of errors at least as large as the
    indicated error.
    The NRSur4d2s\_FDROM
    output converges to the FFT of the NRSur4d2s output
    as the grid size is increased.
  }
  \label{fig:tensor_spline_convg}
\end{figure}

To ensure that
the empirical interpolants and parameter space splines are
sufficiently accurate, we construct many
frequency-domain surrogates for increasingly
large parameter space grids. We monitor the
differences between the frequency domain
surrogate waveforms and the FFT of the tapered NRSur4d2s
waveforms,
and we demand that these differences decrease with increasing
grid size.
We use a different number of grid points in each parameter-space
dimension, since the
waveforms vary more in some dimensions than others.
To determine the number of grid points to use, we construct frequency-domain
surrogates for $1d$ slices of the parameter space, where the other parameters
are fixed at a single intermediate value.
We then arbitrarily choose a value of $N_q$, the number of grid points
covering the dimension of mass ratio,
and we determine the maximum error of the $1d$ surrogate in which only
the mass ratio $q$ is varied.  Call this error $E_q$.  Then
we find the number of points $N_{|\chi_1|}$
for which the $1d$ surrogate for $|\chi_1|$ has an error of approximately
$E_q$, and similarly for the other parameters.  
The resulting grid sizes are listed in Table \ref{tab:gridSizes}.
In Fig.~\ref{fig:tensor_spline_convg},
we see that the errors converge as the grid size increases.

\section{Discussion}
\label{sec:discussion}

We have built the first NR surrogate model
of BBH waveforms that covers a multidimensional
portion of the BBH
parameter space.
This extends the work in~\cite{Blackman:2015pia}, where a
1-dimensional (i.e. zero spin) NR
surrogate served as a proof of principle that surrogate models of NR waveforms
can be made highly accurate.
The non-spinning surrogate model is inappropriate for use in GW
parameter estimation,
as neglecting all spin effects could lead to large parameter biases.
Extending the parameter space to include both aligned spin components
and one precessing component makes the new model presented here the first
NR surrogate suitable for gravitational wave parameter estimation.
While two of the
in-plane spin components are still
neglected by the NRSur4d2s surrogate model,
IMRPhenomPv2 neglects similar information but
obtains parameters for GW150914 that are compatible with
those obtained using
SEOBNRv3, which includes all spin components \cite{Abbott:2016izl}.
We note, however, that for edge-on systems otherwise similar to GW150914
IMRPhenomPv2 can obtain biased parameter estimates~\cite{Abbott:2016wiq}.

To reduce computational cost, the simulations used to build the
NRSur4d2s surrogate were restricted to mass ratios $q \leq 2$ and spin magnitudes
$|\vec{\chi}_i| \leq 0.8$. This limits
the range of GW events for which the surrogate
model could be used.
GW150914 is within this range, while the mass ratio posterior of GW151226 extends
well beyond $q=2$.
Ultimately, a NR surrogate model
covering the fully precessing $7d$ parameter space up to large mass ratios and
spin magnitudes will be needed.

Use of the NRSur4d2s surrogate
is also limited by the length (i.e. number
  of orbits) of the waveforms used to build it.
GW151226 enters the sensitive LIGO band approximately $55$ cycles before merger
\cite{Abbott:2016nmj}, while the NRSur4d2s
surrogate produces waveforms
with between
$30$ and $40$ cycles before merger.
Since these waveforms are tapered before building the faster
NRSur4d2s\_FDROM surrogate,
the latter includes only $25$ to $35$ cycles
before merger.
There are a few ways to build an NR surrogate with
longer waveforms, so that the surrogate is applicable to GW events of
lower total mass.
First, one could build a surrogate model using longer NR waveforms.
A less computational expensive option would be to hybridize
\cite{Bustillo:2015ova, Boyle:2011dy, OhmeEtAl:2011, MacDonald:2011ne} the NR
waveforms with PN or EOB waveforms before building a surrogate model.
A final option would be to use a time domain surrogate which produces waveforms
of moderate length as done here,
to hybridize the surrogate output with PN or
EOB waveforms before transforming them into the frequency domain, and finally
to build
a frequency domain surrogate for the hybrid waveforms.

Phenomenological and semi-analytic waveform modeling approaches have already led
to precessing waveform models suitable for GW parameter estimation from a large
class of GW events. These models have an underlying structure, and are
calibrated by tuning a set of numerical
coefficients such that the model waveforms have good agreement with NR waveforms.
NR surrogate models provide an independent approach. NR surrogate models
make no assumptions about the waveform structure, although
knowledge of the waveform structure may lead to a better
decomposition and smaller errors for a given number of input NR waveforms.
We find our NRSur4d2s
surrogate model to have better agreement with NR waveforms than
other leading waveform models within the range of validity of the surrogate,
although we again note that these other models have not been calibrated
to precessing NR simulations.
As gravitational wave detector sensitivities improve, this increased
waveform accuracy will become important for unbiased measurements of the
parameters from the loudest GW events, as well as when making astrophysical
statements using many GW events.

Since we have
not performed
Cauchy characteristic
extraction~\cite{Handmer:2016,Winicour:2005ge,Babiuc:2010ze,Bishop:1997ik}, but instead have
  extracted waveforms from the simulations
  at a series of finite radii and then extrapolated them to
  infinite radius~\cite{Boyle-Mroue:2008}, the $(2, 0)$ modes of the numerical
waveforms in the coprecessing frame
may not be accurate \cite{Taylor:2013zia}.
In particular, we do not see the expected gravitational wave memory
in the real part of the $(2, 0)$ mode \cite{Pollney:2010hs, Favata:2010zu}.
This should lead to
negligible errors for most LIGO purposes, since the memory
signal is low frequency and has very little contribution within the LIGO band.
However, NRSur4d2s would not be suitable to detect a memory signal with
a method requiring templates that include
memory.
A direct measurement of the memory signal using the method proposed in
\cite{Lasky:2016}, however, could make use of waveforms from
NRSur4d2s,
as they have the $(2, \pm 1)$ and $(3, \pm 3)$ modes in
the coprecessing frame necessary to determine the sign of the memory.

The errors in the NRSur4d2s
surrogate are significantly larger than the resolution of
the NR waveforms used in its construction. An incomplete treatment of
the spin angle $\varphi_\chi$ (see Fig.~\ref{fig:diagram})
is one large source of error, and a complete $7d$
NR surrogate model would not suffer from this issue.
Aligning the rotation of the waveforms (see \S~\ref{sec:alignment})
closer to merger
might reduce the errors, since
$\varphi_-^{2, 2}$ at the empirical nodes
would have less variation across parameter space.
Since the parameters of the NR simulations were chosen such that $\vec{\chi}_2$
is aligned with the orbital angular momentum $4500M$ before merger,
it would be non-trivial to build a surrogate
model from these NR waveforms if the rotation alignment were performed at some other time.
This is another issue which will be resolved by including all
7 dimensions of parameter space.

Incorporating additional NR waveforms into the
NRSur4d2s surrogate should also reduce the surrogate errors, although
Fig.~\ref{fig:errs_vs_size} indicates that with the current surrogate choices
a very large number of additional NR waveforms would be needed for a significant
reduction.
Alternative methods of
fitting empirical nodes could also help. The training
and validation errors in Fig.~\ref{fig:mismatch_hist} and \ref{fig:error_hist}
are nearly identical,
while in \cite{Blackman:2015pia}
the validation errors were roughly a factor of $2$ larger
than the training errors.
This suggests we may be under fitting the data and could use tighter
parameter space fit tolerances.

In addition to model cross-validation, there is a variety of 
informative diagnostics we could monitor to diagnose sources of
surrogate error.
Failing to meet one of these diagnostics
would indicate an unexpected
source of surrogate error that could be improved:
\begin{itemize}
\item {\em Decay of the temporal basis error}.
Smooth models are expected to have an exponentially decaying basis
projection error and empirical interpolation error.
Numerical noise in the NR waveforms means the exponential decay will
not continue to arbitrarily small errors, but if the error curves do
not display a region of exponential decay there is reason to suspect
the basis is not accurate enough.
\item {\em Decay of the parametric fitting error}.
It is known that expanding (with orthogonal projection)
a smooth function with polynomials results in an
exponentially decaying approximation error.
We believe the
waveform data pieces evaluated at empirical nodes
can be described by a smooth function plus (relatively small) noise.
Thus, just as in the case of the basis projection error,
the fitting error is expected to decay
exponentially before the noise sources dominate the approximation.
This can be seen in Fig.~\ref{fig:fit_appendix}, where the exponential decay
only lasts for approximately $10$ coefficients before noise sources cause the
validation errors to flatten and then slowly rise.
\item {\em Robustness to noise}.
We could build surrogates from waveforms with different
NR resolutions.
In our case, since the surrogate errors are larger than the
NR resolution errors, we expect to obtain a surrogate of
comparable quality using slightly lower resolution NR waveforms.
If we use really low resolution NR waveforms, we would expect
the surrogate errors to rise accordingly.
In other cases where we do achieve surrogate errors similar to
the NR resolution errors, comparing surrogates built from NR
waveforms of different resolutions should yield similar differences
to comparing the NR waveforms themselves.
\item {\em Residual structure}.
We could examine the parametric fit residuals and cross-validation residuals
as a function of parameters.
If the surrogate model captures the dominant features of NR waveforms
then these residuals should appear random.
From Fig.~\ref{fig:err_vs_params} we see that the largest errors occur
at large values of $|\vec{\chi}_1|$ and for intermediate values of
$\theta_\chi$, where precession has the largest effect.
This indicates additional highly-precessing NR simulations may help significantly
in reducing the surrogate errors.
\end{itemize}

\section{Acknowledgements}
We thank
    Michael Boyle,
    Alessandra Buonanno,
    Kipp Cannon,
    Maria Okounkova,
    Richard O'Shaughnessy,
    Christian Ott,
    Harald Pfeiffer,
    Michael P{\"u}rrer,
    and Saul Teukolsky
for many useful discussions throughout this project.
We also thank
    Andy Bohn,
    Nick Demos,
    Alyssa Garcia,
    Matt Giesler,
    Maria Okounkova,
    and Vijay Varma
for helping to carry out the SpEC simulations used in this work.
This work was supported in part by the Sherman Fairchild Foundation
and NSF grant PHY-1404569 at Caltech.
J.B. gratefully acknowledges support from NSERC of Canada.
Computations were performed
on NSF/NCSA Blue Waters under allocation PRAC ACI-1440083;
on the NSF XSEDE network under allocation TG-PHY990007;
on the Zwicky cluster at Caltech,
which is supported by the Sherman Fairchild
Foundation and by NSF award PHY-0960291;
and on the ORCA cluster at California State University at Fullerton,
which is supported by NSF grant PHY-1429873,
the Research Corporation for Science Advancement, and
California State University at Fullerton.

\appendix
\section{Forward-stepwise greedy fit algorithm}
\label{sec:fitappendix}

Here we describe in more detail the algorithm we use in
Sec.~\ref{sec:param_space_fits} used to fit the waveform data pieces
evaluated at the empirical time nodes.
Given $N$ numerical relativity simulations at parameters
$\pmb{\lambda}_{\mathrm NR}=\{\pmb{\lambda}_i\}_{i=1}^N$
where $\pmb{\lambda} = (q, |\chi_1|, \theta_\chi, \chi_2^z) =
(\lambda^1, \lambda^2, \lambda^3, \lambda^4)$,
we obtain each
waveform data piece
$X = \{X(t;\pmb{\lambda}_i)\}_{i=1}^N$.
Evaluating the surrogate model requires
predicting $X_m(\pmb{\lambda}) = X(T_m,\pmb{\lambda})$ for each empirical time
node $T_m$ and for $\pmb{\lambda} \notin \pmb{\lambda}_{\mathrm NR}$.
Denoting the model prediction
as $X_{mS}(\pmb{\lambda})$, we need not restrict to
an interpolation scheme where $X_{mS}(\pmb{\lambda}_i) = X_m(\pmb{\lambda}_i)$
because the data contain numerical noise.
Instead, we use linear fits such that
\begin{equation}
X_{mS}(\pmb{\lambda}) = \sum_{i=1}^M c_i B^i(\pmb{\lambda})
\end{equation}
for some set of basis functions $\{B^i\}_{i=1}^M$.

For simplicity, we choose all multivariate basis functions to be products
of one-dimensional basis functions; that is, we choose
$B^i \in \{B^{\vec{\alpha}}\}$ where
\begin{equation}
B^{\vec{\alpha}}(\pmb{\lambda}) = \prod_{l=1}^d B_l^{\alpha^l}(\lambda^l).
\end{equation}
Here $d=4$ is the dimension of the parameter space,
$\vec{\alpha} = (\alpha^1, \ldots, \alpha^d)$ labels which
univariate basis functions enter the product,
and we choose
\begin{center}
  \begin{itemize}
    \item $B_1^k(q) = T_k(2q - 3)$
    \item $B_2^k(|\chi_1|) = \left(\frac{|\chi_1|}{0.8}\right)^k$
    \item $B_3^k(\theta_\chi) = \mathrm{cos}(k\theta_\chi)$
    \item $B_4^k(\chi_2^z) = T_k(\frac{\chi_2^z}{0.8})$
  \end{itemize}
\end{center}
where the $T_k$ are Chebyshev polynomials of the first kind.
We restrict the maximum order of the basis functions so that
$\alpha^l \leq k^l_\mathrm{max}$ where
$\vec{k}_\mathrm{max} = (5, 6, 6, 4)$.
We also restrict $\alpha^3 \leq \alpha^2$ to ensure $\theta_\chi$ does
not affect the surrogate output when $|\chi_1|=0$.

The above choices are made for all waveform data pieces $X$ except
for $X=\varphi_p$. If the waveform data piece is $\varphi_p$ we do
the same as above except
we instead choose
\begin{equation}
B_3^k(\theta_\chi) = \mathrm{sin}((k+1)\theta_\chi),
\end{equation}
and we restrict $1 \leq \alpha^2 \leq 6$
and allow all $0 \leq \alpha^3 \leq 6$.
We treat $\varphi_p$ differently because the amount of precession
is approximately proportional to the spin component orthogonal to the
orbital angular momentum, while other waveform data pieces depend
more strongly on the parallel component.

The above choices yield
$1008$ possible basis functions
($1512$ for $\varphi_p$),
which is more than $N \leq 300$, so we will use only a subset
  of the possible basis functions.  We determine elements
$B^i \in \{B^{\vec{\alpha}}\}$
of this subset in a greedy manner
with a forward-stepwise least-squares fit~\cite{Hocking1976}.
We proceed by iteratively updating two quantities: $r_j^n$,
  which is the $j$th fit residual at the $n$th iteration, and
  $b^{\vec{\alpha}, n}_j$, which is the orthogonal component of
  the basis function $B^{\vec{\alpha}}$
  at the $n$th iteration evaluated at parameters $\pmb{\lambda}_j$.
For the zeroth iteration we
begin with
\begin{align}
    r_j^0 &= X_m(\pmb{\lambda}_j) \\
    b^{\vec{\alpha}, 0}_j &= B^{\vec{\alpha}}(\pmb{\lambda}_j).
\end{align}
At the $n$th iteration, we compute the inner product of the residuals
with the basis functions
\begin{equation}
    d_n^{\vec{\alpha}} = \sum_j r_j^n b^{\vec{\alpha}, n}_j.
\end{equation}
We then select the next most relevant basis function as the one
with the largest magnitude inner product with the residuals
\begin{equation}
    \vec{\alpha}^*_n = \argmax_{\vec{\alpha}} |d_n^{\vec{\alpha}}|
\end{equation}
and choose $B^n = B^{\vec{\alpha}^*_n}$.
We compute the new residuals by subtracting the projection onto the
newly chosen basis function
\begin{equation}
    r_j^{n+1} = r_j^n - d_n^{\vec{\alpha}^*_n} b^{\vec{\alpha}^*_n, n}_j
\end{equation}
and also orthogonalize the basis functions with respect to the new
basis function
\begin{align}
    b^{\vec{\alpha}, n+1}_j &= b^{\vec{\alpha}, n}_j - e^{\vec{\alpha}, n} b^{\vec{\alpha}^*_n, n}_j \\
    e^{\vec{\alpha}, n} &= \sum_j b^{\vec{\alpha}, n}_j b^{\vec{\alpha^*_n}, n}_j.
\end{align}
We continue until we have performed
$m \leq N$ iterations.
We can then perform a least-squares fit using the $m$
selected basis functions to find the coefficients $c_i$.
In practice this is done during the greedy iteration by
keeping track of the matrix of transformations relating
$B^{\vec{\alpha}}(\vec{x}_j)$ and $b^{\vec{\alpha}, n}_j$
as well as the coefficients $d_n^{\vec{\alpha}^*_n}$.

\begin{figure}
  \includegraphics[width=\linewidth]{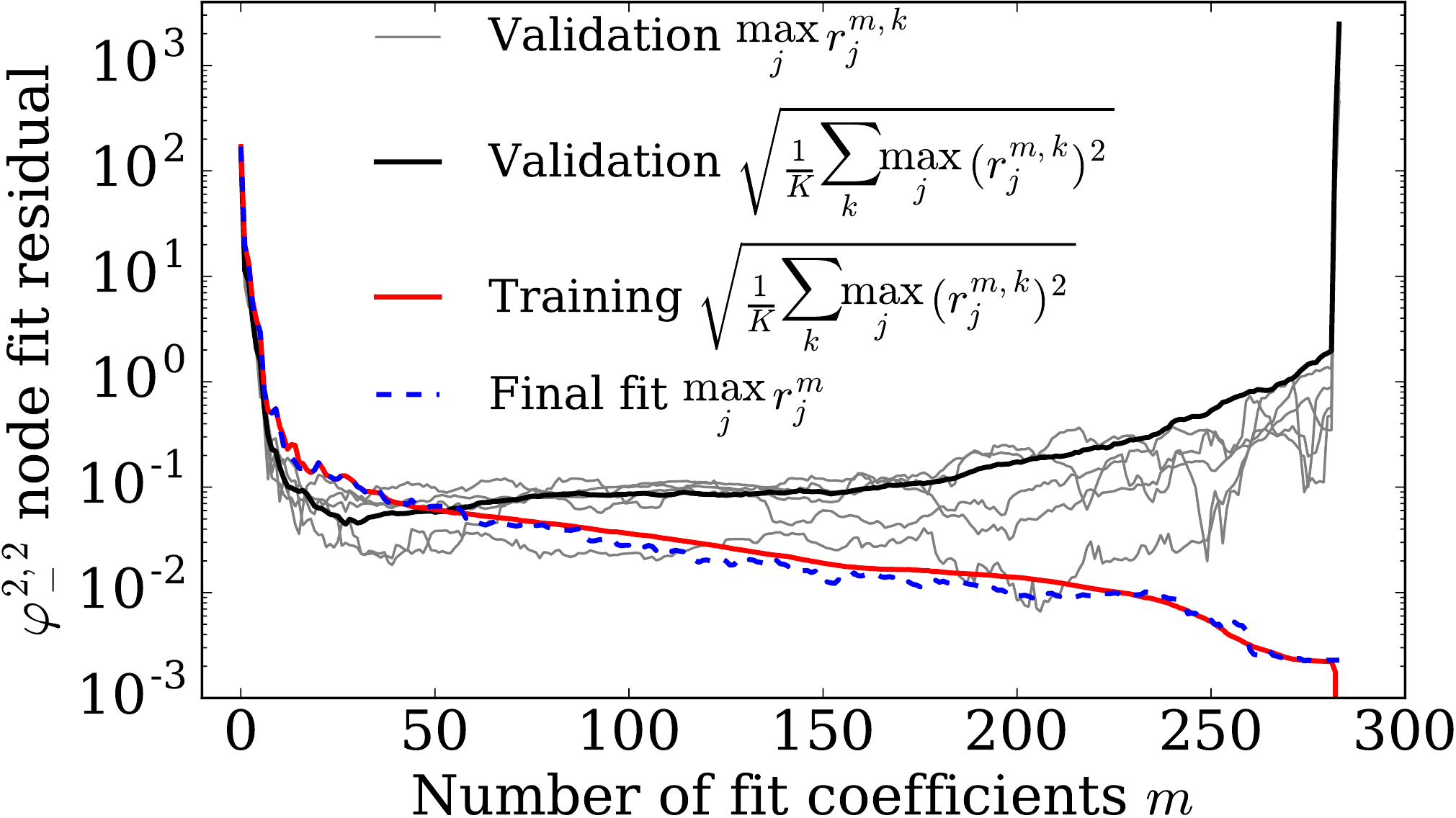}
  \caption{
    Fit residuals for the second empirical node of $\varphi_-^{2, 2}$ at
    $t=-806.5M$. Blue dashed: The maximum fit residual using all data.
    Thin grey lines: Maximum validation residual for individual trials.
    Thick black line: The root mean square (RMS)
    of the validation residuals for $K=50$ trials.
    It takes its minimum value at $m=30$, which determines the number of
    fit coefficients to use for this node in the model.
    Red: The RMS of the training residuals for $K=50$ trials.
  }
  \label{fig:fit_appendix}
\end{figure}

This procedure does not indicate which value of $m$
(the number of fit coefficients) to use.
Using $N$ fit coefficients would be overfitting the data, and setting
individual fit
tolerances by hand for each empirical node of each data component would
be time consuming
and error prone.
So instead, we repeat the above procedure for different values of
$m$, we perform cross-validation studies on the resulting fits, we
find the value of $m$ that leads to the smallest
validation errors (call this value $m^*$), and we choose $m=m^*$.
For each
trial $k=1, \ldots, K = 50$ of this cross-validation procedure,
we randomly divide the $N$ data points into
$N_v = 5$ validation points and $N_t = N - N_v$ training points.
Using only the training data, we perform the above greedy forward-stepwise fitting
procedure. For values of $m \in [0, N_t]$, we obtain a least-squares fit with $m$
coefficients using the training data and evaluate the fit residuals
$r_j^{m, k}$ for the
validation data.
We choose
\begin{equation}
m^* = \argmin_m \sum_{k=1}^K \max_{j=1}^{N_v} \left( r_j^{m, k} \right)^2.
\end{equation}
We use the maximum over $j$ because we seek to minimize the largest
fit residuals, and we sum in quadrature over $k$ rather than maximize to
account for cases where data points with large errors or corner cases
are selected as validation points, which can lead to large fit residuals.
The dependence of the residuals on $m$
for one case is shown in
Fig.~\ref{fig:fit_appendix}.

\section{Comparing reduced basis constructions} \label{sec:RB_SVD}

We compare two commonly used methods to generate a reduced basis in gravitational waveform reduced-order modeling. 
The first uses a singular value decomposition (SVD) of a data set whose output consists of a set of basis vectors ranked by their ``singular values'', which are eigenvalues when the input data is square. 
The SVD reduced basis follows by truncating the output basis beyond a selected singular value. 
The resulting basis is accurate up to that singular value as measured in a root-mean-square norm.
The second method uses a greedy algorithm, which is iterative and nested, to expose the most relevant elements of the input (or {\it training}) data
set~\cite{Cormen:2009, Binev10convergencerates}. 
The greedy algorithm selects the element
with the largest current projection error (as measured by a specified norm), orthonormalizes the selected element
with respect to the current basis, and 
  adds this orthonormalized element to the set of basis vectors. 
In practice, one uses an iterated, modified Gram-Schmidt process~\cite{Hoffmann1989}
for orthonormalization, which is robust to the accumulation of numerical round-off effects from subtraction until very large basis sizes. 
The algorithm ends when the largest projection error is below a specified tolerance; it also ends if a previously-selected training data element is selected again, which, if it were allowed to occur, would introduce a linearly dependent element to the basis. 
The output includes a (greedy) reduced basis and a set of parameters or labels that indicate the most relevant elements of the training data from which the basis is built.

Both SVD and greedy methods output a reduced basis that accurately represents the 
training data to the requested singular value or tolerance.
  The output of the SVD algorithm depends only on the training data.  The
  greedy algorithm, on the other hand, begins by choosing one of the
  training data elements as the first basis vector, so its output
  depends also on that choice.
How that choice is made is often arbitrary and may depend on the application. 
For example, one may seed the greedy algorithm with an arbitrary element from the training set or choose the element that has the largest absolute value or norm. 
However, it has been shown that the choice of seed is largely irrelevant as the greedy algorithm seeks to minimize the maximum projection error across the entire training set, no matter what the seed. 
The resulting variations in the size of the greedy reduced basis due to arbitrary seed choices are marginal and typically span a few percent about the mean size \cite{Field:2011mf, Caudill2012, Galley:2016mvy}.

Practical implementations of the SVD algorithm can be found rather easily because of its broad use across many disciplines.
Therefore, building an SVD reduced basis for a training set of waveforms is as straightforward as calling the appropriate programmed function.
However, if the training data contains $N$ waveforms with $L$ time or frequency samples then the SVD algorithm is ${\cal O}(N^2 L)$, which can be intensive in both time and physical memory.
For this reason, the authors in \cite{Cannon:2010qh} divide the full training space into narrow strips in one
direction of the parameter space.
Dividing the training space into smaller subsets results in a direct product of reduced bases, one basis for each subset.
Unfortunately, the total number of the basis elements tends to be larger than if one had performed a SVD on the full training data (if it can be done).
Consequently, the reduction of the data is not maximized.

One often has considerable flexibility in designing a greedy algorithm for a specific application.
If the training set remains fixed throughout the course of the greedy algorithm
(see ~\cite{Blackman:2014maa} for an example where this is not the case)
then each iteration step can be performed in constant time so that the
totality scales as ${\cal O}(n N)$ if $n$ is the number of reduced basis
elements needed to reach the specified tolerance. 
Typically, $n \ll N$ so that greedy algorithms tend to terminate more quickly than an application of SVD on the same training data, though there is some additional influence from implementation details.
The greedy algorithm can be parallelized to break up the computation of expensive integrals across different processes \cite{GreedyCPP}.
In addition, the size and memory requirements of a very large training set pose little problem for greedy algorithms. 
The training space can be divided into subsets so that a reduced basis is built for each with a tolerance up to numerical round-off as measured in the $L_\infty$ norm (to have point-wise accuracy for the data). 
Then, 
one may apply a second greedy algorithm on the full training data by using instead the basis data on each subset to represent the original data of each subset. 
In this way, one can generate a reduced basis that spans all the subsets and maximizes the reduction of the full training set~\cite{Galley:unpub}.
Combining this two-step greedy algorithm with the parallelization of the projection integrals discussed above provides a viable and practical strategy for building a reduced basis for training sets of virtually any size. Another strategy is to randomly repopulate the training set at each iteration of the greedy algorithm~\cite{Hesthaven:2014, Blackman:2014maa}. This approach requires that the training data can be generated at will for any parameter values but also avoids storing prohibitively large amounts of data at any step in the greedy algorithm.

Finally, greedy algorithms allow one to use any measure for determining the projection errors.
This includes choosing among $L_2$, $L_\infty$, and $L_n$ error norms or any combination thereof. 
In addition, computing the integrals for projecting the training data onto the basis can be achieved with any quadrature rule one wishes.
However, implementations of the SVD algorithm are restricted to the $L_2$ measure and
 the reduced basis will depend on how the training data
is sampled in time or frequency.

Let us next investigate a toy problem to facilitate a comparison of the outputs of a basic greedy algorithm and SVD. 
We consider a function
\begin{align}
	X(t; \lambda) = 
			\sin (\lambda t)
			+ 10^{-5} \sin(10 \lambda t) 
			+ 10^{-10} \xi(t)
	\label{eq:function}
\end{align}
where $t \in [0,10]$ with a parameter $\lambda \in [1,20]$. 
There is a relatively high frequency component with an amplitude of $10^{-5}$.
The quantity $\xi(t)$ is a random variable drawn from a normal distribution with zero mean and variance of one.
This stochastic term has an amplitude of only $10^{-10}$. 

\begin{figure}
  \includegraphics[width=\linewidth]{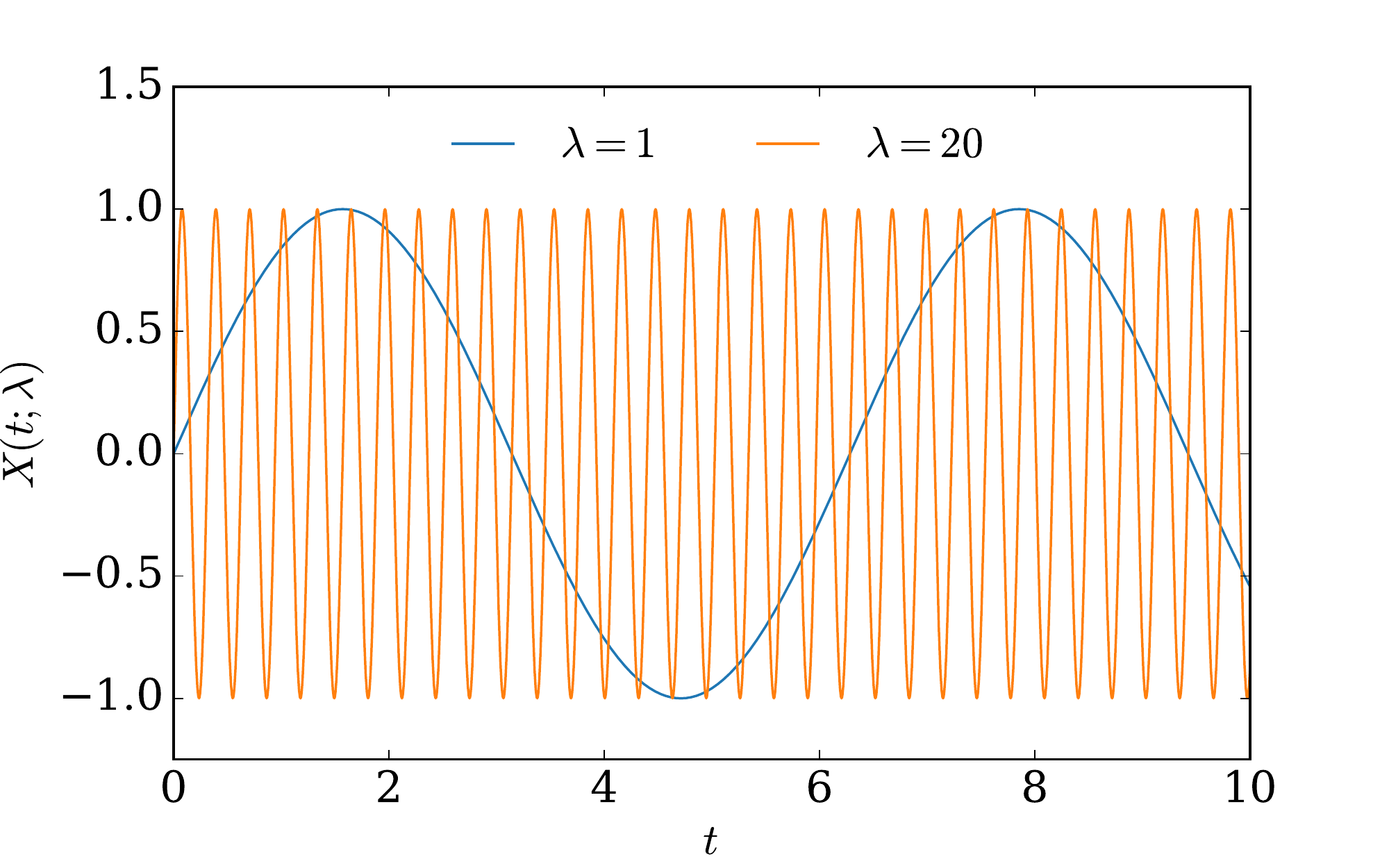}
  \caption{
    Plots of $X(t; \lambda)$ for our toy problem evaluated at the smallest
    and largest values of $\lambda$ in the training set.
  }
  \label{fig:basis_toy_solution}
\end{figure}

Our training set will consist of $N=1000$ uniformly spaced values of $\lambda$.
Figure~\ref{fig:basis_toy_solution} shows training data for the smallest and largest parameter values considered here.
We sample the function in~\eqref{eq:function} at $10,\!000$ uniformly spaced times.

\begin{figure}
  \includegraphics[width=\linewidth]{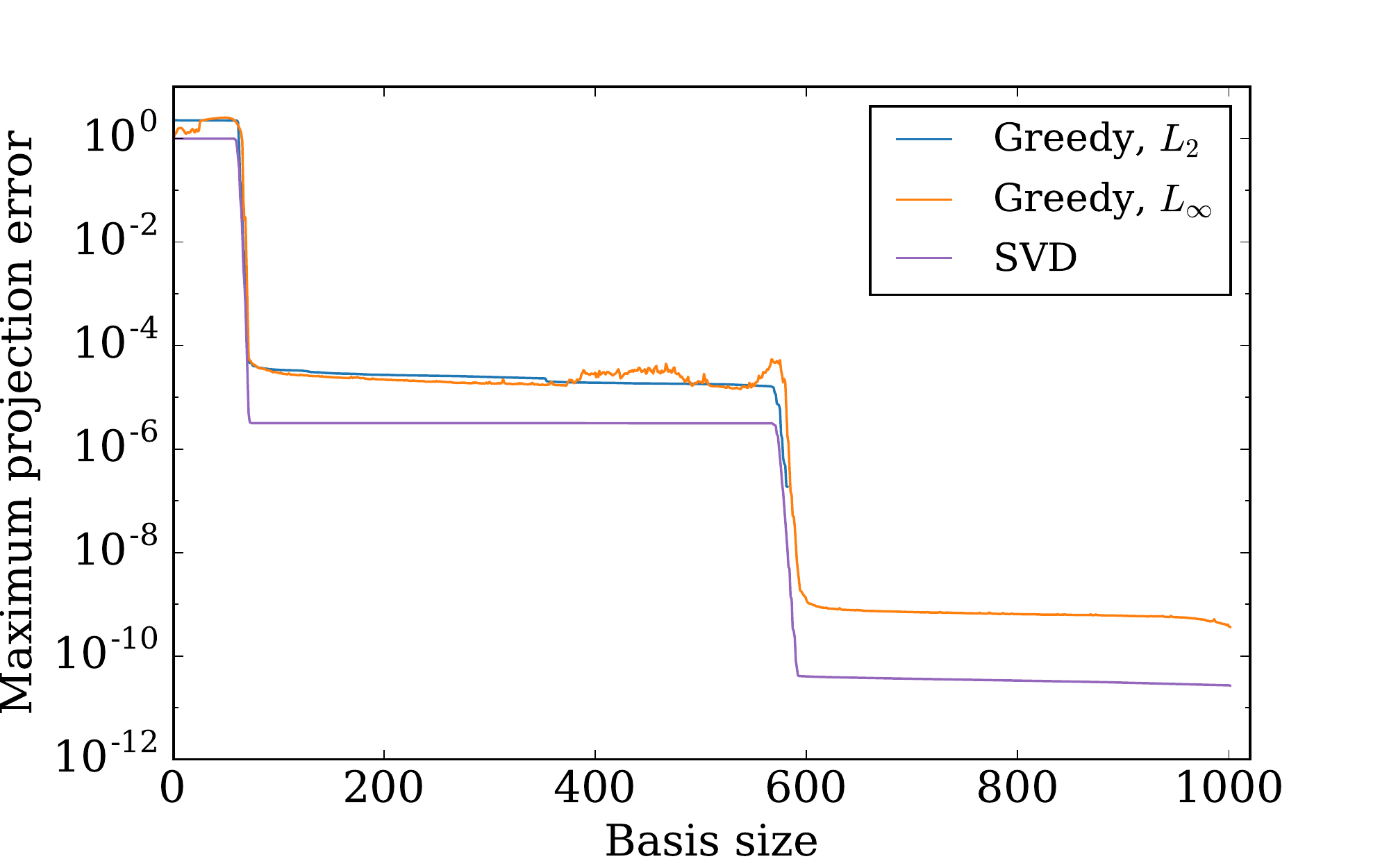}
  \caption{
  	Maximum projection errors of all three reduced bases (see text for a description) versus the size of the basis.
  }
  \label{fig:basis_toy_rb_estimate}
\end{figure}

We construct three reduced bases. 
The first is built from an SVD on the training data. 
The second uses a greedy algorithm to generate a reduced basis and a corresponding set of parameters; here we use the $L_2$ norm to measure the difference between each training set element and its projection onto the basis. 
The third is built in the same greedy
manner as the second but uses the $L_\infty$ norm to measure the projection error. 
Recall that the $L_2$ error constitutes a kind of average as it involves an integration in time whereas the $L_\infty$ error measures the largest, point-wise, absolute difference and is thus more stringent.
Figure~\ref{fig:basis_toy_rb_estimate} shows the maximum projection errors, as measured with their respective norms, associated with these three methods as a function of the size of the basis.
The absolute tolerance on the greedy algorithm bases is $10^{-14}$ while the smallest singular value kept is $10^{-14}$ relative to the largest.
We observe three plateaus for each of the cases, which can be attributed to each algorithm trying to resolve the features at the ${\cal O}(1)$, ${\cal O}(10^{-5})$, and ${\cal O}(10^{-10})$ scales in the data; see~\eqref{eq:function}. 
In fact, none of the algorithms are able to completely resolve the very low-amplitude stochastic features until the training set has been exhausted and all data has been used to build the reduced bases. 
Notice that the error curve is somewhat noisy for the $L_\infty$ case while the other two are smooth.
Also, the maximum projection error for the $L_2$ case ends at about $10^{-7}$ due to a parameter being selected a second time.

\begin{figure}
  \includegraphics[width=\linewidth]{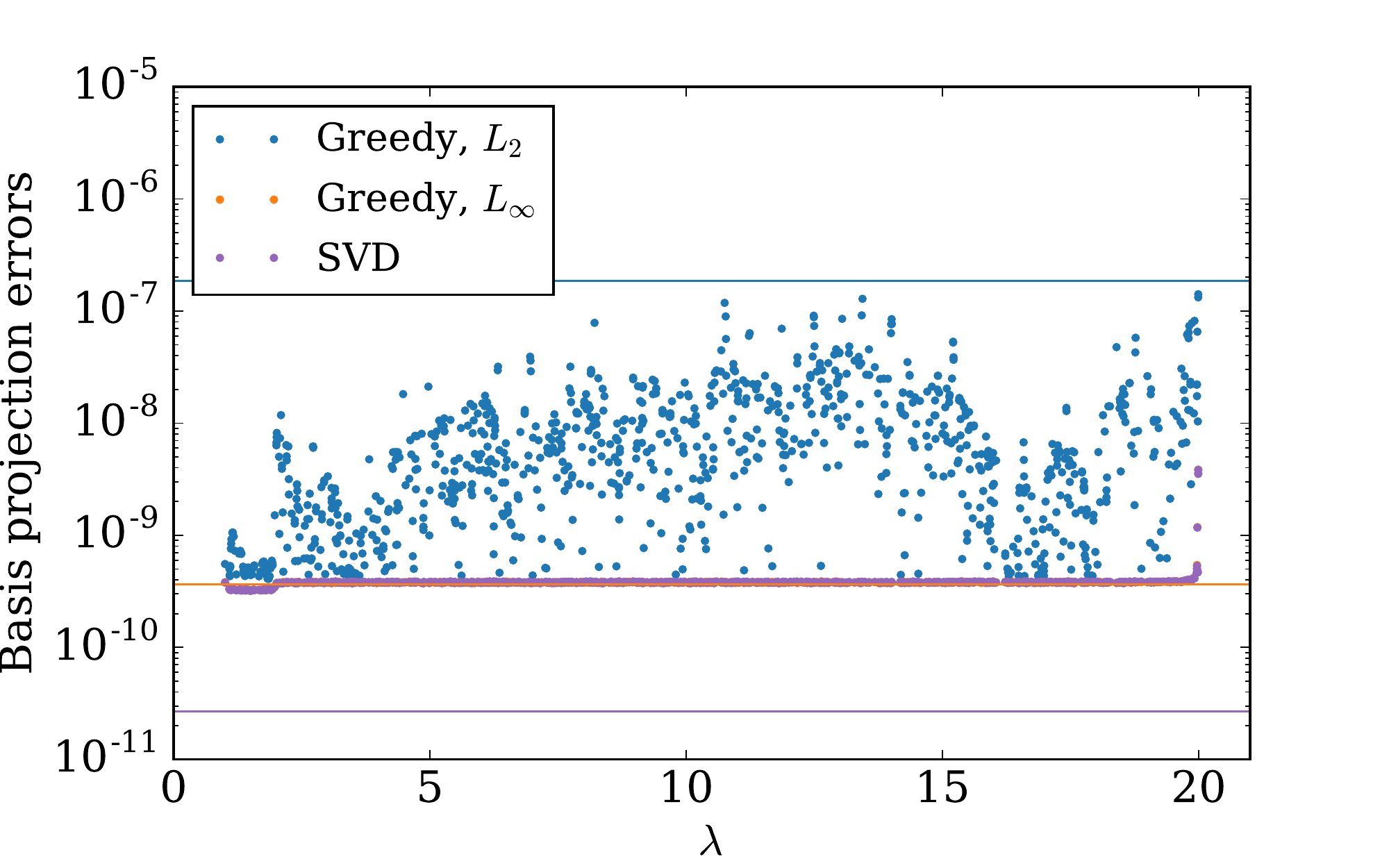}
  \caption{
  	Projection errors, measured in the $L_2$ norm for the three reduced bases described in the text, computed for test data generated from $1000$ randomly selected parameters $\lambda$ in $[1,20]$.
	The corresponding colored lines indicate the smallest projection errors on the training sets shown in Fig.~\ref{fig:basis_toy_rb_estimate}.
	The errors for ``Greedy, $L_\infty$'' and ``SVD'' lie nearly on top of each other.
	However, the maximum projection error implied by SVD (purple line) underestimates the true errors (dots) by an order of magnitude.
  }
  \label{fig:basis_toy_rb_errs}
\end{figure}

Figure~\ref{fig:basis_toy_rb_errs} shows the projection errors (as measured in the $L_2$ norm) onto each of the three reduced bases for test data generated by randomly selecting $1000$ values of $\lambda$ in the training interval $[1,20]$.  
The errors for ``Greedy, $L_\infty$'' and ``SVD'' lie nearly on top of each other while those for ``Greedy, $L_2$'' are relatively large because the effective greedy algorithm tolerance for this basis is only $10^{-7}$ as discussed above. 
In all cases, the small-amplitude stochastic noise in the data prevents the projection errors of the test data from being less than a few times $10^{-10}$; see~\eqref{eq:function}.

Finally, the SVD method is able to produce a reduced basis with
elements that smooth many uncorrelated features 
    manifest in the training data. Such smoothing is useful for
surrogate model building because the resulting basis elements tend to
exhibit smoother variation in time or frequency; this
translates into
smoother variations across parameters, thereby yielding more accurate
fits for the parametric variation at the empirical interpolation
nodes. The reduced bases produced by greedy methods tend to not to share this
smoothing ability of the SVD method.

\begin{figure}
  \includegraphics[width=\linewidth]{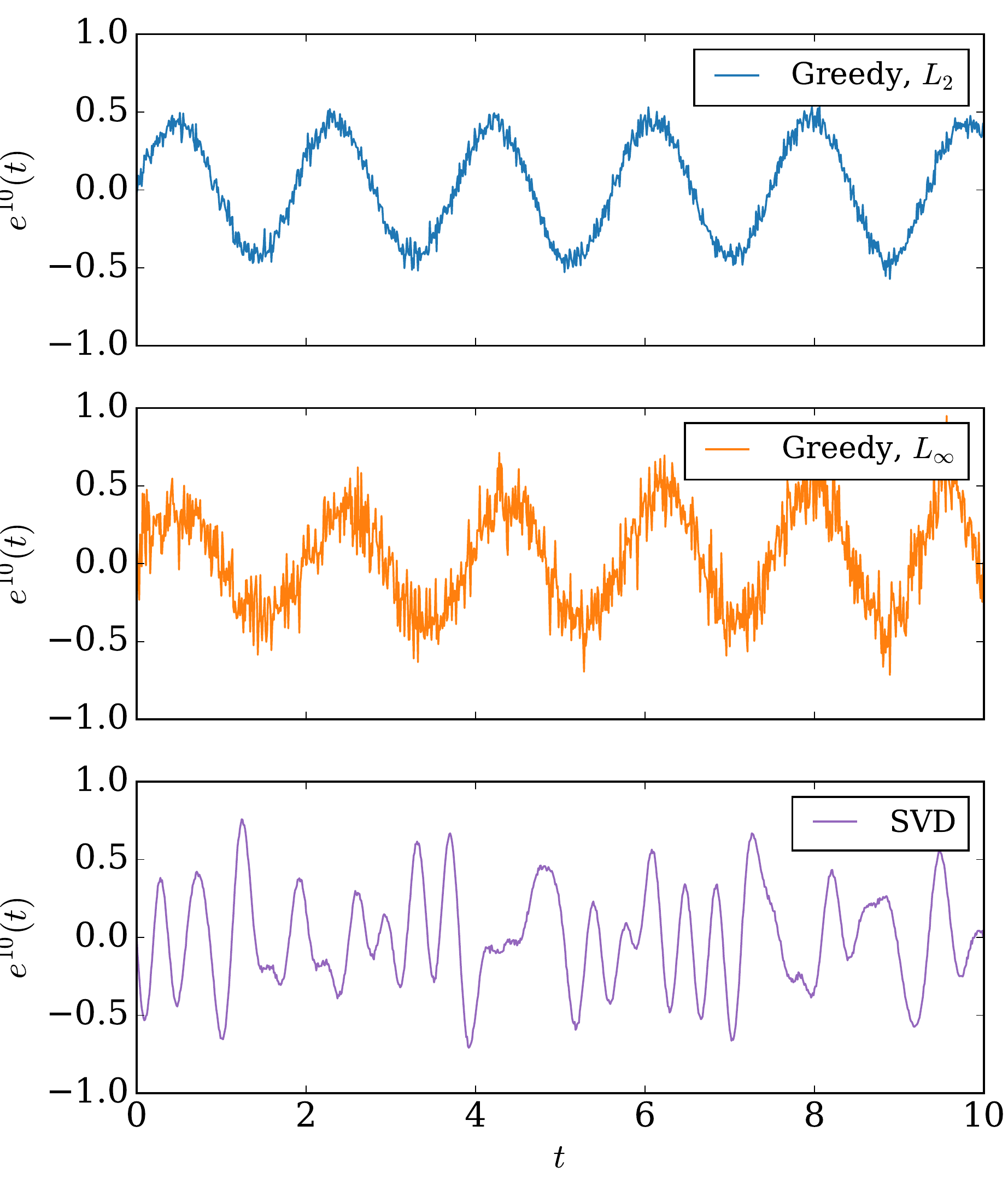}
  \caption{
    The tenth basis element as a function of $t$ from the three reduced bases elements 
    described in the text. The training data used is given by the parameterized function
    in \eqref{eq:function2} and exhibits relatively large amplitude fluctuations. 
    Whereas the top two plots show significant
    noise in the basis element, the SVD method smooths
    away, almost completely, the uncorrelated stochastic features to generate
    a basis element that is smooth in $t$.
  }
  \label{fig:basis_vecs}
\end{figure}

To demonstrate SVD's smoothing abilities, we replace the function in \eqref{eq:function} with a smooth oscillating term plus a stochastic term with amplitude of $10\%$ of the first so that the noise is visible to the naked eye,
\begin{equation}
	X(t; \lambda) = \sin (\lambda t) + 0.1 \, \xi(t) .
	\label{eq:function2}
\end{equation}
We build three reduced bases on the corresponding training sets (with
the same $t$ and $\lambda$ intervals and samples) using the same methods
as
before. Figure~\ref{fig:basis_vecs} shows the tenth basis element as a
function of $t$ for each of the three reduced basis building strategies. The two
bases built from a greedy method exhibit the noise found in the
training data. However, the SVD basis element in the bottom panel
reveals a smooth function with very low amplitude noise, much lower
than appears in the training data amplitudes.

In the case of the NRSur4d2s surrogate discussed here,
  note that data from each of the NR simulations contains spurious
  oscillations on
  the orbital timescale; these oscillations are caused by residual
  orbital eccentricity and by nutation effects that we have not filtered out
  (\S~\ref{sec:filtCoPrecessing}), and because these oscillations are
  uncorrelated from one simulation to another, they appear as stochastic
  noise.  To smooth this noise, we therefore use the SVD method
  to obtain basis vectors for emperical interpolation when
  building NRSur4d2s (\S~\ref{sec:empir-interp}). This smoothing significantly
  improves the accuracy of our fits of the waveform
  quantities at the empirical interpolation nodes.
However, note also that for NRSur4d2s we use the
greedy method to expose the BBH parameters for performing expensive NR
simulations (\S~\ref{sec:popul-train-set}).
Therefore, we use the benefits of both the greedy and SVD
methods in building NRSur4d2s.

\section{Motivating the use of $\mathcal{E}$} 
\label{app:TFErrors}

A commonly used measure of the difference between waveforms
$h_1(t, \theta_1, \phi_1; \pmb{\lambda}_1)$
and $h_2(t, \theta_2, \phi_2; \pmb{\lambda}_2)$ is the overlap error
\begin{equation}
1 - \mathcal{O} = 1 - \frac{\langle h_1, h_2 \rangle}{
        \sqrt{\langle h_1, h_1 \rangle
              \langle h_2, h_2 \rangle}},
\end{equation}
where $\langle \cdot , \cdot \rangle$ is often chosen
to be the frequency domain noise-weighted inner product~\cite{Cutler:1994ys}
\begin{equation}
\langle a, b \rangle_{f} = 4 \mathrm{Re} \int_0^\infty
        \frac{\tilde{a}(f) \tilde{b}^*(f)}{S_n(f)} df.
\end{equation}
Here $S_n(f)$ is the power spectral density of noise in a gravitational wave detector
and tildes are used to represent a Fourier transform.

If we use a flat (frequency-independent) power spectral density,
we may instead perform the integration in the time domain and use
\begin{equation}
\langle a, b \rangle_{t} = \mathrm{Re} \int_{t_\mathrm{min}}^{t_\mathrm{max}} a(t) b^*(t) dt
\end{equation}
to obtain the same overlap error.
While a completely flat power spectral density is unphysical, the design sensitivity
of aLIGO~\cite{LIGO-aLIGODesign-NoiseCurves} varies only by a factor of $\sim2$ between $50\mathrm{Hz}$ and
$1000\mathrm{Hz}$.
Putting rigorous limits on weighted frequency domain errors based on unweighted
time domain errors is not straightforward~\cite{Lindblom2008, Lindblom2009b},
but the time domain errors are computationally cheap to compute,
useful for quantifying time domain waveform models,
and (like NR waveforms and our surrogate model NRSur4d2s)
independent of the total binary mass $M$.

We can relate the time domain overlap error to $\delta h$ by performing a
weighted average over the sphere and using
\begin{equation}
\int_{S^2} a(\theta, \phi) b^*(\theta, \phi) d\Omega =
        \sum_{\ell, m} a^{\ell, m} b^{\ell, m *}
\end{equation}
due to the orthonomality of the SWSHs.
Using $\| a \|^2_t = \langle a, a\rangle_t$, we have
\begin{align}
\delta h^2  &= \frac{1}{T} \sum_{\ell, m} \| \delta h^{\ell, m} \|^2_t \\
    &= \frac{1}{T}\int_{S^2} \| h_1(t, \theta, \phi; \pmb{\lambda}_1) - 
                     h_2(t, \theta, \phi; \pmb{\lambda}_2) \|^2_t d\Omega\\
    &= \frac{1}{T}\int_{S^2} \left( \|h_1\|^2_t + \|h_2\|^2_t -
                        2\langle h_1 h_2 \rangle_t \right) d\Omega
\end{align}
where in the last line
we have omitted arguments to $h_1$ and $h_2$.
If $\|h_1(t, \theta, \phi; \pmb{\lambda}_1)\|_t = \|h_2(t, \theta, \phi; \pmb{\lambda}_2)\|_t$
for all $\theta, \phi$ then we would have
\begin{equation}
\frac{\delta h^2}{\sum_{\ell, m}\|h_1^{\ell, m}\|^2_t} =
        \frac{ 2 \int_{S^2} w(\theta, \phi) (1 - \mathcal{O}(\theta, \phi)) d\Omega}{
                 \int_{S^2} w(\theta, \phi) d\Omega}
\end{equation}
where $w(\theta, \phi) = \|h_{i}(t, \theta, \phi; \pmb{\lambda}_i)\|^2_t.$
Denoting $\|h\|^2 \equiv \sum_{\ell, m}\|h^{\ell, m}\|^2_t$,
this motivates the use
of the relative error measure
\begin{equation}
\mathcal{E} \equiv \frac{1}{2} \frac{\delta h^2}{\|h_1\|^2}
\end{equation}
as it is similar to a sphere-weighted average of overlap errors,
where the weighting emphasizes directions with a larger amount of
gravitational wave emission.
We note, however, that while the overlap error vanishes if $h_1$
and $h_2$ are identical except for normalization, $\mathcal{E}$ does not
and vanishes only when $h_1$ and $h_2$ are identical.
This is important as a different normalization will lead to a bias
when measuring the distance to the source of a gravitational wave.

\section{Mismatches optimized over time and polarization shifts}
\label{app:mismatches}
Given gravitational waveform polarization signals $h_+(t)$ and $h_\times(t)$,
each gravitational wave detector in a detector network will observe a linear
combination of $h_+(t)$ and $h_\times(t)$ depending on their orientation with
respect to the direction of propagation and polarization axes.
For the purposes of building gravitational wave models, we are interested
in the best case scenario when both polarizations are measured.
Including ``blind spots" in the detector network could lead to artificially
large relative errors, so we assume a network of two detectors where one
measures $h_+(t)$ and the other measures $h_\times(t)$.
Given model predictions $h_+^m(t)$ and $h_\times^m(t)$ for the two polarizations,
we compute the two-detector overlap
\begin{equation*}
\mathcal{O} = \frac{
    \langle h_+, h_+^m \rangle + \langle h_\times, h_\times^m \rangle}{
    \sqrt{
        \left(
            \langle h_+, h_+ \rangle +
            \langle h_\times, h_\times \rangle \right)
        \left(
            \langle h_+^m, h_+^m \rangle +
            \langle h_\times^m, h_\times^m \rangle \right)}}
\end{equation*}
with a real inner product given by
\begin{align}
\langle a, b \rangle &= \mathrm{Re}\left[\langle a, b \rangle_C\right] \\
\langle a, b \rangle_C &= \int \frac{\tilde{a}(f) \tilde{b}^*(f)}{S_n(|f|)} df.
\end{align}
As in Eq.~\ref{eq:match}, a tilde denotes a frequency domain signal,
which is computed
by using an FFT after tapering the ends of the time domain signal.
In this case, the complex inner product $\langle \cdot, \cdot \rangle_C$
is integrated over the negative and positive frequency intervals
$[-f_\mathrm{max}, -f_\mathrm{min}]$ and $[f_\mathrm{min}, f_\mathrm{max}]$
for some positive $f_\mathrm{min}$ and $f_\mathrm{max}$.
Note that for any two real functions $a(t)$ and $b(t)$,
we have
\begin{equation}
\tilde{a}(-f)\tilde{b}^*(-f) = \left(\tilde{a}(f)\tilde{b}^*(f)\right)^*
\end{equation}
and so $\langle a, b\rangle_C$ is real.

Defining complex gravitational wave signals
\begin{align}
h(t) &= h_+(t) - i h_\times(t) \\
h^m(t) &= h_+^m(t) - i h_\times^m(t),
\end{align}
we can compute a complex overlap
\begin{align*}
\mathcal{O}_C
    &= \frac{\langle h, h^m\rangle_C}{
        \sqrt{\langle h, h \rangle_C \langle h^m, h^m\rangle_C}}\\
    &= \frac{\langle h_+, h_+^m \rangle +
             \langle h_\times, h_\times^m \rangle +
             i\left(\langle h_+, h_\times^m\rangle -
                    \langle h_\times, h_+^m\rangle\right)
            }{\sqrt{
                \left(
                    \langle h_+, h_+ \rangle +
                    \langle h_\times, h_\times \rangle \right)
                \left(
                    \langle h_+^m, h_+^m \rangle +
                    \langle h_\times^m, h_\times^m \rangle \right)}}.
\end{align*}
Since the time domain polarization signals are all real,
we have
\begin{equation}
\mathcal{O} = \mathrm{Re}[\mathcal{O}_C].
\end{equation}

A polarization angle shift of $\psi$ and time shift of $\delta t$
in the model waveform
results in the transformations
\begin{align}
h^m(t) &\rightarrow h^m_T(t) = h^m(t + \delta t)e^{2i\psi},\\
\tilde{h}^m(f) &\rightarrow \tilde{h}^m_T(f) =
    \tilde{h}^m(f)e^{2i\psi}e^{2\pi i\delta t}
\end{align}
where $h^m_T$ is the transformed model waveform.
The overlap of the signal waveform with the transformed model waveform is then
\begin{align*}
&\mathcal{O}(\psi, \delta t)
    = \mathrm{Re}\left[\frac{\langle h, h^m_T\rangle_C}{
        \sqrt{\langle h, h\rangle_C \langle h^m_T, h^m_T\rangle_C}}\right]\\
    &= \mathrm{Re}\left[\frac{e^{-2i\psi}}{
        \sqrt{\langle h, h\rangle_C \langle h^m, h^m\rangle_C}}
            \int\frac{\tilde{h}(f)\tilde{h}^{m*}(f)}{S_n(|f|)}
            e^{-2i\pi\delta t} df \right].
\end{align*}
The above integral can be evaluated efficiently for many values of $\delta t$
using an FFT.
We can then compute the mismatch
\begin{equation}
\mathrm{mismatch} = 1 - \max_{\psi, \delta t}\mathcal{O}(\psi, \delta t)
\end{equation}
by taking the absolute value of the complex overlap for each $\delta t$
to maximize over $\psi$, and taking the maximum over all available values
of $\delta t$.
In practice, the true maximum over $\delta t$ will lie between available samples,
so we fit the overlap peak to a quadratic function in $\delta t$ using
the largest overlap sample and the neighboring value on either side.
We also pad with zeros before taking the FFT to obtain a finer sampling in
$\delta t$.

\section{Post-Newtonian surrogate waveform decomposition}
\label{app:pn_surrogates}
The second greedy algorithm described in Sec.~\ref{sec:greedyselection}
makes use of surrogate models of Post-Newtonian (PN) waveforms.
At each greedy step, a new PN surrogate model is built from PN waveforms
evaluated at the currently known greedy parameters $G$.
This surrogate is evaluated for each training point
$\pmb{\lambda} \in \mathcal{T}_\mathrm{TS}^i$ and the surrogate waveform is
compared to the actual PN waveform.
Here, we describe the differences between how the PN surrogates were built
compared to the NR surrogate NRSur4d2s described in the main body.

PN waveforms do not contain a merger phase, so we cannot use the peak
amplitude to align the waveforms in time.
We instead choose $t=0$ to correspond to an
orbital angular frequency of $0.09$.
This frequency is computed from the waveform~\cite{Boyle:2013a}.
We choose $t_\mathrm{min}=-5000M$, $t_0=-4500M$, $t_f=-100M$, and $t_\mathrm{max}=0$.
The PN waveforms used to build the PN surrogate then have domain
$t \in [-5000M, 0]$, and the PN surrogate waveforms
have domain
$t \in [-4500M, -100M]$.
The parameters of the PN waveforms are given at $t=t_0$.
The rotation alignment at $t=t_0$ is the same as for the NR waveforms,
described in Sec.~\ref{sec:alignment}.

The waveform decomposition used for the PN surrogates was slightly different
from the one described in Sec.~\ref{sec:decomposition}.
We limited the PN waveforms to contain only the $\ell=2$ modes
(with all $5$ values of $m$).
Additionally, since we were able to obtain the
desired values of $\phi_\chi$ at $t=t_0$
with PN waveforms, there was no need to make any transformations
related to $\phi_\chi$.

The number of coefficients used in the parametric fits of the empirical nodes
was determined differently for PN surrogates than for NRSur4d2s.
Instead of the cross-validation
method described in Appendix~\ref{sec:fitappendix},
coefficients were added until the fit residuals fell below a specified tolerance,
given in Table~\ref{tab:pnFitTols}.
To prevent overfitting, the number of fit coefficients was also limited to
be at most $75\%$ of the number of data points used in the fit.
The basis functions in $|\chi_1|$ used for the fits were also different, with
$B_2^k(|\chi_1|) = T_k(2.5|\chi_1| - 1)$.
For the PN surrogates, we did not make the restriction $\alpha^3 \leq \alpha^2$
so $\theta_\chi$ affected the PN surrogate output when
$|\vec{\chi}_1|=0$.

\begin{table}
\begin{tabular}{c | c || c | c || c | c || c | c}
Data & Tol & Data & Tol & Data & Tol & Data & Tol \\
\hline
$\varphi_p$ & $0.01$ & $\varphi_d$ & $0.1$ &
    $\varphi_-^{2, 2}$ & $0.01$ & $\varphi_-^{2, 1}$ & $0.1$ \\
$\varphi[H[X]]$ & $0.1$ & $|H[\varphi_+^{2, 2}]|$ & $0.0001$ &
    $|H[\varphi_+^{2, 1}]|$ & $0.0001$ & & \\
\end{tabular}
\caption{Fit tolerance for the empirical node parametric fits of PN surrogates.
         Fit coefficients were added until the maximum fit residual fell
         below the tolerance. A tolerance of $0.001$ was used for unlisted
         waveform data pieces.}
\label{tab:pnFitTols}
\end{table}

\section*{References}
\bibliography{../References/References}

\end{document}